%% file: arxiv_main.tex
\documentclass[journal=jctcce,manuscript=article]{achemso}

\usepackage{amsmath, amssymb, amsfonts}
\usepackage{array, booktabs, tabularx}
\usepackage{xcolor, graphicx, tikz}
\usepackage{subcaption}
\usepackage{hyperref}  
\usepackage{braket}
\usepackage{bm}
\usepackage{cancel}
\usepackage[ruled]{algorithm2e}
\usepackage[normalem]{ulem} 
\usepackage{qcircuit}

\hypersetup{colorlinks = True, allcolors = {blue}}

\def\kVacP{ \ket{-}_F }
\def\kVacF{ \ket{0, \cdots, N - 1}_F }
\def\kVacB{ \ket{0, \cdots, 0}_B }
\newcommand{\KetBra}[1]{\ket{#1} \bra{#1}}
\newcommand{\kIndF}[2]{ \ket{ {#1}_1, \cdots, {#1}_{#2} }_F}
\newcommand{\kIndB}[2]{ \ket{ {#1}_1, \cdots, {#1}_{#2} }_B}

\newcommand{\FC}[1]{ f_{#1}^\dagger }
\newcommand{\FA}[1]{ f_{#1} }
\newcommand{\BiF}[2]{ E_{#2}^{#1} }

\newcommand{\BC}[1]{ \hat{b}_{#1}^\dagger }
\newcommand{\BA}[1]{ \hat{b}_{#1} }
\newcommand{\BN}[1]{ \hat{n}_{#1} }
\newcommand{\BX}[1]{ \hat{q}_{#1} }
\newcommand{\BP}[1]{ \hat{p}_{#1} }
\newcommand{\NBC}[1]{ \sigma_{#1}^\dagger }
\newcommand{\NBA}[1]{ \sigma_{#1} }
\newcommand{\BigP}[1]{ \mathcal{P}_{#1} }
\newcommand{\BigC}[1]{ \mathcal{C}_{#1} }
\newcommand{\BigT}[2]{ \mathcal{T}_{#2}^{#1} }
\newcommand{\DCoeff}[1]{ d_{#1} }

\newcommand{\UniAn}[1]{ U_A^{(#1)} }
\newcommand{\UniGate}[2]{ U_{#2}^{(#1)} }
\def\UniECD{ U_{ER} }
\def\UniSNAP{ U_{SD} }

\newcommand{\Pauli}[2]{ \mathbb{P}_{#2}^{(#1)} }
\newcommand{\QubitOp}[2]{ W_{#2}^{(#1)} }
\newcommand{\SX}[1]{ \sigma_x^{#1} }
\newcommand{\SY}[1]{ \sigma_y^{#1} }
\newcommand{\SZ}[1]{ \sigma_z^{#1} }

\def\HElec{ H_{\text{elec}} }
\def\HamF{ H_F }
\def\HamB{ H_B }
\def\HamQ{ H_Q }

\newcommand{\HOne}[2]{ h_{#2}^{#1} }
\newcommand{\HTwo}[2]{ v_{#2}^{#1} }
\newcommand{\HATwo}[2]{ \tau_{#2}^{#1} }
\newcommand{\HBCoeff}[1]{ g_{#1} }
\newcommand{\HQCoeff}[1]{ g_{#1} }

\def\EYE{ \mathbb{I} }
\def\HermConj{\text{h.c.}}
\newcommand{\ComCom}[1]{\mathcal{O}(#1)}
\newcommand{\BigEYE}[1]{ \mathcal{I}_{#1} }

\newcommand{\Eq}[1]{Eq.~({#1})}
\newcommand{\Fig}[1]{Figure~{#1}}

\newcommand{\Sec}[1]{Section~{#1}}
\newcommand{\Table}[1]{Table~{#1}}
\newcommand{\Reference}[1]{Ref.~{#1}} 
\newcommand{\Appx}[1]{Appendix~{#1}}
\def\SuppInfo{\text{ Supplementary Information }}
\def\MainText{\text{Main Text}~}

\newcommand{\st}[1]{\ifmmode\text{\sout{\ensuremath{#1}}}\else\sout{#1}\fi}

\title{Simulating electronic structure on bosonic quantum computers}

\author{Rishab Dutta}
\affiliation{Department of Chemistry, Yale University, New Haven, CT, USA 06520}

\author{Nam P. Vu}
\affiliation{Department of Chemistry, Yale University, New Haven, CT, USA 06520}
\alsoaffiliation{Department of Chemistry, Lafayette College, Easton, PA, USA 18042}
\alsoaffiliation{Department of Electrical and Computer Engineering, Lafayette College, Easton, PA, USA 18042}

\author{Chuzhi Xu}
\affiliation{Department of Chemistry, Yale University, New Haven, CT, USA 06520}

\author{Delmar G. A. Cabral}
\affiliation{Department of Chemistry, Yale University, New Haven, CT, USA 06520}

\author{Ningyi Lyu}
\affiliation{Department of Chemistry, Yale University, New Haven, CT, USA 06520}

\author{Alexander V. Soudackov}
\affiliation{Department of Chemistry, Yale University, New Haven, CT, USA 06520}

\author{Xiaohan Dan}
\affiliation{Department of Chemistry, Yale University, New Haven, CT, USA 06520}

\author{Haote Li}
\affiliation{Department of Chemistry, Yale University, New Haven, CT, USA 06520}

\author{Chen Wang}
\affiliation{Department of Physics, University of Massachusetts-Amherst, Amherst, MA, USA 01003}

\author{Victor S. Batista}
\affiliation{Department of Chemistry, Yale University, New Haven, CT, USA 06520}
\alsoaffiliation{Yale Quantum Institute, Yale University, New Haven, CT, USA 06511}
\email{victor.batista@yale.edu}

\begin{document}


\begin{abstract}

Quantum harmonic oscillators, or qumodes, provide a promising and versatile framework for quantum computing. 
Unlike qubits, which are limited to two discrete levels, qumodes have an infinite-dimensional Hilbert space, making them well-suited for a wide range of quantum simulations. 
In this work, we focus on the molecular electronic structure problem. 
We propose an approach to map the electronic Hamiltonian into a qumode bosonic problem that can be solved on bosonic quantum devices using the variational quantum eigensolver (VQE). 
Our approach is demonstrated through the computation of ground potential energy surfaces for benchmark model systems, including \( \text{H}_2 \) and the linear \( \text{H}_4 \) molecule. 
The preparation of trial qumode states and the computation of expectation values leverage universal ansatzes based on the echoed conditional displacement (ECD), or the selective number-dependent arbitrary phase (SNAP) operations. 
These techniques are compatible with circuit quantum electrodynamics (cQED) platforms, where microwave resonators coupled to superconducting transmon qubits can offer an efficient hardware realization. 
This work establishes a new pathway for simulating many-fermion systems, highlighting the potential of hybrid qubit-qumode quantum devices in advancing quantum computational chemistry.

\end{abstract}


\section{Introduction} \label{sec: intro}

Understanding the ground and excited states of many-fermion systems is one of the fundamental problems in chemistry and physics.
Accurate simulation of molecular electronic structure, a many-fermion problem, is critical in understanding chemical reaction mechanisms or designing new molecules and materials with novel properties.
Classical computers are fundamentally restricted in simulating exact molecular electronic structure problems beyond a certain system size, \cite{Vogiatzis2017} and approximate classical computing methods fail to simulate a range of electronic structure systems with strong electron correlation. \cite{Chan2024spiers}
The recent interest in developing algorithms based on quantum computers can potentially address this issue.

The current era of noisy intermediate scale quantum (NISQ) computers relies on the quantum information unit known as qubits which are two-level quantum systems.
NISQ computers have inherent limitations due to the decoherence associated with qubits and the quantum operators acting on them. 
Nevertheless, several hybrid quantum-classical algorithms have been developed to simulate molecular electronic structure, that combines resources from both classical and quantum devices. \cite{Peruzzo2014,Grimsley2019,Mcardle2019,Motta2020,Smart2021,Kyaw2023} 
One of the steps in all these algorithms involves mapping the fermionic Hamiltonian of the molecule of interest to a qubit Hamiltonian. \cite{Whitfield2011,Seeley2012}

The development of bosonic quantum devices introduces a fundamentally novel approach to quantum computing.
Bosonic quantum computing can be conceptually understood as computations with quantum harmonic oscillators (QHOs), also known as \textit{qumodes}, instead of qubits. 
Qumodes can store quantum information in the unbounded Hilbert space of QHOs and naturally support continuous variable bases due to the position and momentum operators associated with oscillator modes.
A range of applications has been demonstrated using bosonic quantum devices for chemistry, \cite{Dutta2024perspective} including simulation of molecular vibronic spectra, \cite{Huh2015,Wang2020,Malpathak2024simulating} 
understanding conical intersections, \cite{Wang2023} 
and implementing quantum dynamics for modeling chemical processes. \cite{Lyu2023,Cabral2024roadmap}

Qumodes can be realized with different hardware approaches, \cite{Copetudo2024} including but not limited to electromagnetic fields inside resonators, \cite{Deleglise2008,Hacker2019} 
and the motion of trapped ions. \cite{Bruzewicz2019,Araz2024toward}
A promising and rapidly evolving hardware platform for realizing bosonic quantum computation is the circuit quantum electrodynamics (cQED) approach. \cite{Joshi2021,Blais2021,Liu2024qumodequbitreview,Crane2024hybrid} 
The cQED hardware comprises microwave resonators as the qumodes and superconducting qubits based on Josephson junctions acting as the non-linear element that controls and measures the quantum information.
Bosonic cQED devices with 3D resonator geometries can have a photon lifetime of up to two seconds. \cite{Romanenko2020}
Conceptually, the cQED devices are a hybrid qubit-qumode hardware approach that promises a new quantum computing paradigm. \cite{Liu2024qumodequbitreview} 

Quantum algorithms for molecular electronic structure tailored for qubits, however, can not be trivially applied to qumode hardware due to the fundamental difference between qubits which are spin-$1/2$ systems, and qumodes which are bosonic.
An important step in simulating molecular electronic structure on bosonic quantum computers would be to map the corresponding fermionic Hamiltonian to a bosonic one. 
There has been substantial past work on fermion to boson mapping, including exact and approximate transformations. 
\cite{RingBook,Garbaczewski1975,Garbaczewski1978,Klein1991,Ginocchio1996,VonDelft1998,Scuseria2013,Liu2016,Montoya2018}
An exact state mapping between fermionic Slater determinants and bosonic Fock states of QHOs was established by Ohta based on the fact that particle-hole excitations from the Fermi vacuum can be represented as photon transitions. \cite{Ohta1998}
An exact operator mapping between a number-conserving bilinear fermionic operator and oscillator projection operators can be derived from this state mapping, as shown by Dhar, Mandal, and Suryanarayana, \cite{Dhar2006} which we will call the \textit{direct} mapping.
Although the direct mapping is conceptually appealing, it may lead to an impractical number of bosonic operator terms in the mapped Hamiltonian for larger electronic systems.


\begin{figure}[b!]
    \centering
    \begin{subfigure}[c]{0.45\textwidth}
    
        \centering
        \includegraphics[width=1.0\linewidth]{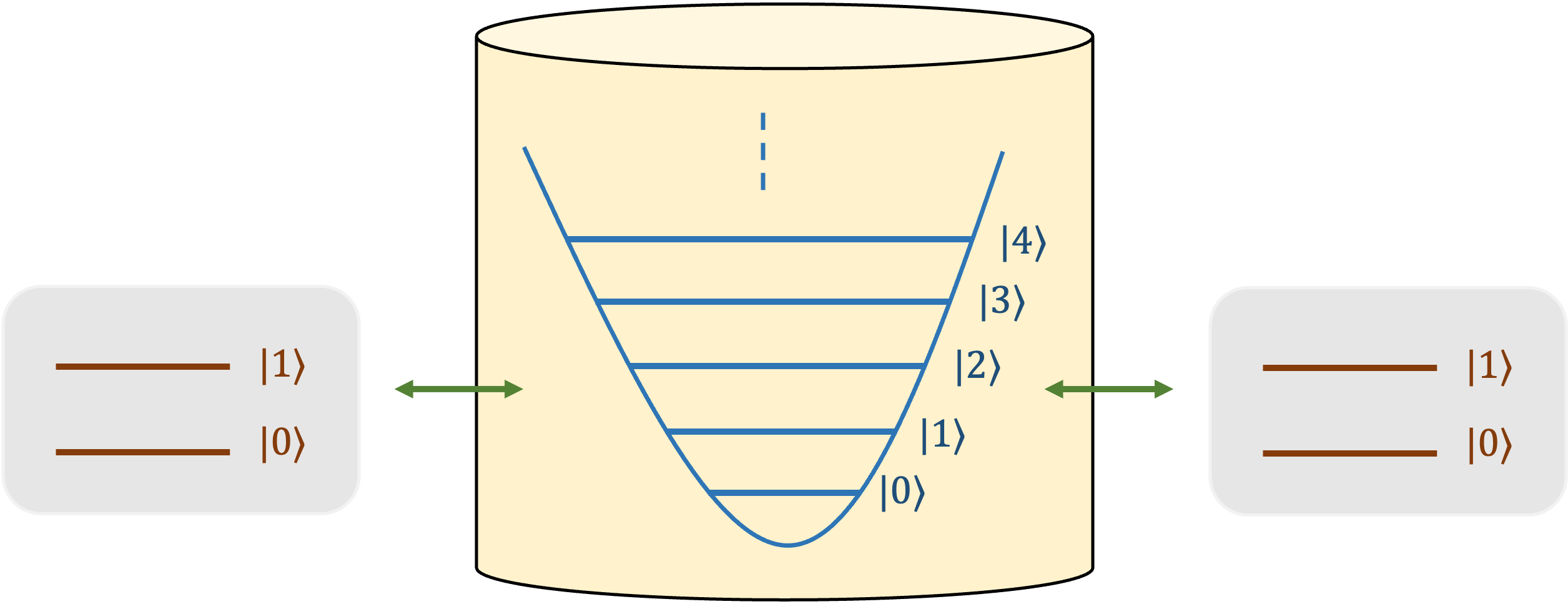}
        
    \end{subfigure}
    \hskip1ex
    \begin{subfigure}[c]{0.45\textwidth}
    
        \centering
        \includegraphics[width=0.75\linewidth]{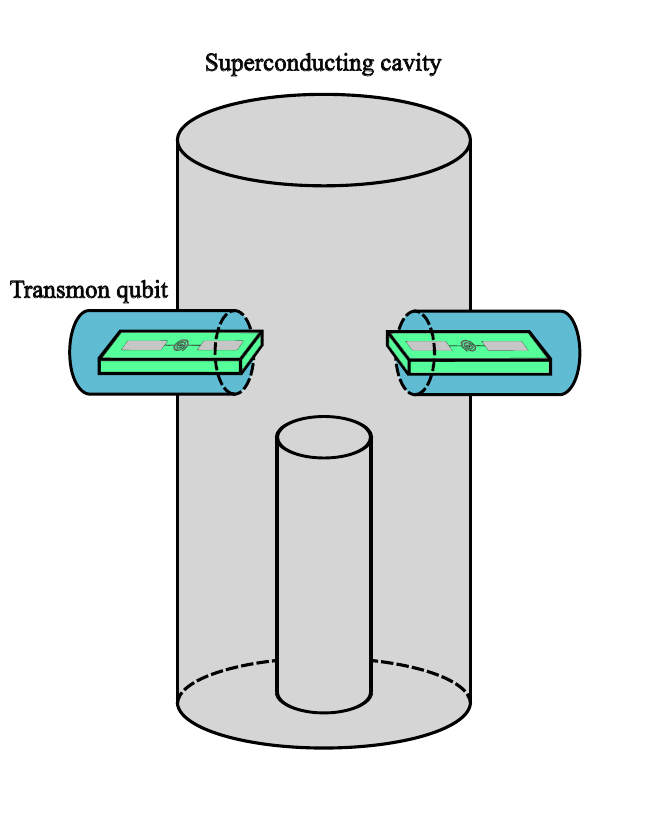}
        
    \end{subfigure}
    
    \caption{
     Illustration of a quantum device consisting of one qumode and two qubits. 
     (Left) A conceptual illustration of a qumode coupled with two qubits. 
     (Right) A schematic of the quantum hardware where two transmon qubits are connected to a microwave resonator that acts as the qumode in the cQED approach. 
     }
     \label{fig: qubits_qumode}
\end{figure}


In this work, we introduce a \textit{qubit-assisted} fermion to boson mapping, where the fermionic operators will be first mapped to qubit operators followed by a qubit to qumode mapping.
We will show how the direct and qubit-assisted fermion to boson mappings can be applied to transform the molecular electronic structure Hamiltonian to a system of qumodes.
To the best of our knowledge, this is the first time a molecular electronic structure Hamiltonian has been simulated as a bosonic system, which allows us to develop bosonic variational quantum eigensolver (VQE) algorithms for finding the ground state of the molecular electronic system using trial states generated by a qubit-qumode device. 
Specifically, the bosonic VQE methods can take advantage of the unique universal gate sets native to the hybrid qubit-qumode device, such as the echoed conditional displacement (ECD) gates combined with qubit rotations or the selective number-dependent arbitrary phase (SNAP) combined with qumode displacement gates. 
We apply the resulting ECD-VQE and SNAP-VQE approaches to show that the expectation value of a Hamiltonian of four qubits representing the electronic structure of the dihydrogen molecule can be computed using quantum hardware consisting of two transmon qubits and one microwave resonator acting as a qumode, as illustrated in \Fig{\ref{fig: qubits_qumode}}, while the trial energy is optimized on a classical computer.
We also generalize our approach for larger systems where many qubits can be mapped to a few qumodes with the operations modularized and optimized for hardware efficiency. 
We illustrate our multi-qumode generalization on 
the ground state of the linear H$_4$ molecule to exemplify how our proposed method can outperform traditional qubit-based variational approaches. 
Although this work focuses specifically on the molecular electronic structure Hamiltonian, the techniques presented here can be applied to any many-qubit and many-fermion Hamiltonians such as systems studied in condensed matter \cite{Qin2022} or nuclear physics. \cite{RingBook}


\section{Background}

The electronic Hamiltonian under the Born--Oppenheimer approximation can be represented in the second quantization form as \cite{SzaboBook,HelgakerBook} 
\begin{equation} \label{eq: molecular_ham}
\HElec
= \sum_{pq} \HOne{p}{q} \: \FC{p} \FA{q}
+ \frac{1}{2} \: \sum_{pqrs} 
\HTwo{pq}{rs} \: \FC{p} \FC{q} \FA{r} \FA{s},
\end{equation}
where $p, q, r, s$ indices represent molecular spin-orbitals $\{ \chi_p \}$ and $\{ \FC{p}, \FA{q} \}$ are the corresponding fermionic creation and annihilation operators, respectively. 
The scalars $ \{ \HOne{p}{q} \} $ and $ \{ \HTwo{pq}{rs} \} $ are the one-electron and the two-electron integrals, 
\begin{subequations} \label{eq: electron_integrals}
\begin{align}
\HOne{p}{q}
&= \int d \textbf{x} \: \chi_p^* (\textbf{x}) \: \Big( - \frac{1}{2} \nabla^2 
-  \sum_A \: \frac{ Z_A }{r_A } \Big) \: \chi_q (\textbf{x}),   
\\
\HTwo{pq}{rs}
&= \int d \textbf{x}_1 \: d \textbf{x}_2 \:
\frac{ \chi_p^* (\textbf{x}_1) \chi_q^* (\textbf{x}_2) \: \chi_r (\textbf{x}_2) \chi_s (\textbf{x}_1) }{ r_{12} },
\end{align}
\end{subequations}
where $\nabla$ is the Laplacian operator representing differentiation with respect to the coordinates of each electron,
$Z_A$ is the nuclear charge of the $A\textsuperscript{th}$ nucleus,
$ r_{A} = | \mathbf{r} - \mathbf{R}_A | $ is the distance between an electron and $A\textsuperscript{th}$ nucleus,  
$ r_{12} = | \mathbf{r}_1 - \mathbf{r}_2 | $ is the distance between two electrons.
These elementary fermionic operators follow the canonical anticommutation relation (CAR) 
\begin{subequations} \label{eq: car}
\begin{align}
\{ \FC{p}, \FC{q} \} 
&= \FC{p} \FC{q} + \FC{q} \FC{p} 
= 0, \quad  
\\
\{ \FA{p}, \FC{q} \} 
&= \FA{p} \FC{q} + \FC{q} \FA{p} 
= \delta_{pq}, 
\end{align}
\end{subequations} 
where the fermionic mode indices span the $M$ number of spin-orbital functions $\{ \chi_p \}$. 
The Pauli exclusion principle is then equivalent to the relation
$ ( \FC{p} )^2 = 0 $, which is simply a consequence of the CAR in \Eq{\ref{eq: car}}. 

We assume the number of spin-orbitals $M$ to be an even integer since there is an underlying $M / 2$ number of spatial functions $\{ \phi_p (\mathbf{r}) \}$ which can associate with either up-spin $\alpha (\omega)$ or down-spin $\beta (\omega)$ functions.
Thus, $N$ electrons in $M$ molecular spin-orbitals give rise to ${M \choose N}$ number of many-electron basis states, each of which is an antisymmetrized product state called the Slater determinants, defined as
\begin{equation} \label{eq: free_fermion_state}
\kIndF{p}{N} 
\equiv \FC{p_1} \cdots \FC{p_N} \kVacP,
\end{equation}
where $\kVacP$ is the physical vacuum representing the state with $N = 0$ electrons and any $\FA{p} \kVacP = 0$. 
We provide further context to the electronic structure problem within the Supplementary Information. 
We note that the combinatorial number of many-electrons basis states scales exponentially with the problem size and thus proves challenging to enumerate and perform operations on. 
This is where quantum computer promises to be useful, as it can potentially address the problem of finding stationary states and energies at a reduced computational cost.


\section{Qubit-qumode circuits} \label{sec: mapping_ecd}

It is possible to transform the electronic Hamiltonian of \Eq{\ref{eq: molecular_ham}} 
to a bosonic Hamiltonian with an algebraic map of 
$ \BiF{p}{q} = \FC{p} \FA{q} $ operators to bosonic operators based on its Fock basis. 
However, it may not be an efficient approach for mapping systems with a large number of electrons. 
We present this direct mapping in \Appx{\ref{app: direct_mapping}}. 
In this section, we focus on a more systematic approach to transform the electronic Hamiltonian in terms of a linear combination of qubit-qumode operators, with the help of a fermion to qubit mapping. 
We discuss this qubit-assisted mapping for the rest of this article. 

\subsection{Fermion to qubit mapping}

Let us review the basic concepts related to qubits which are two-level quantum systems. 
The elementary one-qubit operators are the Pauli matrices 
\begin{equation}
X 
= \begin{pmatrix}
0 & 1 \\ 1 & 0
\end{pmatrix}, \quad
Y 
= \begin{pmatrix}
0 & -i \\ i & 0
\end{pmatrix} , \quad
Z 
= \begin{pmatrix}
1 & 0 \\ 0 & -1
\end{pmatrix},
\end{equation}
which can be generalized to multi-qubit operators by taking tensor products of Pauli matrices (also known as a Pauli word), e.g., 
$ X_1 \: Y_2 \: Z_3 = X \otimes Y \otimes Z $ represents a three-qubit operator.
We will also use $\{ \sigma_j \}$ to denote Pauli matrices where $j \in \{ x, y, z\}$ represent $X, Y$ and $Z$ matrices.
Any qubit Hamiltonian can be represented as a linear combination of Pauli words  
\begin{equation} \label{eq: qubit_ham}
\HamQ 
= \sum_{\mu = 1}^{N_H} 
\Big( \HQCoeff{\mu} \: 
\bigotimes_{p = 1}^{N_Q} \: \sigma_{j_p} 
\Big)
= \sum_{\mu = 1}^{N_H} \: \HQCoeff{\mu} \: \Pauli{N_Q}{\mu} 
\end{equation}
where $\{ \HQCoeff{\mu} \}$ are scalar Hamiltonian coefficients and $N_Q$ is the number of qubits. 
The number of terms $N_H$ is usually a computationally manageable integer for a physical Hamiltonian.

The molecular electronic Hamiltonian of 
\Eq{\ref{eq: molecular_ham}} 
can be transformed to a qubit Hamiltonian of the form in \Eq{\ref{eq: qubit_ham}} by applying a fermion to qubit mapping. 
There are many fermion to qubit maps that have been explored recently.  
For example, the Jordan--Wigner transformation (JWT) \cite{Jordan1935} maps the fermionic creation and annihilation operators to the following Pauli words
\begin{subequations}
\begin{align}
\FC{p}
&\mapsto \frac{1}{2} \: ( X_p - i Y_p ) \: 
\bigotimes_{q < p} \: Z_q,
\\
\FA{p}
&\mapsto \frac{1}{2} \: ( X_p + i Y_p ) \: 
\bigotimes_{q < p} \: Z_q,
\end{align}    
\end{subequations}
where the qubit indices represent the spin-orbital indices of the fermionic operators. 
This means $\HElec$ is transformed by JWT to a qubit Hamiltonian of \Eq{\ref{eq: qubit_ham}} with $N_H = \ComCom{M^4}$ and $N_Q = M$, where $M$ is the number of spin orbitals. 
We will focus on JWT in this paper, but we note that recent developments on fermion to qubit maps that go beyond JWT can reduce the scaling of both $N_H$ and $N_Q$ in terms of the number of spin orbitals. \cite{Liu2024fermihedral,Babbush2018}

\subsection{Qubit to qumode mapping} \label{sec: qubit_to_qumode_ecd}

Usually, some of the $\HQCoeff{j}$ coefficients for different $ \Pauli{N_Q}{j} $ operators in \Eq{\ref{eq: qubit_ham}} may be the same, in which case these operators can be grouped and $\HElec$ can be written as
\begin{equation} \label{eq: qubit_ham_grouped}
\HamQ 
= \sum_{\mu = 1}^{N_H'} \: \HQCoeff{\mu} \: \QubitOp{N_Q}{\mu},
\end{equation}
where $\{ \QubitOp{N_Q}{\mu} \}$ are either a single or a summation of $N_Q$-qubit Pauli words and $N_H' < N_H$.
Each $\QubitOp{N_Q}{\mu}$ operator can be represented in the computational basis by a matrix of dimensions $2^{N_Q} \times 2^{N_Q}$. 
We want to map the multi-qubit $\QubitOp{N_Q}{\mu}$ operator to a parameterized qumode operator with the help of an ancilla qubit, as implementable in a cQED transmon-resonator device. 
Since the target multi-qubit operator can be arbitrary, the implemented qumode operator must be represented by a set of universal qubit-qumode gates.

\begin{figure}[t]

\includegraphics[width=0.9\columnwidth]{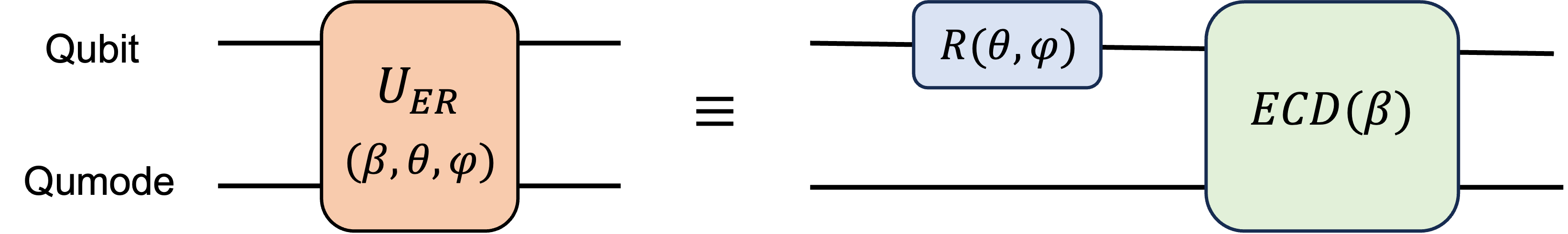}

\caption{
    Qubit-qumode gate consisting of one qubit rotation as defined in \Eq{\ref{eq: qubit_rotation}} and an ECD gate as defined in \Eq{\ref{eq: ecd}}.
}
\label{fig: single_ecd_rot}
\end{figure}

There are different universal qubit-qumode gates that have been explored in the past. \cite{Eickbusch2022,You2024Crosstalk,Krastanov2015,Diringer2024,Zhang2023energy,Job2023efficient}
We focus here on a qubit-qumode universal unitary circuit $\UniAn{N_d}$ based on $N_d$ unitary blocks of the following form \cite{Eickbusch2022,You2024Crosstalk,Zhang2023energy}
\begin{subequations} \label{eq: ecd_rot_ansatz}
\begin{align}
\UniAn{N_d} ( \bm{\beta}^G, \bm{\theta}^G, \bm{\varphi}^G )
&= \UniECD (\beta_{N_d}^G, \theta_{N_d}^G, \varphi_{N_d}^G) \: \cdots \: 
\UniECD (\beta_1^G, \theta_1^G, \varphi_1^G)
\\
\UniECD (\beta, \theta, \varphi)
&= ECD (\beta) \: 
\Big[ \EYE \otimes R (\theta, \varphi) \Big],
\end{align}   
\end{subequations}
where the symbols `A' and `G' represent the ansatz unitary and the parameters for the target unitary gates, respectively.
We will denote $N_d$ as the circuit depth of the universal ansatz from now on.
Each unitary gate $\UniECD$, schematically represented in \Fig{\ref{fig: single_ecd_rot}} consists of a qubit rotation 
\begin{equation} \label{eq: qubit_rotation}    
R (\theta, \varphi)
= \exp \big[ - i \: (\theta/2) \: 
( \cos \varphi \: X
+ \sin \varphi \: Y ) \big],
\end{equation}
and an echoed conditional displacement (ECD) operator
\begin{equation} \label{eq: ecd}
ECD (\beta) 
= \ket{1} \bra{0} \otimes D (\beta/2) 
+ \ket{0} \bra{1} \otimes D (- \beta/2),  
\end{equation}
with the displacement operator defined as 
\begin{equation} \label{eq: displacement_operator}
D (\beta) 
= \exp ( \beta \BC{} - \beta^* \BA{} ) ,   
\end{equation}
where $\BC{}$ and $\BA{}$ are the bosonic creation and annihilation operators.
We refer the reader to \Appx{\ref{app: ecd_rotation}} for more details on the ECD with qubit rotation ansatz.


Before discussing further, let us first formalize a general multi-qubit to qubit-qumode mapping approach. 
Let us assume we have a target operator $W_T$ corresponding to $N_Q$ qubits, which means it can be also understood as a qumode operator with Fock cutoff $L = 2^{N_Q}$.
There is no constraint on the $W_T$ except that it is a quantum operation, i.e., $W_T$ can be non-unitary. 
We want to find a parametrized qubit-qumode operator $V$ such that the following is true 
\begin{equation}
( \EYE_2 \otimes W_T ) \ket{0}_Q \otimes \ket{\psi}_R
\approx V \ket{0}_Q \otimes \ket{\psi}_R,
\end{equation}
where `Q' and `R' represent the states of the qubit and resonator and $\ket{\psi}$ is an arbitrary qumode state.
We have fixed the qubit state to $\ket{0}$ since our mapping is focused only on the qumode space.
This also allows us to optimize in the $\ket{0}_Q$ subspace of the full Hilbert space spanned by the combined qubit-qumode basis states, thus making the numerical optimization for finding the parameters more robust, as discussed below.
We express the qubit-qumode operator $V$ as a linear combination of ECD and rotation operators 
\begin{subequations}
\begin{align}
V ( \bm{\lambda}, \bm{\beta}^G, \bm{\theta}^G, \bm{\varphi}^G )  
&= \sum_{j = 1}^{N_t} \: \lambda_j \: 
\UniGate{N_d}{j} ( \bm{\beta}_j^G, \bm{\theta}_j^G, \bm{\varphi}_j^G ),
\\
\UniGate{N_d}{j} ( \bm{\beta}_j^G, \bm{\theta}_j^G, \bm{\varphi}_j^G )  
&= \UniECD (\beta_{j, N_d}^G, \theta_{j, N_d}^G, \varphi_{j, N_d}^G) \: \cdots \: 
\UniECD (\beta_{j, 1}^G, \theta_{j, 1}^G, \varphi_{j, 1}^G),
\end{align}
\end{subequations}
where $N_d$ is the depth of each unitary gate and $N_t$ is the number of terms in the linear expansion.
We have to solve the following optimization problem to find the parameters for a given target matrix $W_T$
\begin{equation} \label{eq: opt_cost_fun}
\min_{ \bm{\lambda}, \bm{\beta}^G, \bm{\theta}^G, \bm{\varphi}^G } F 
= \frac{1}{L^2} \: \sum_{n, m = 0}^{L - 1} \: | 
\braket{0, n | ( \EYE_2 \otimes W_T ) |0, m }
- \braket{0, n | V ( \bm{\lambda}, \bm{\beta}^G, \bm{\theta}^G, \bm{\varphi}^G ) |0, m } |^2,
\end{equation}
where $\ket{0, n} \equiv \ket{0}_Q \otimes \ket{n}_R$ and $\{ \ket{n} \}$ are the qumode Fock basis states.
The Fock cutoff $ L = 2^{N_Q} $ is the dimension of the target $N_Q$-qubit operator $W_T$.
In the context of mapping the Hamiltonian in \Eq{\ref{eq: qubit_ham_grouped}} into a qubit-qumode system, each of the $\QubitOp{N_Q}{j}$ represents the $W_T$ operator in \Eq{\ref{eq: opt_cost_fun}}.
The qubit Hamiltonian in \Eq{\ref{eq: qubit_ham_grouped}} can then be approximated as 
\begin{equation} \label{eq: qubit_ham_mapped}
\HamQ 
\approx \sum_{\mu = 1}^{N_H'} \: \HQCoeff{\mu} \: 
\sum_{j = 1}^{N_t} \: \lambda_{\mu, j} \: 
\UniGate{N_d}{\mu, j} 
( \bm{\beta}_{\mu, j}^G, \bm{\theta}_{\mu, j}^G, \bm{\varphi}_{\mu, j}^G ),
\end{equation}
where the parameters for the ECD and rotation gates can be represented by the tensors 
$ \bm{\beta}^G, \bm{\theta}^G, \bm{\varphi}^G $. 
After the optimization of \Eq{\ref{eq: opt_cost_fun}} is achieved, these tensors may be stored in the memory and can be reused for further calculations involving $\HamQ$.

\subsection{Computation of expectation values} \label{sec: exp_val_ecd}

Our goal is to find the ground state energy of the Hamiltonian $\HamQ$ by variationally updating a normalized trial state $\ket{\psi}$ while we compute the trial energy
\begin{equation} \label{eq: vqe}
\min_{\psi} E (\psi)
= \frac{\braket{\psi | \HamQ | \psi}}{\braket{\psi | \psi}}
= \braket{\psi | \HamQ | \psi}
\end{equation}
with the help of a quantum device, following the variational quantum eigensolver (VQE) approach. \cite{Peruzzo2014}
Since the mapped Hamiltonian in \Eq{\ref{eq: qubit_ham_mapped}} can now be understood as a Hamiltonian of one qumode, we explore the space of qumode states $\{ \ket{\psi} \}$ as the trial state. 
Then the trial qumode states $\ket{\psi}$ can be generated from a universal qubit-qumode gate involving ECD and qubit rotations. 
In other words, we prepare a parameterized qubit-qumode state 
\begin{equation} \label{eq: ecd_rot_ansatz_state}
\ket{\Psi}
= \UniAn{D} \: \big( \ket{0}_Q \otimes \ket{0}_C \big),
\end{equation}
where the unitary is written as 
\begin{subequations} \label{eq: ecd_rot_ansatz_for_trial}
\begin{align}
\UniAn{D} ( \bm{\beta}^\psi, \bm{\theta}^\psi, \bm{\varphi}^\psi )
&= \UniECD (\beta_{D}^\psi, \theta_{D}^\psi, \varphi_{D}^\psi) \: \cdots \: 
\UniECD (\beta_1^\psi, \theta_1^\psi, \varphi_1^\psi),
\\
\UniECD (\beta, \theta, \varphi)
&= ECD (\beta) \: 
\Big[ \EYE \otimes R (\theta, \varphi) \Big],
\end{align}   
\end{subequations}
and $D$ is the circuit depth of the ansatz unitary circuit for the trial state preparation.
The qumode state $\ket{\psi}$ can be projected from the qubit-qumode state $\ket{\Psi}$ by measuring the qubit part and continuing with qumode circuit if the qubit measurement results in state $\ket{0}$.
In other words, the disentangled qubit-qumode state can be represented as 
\begin{equation}
\ket{0}_Q \otimes \ket{\psi}_C
= \frac{ P_0 \ket{\Psi} }{ \braket{ \Psi | P_0 | \Psi } },
\end{equation}
where $ P_0 = \KetBra{0} \otimes \EYE $ and $\ket{\Psi}$ is defined in \Eq{\ref{eq: ecd_rot_ansatz_state}}.
We note that bosonic ansatz has also been recently explored for molecular electronic structure in \Reference{\citenum{Shang2024boson}}.

\begin{figure}[h!]

\includegraphics[width=0.9\columnwidth]{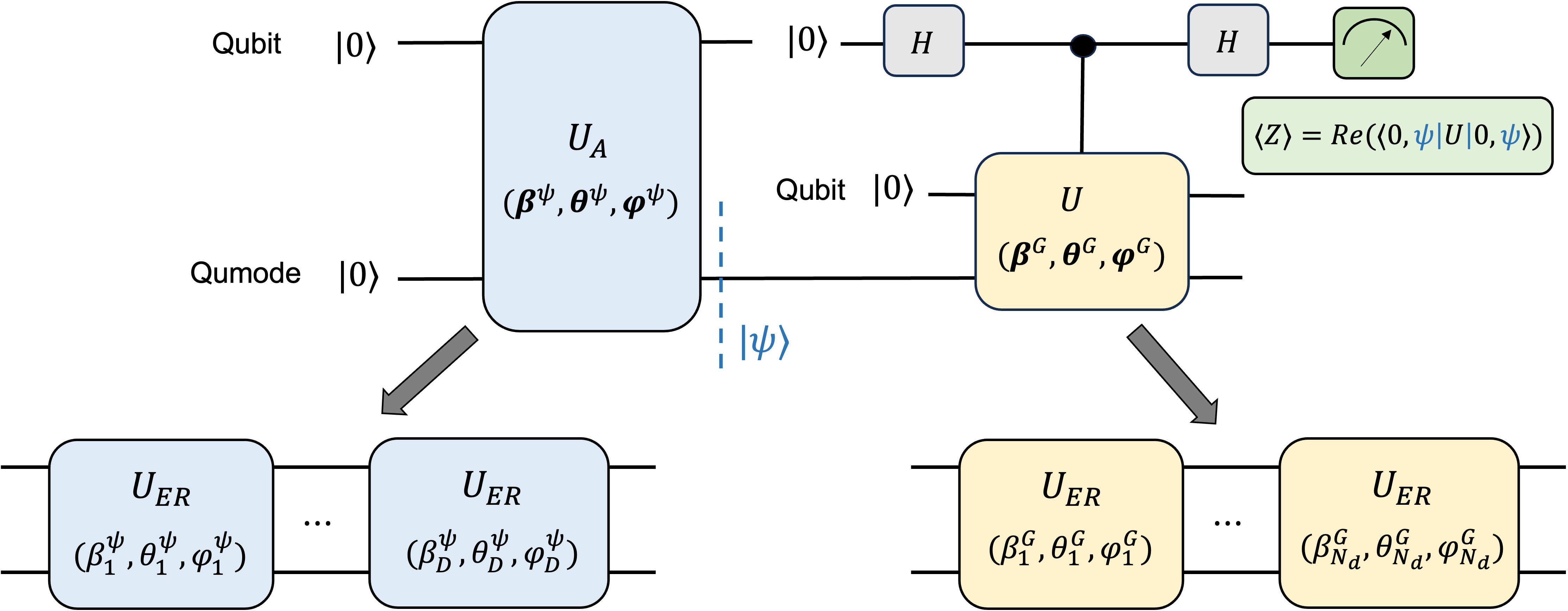}

\caption{
    Full circuit involving one qumode with two qubits for computing the expectation value of a qubit-qumode unitary $U$ for a qumode state $\ket{\psi}$, where the  $\UniECD$ operation is defined in \Fig{\ref{fig: single_ecd_rot}}.
    After generating a qumode state $\ket{\psi}$, the Hadamard test computes $\text{Re} ( \braket{0, \psi| U |0, \psi} )$.
    The controlled-ECD gates can be decomposed further in terms of CNOT and conditional displacement gates, as discussed in \Appx{\ref{app: controlled_ecd}} and \Fig{\ref{fig: compiled-cECD}}.
}
\label{fig: full_circuit_ev}
\end{figure}

The expectation value of $\HamQ$ for a trial state $\ket{\psi}$ can now be written as
\begin{align} \label{eq: qubit_ham_ev}
\braket{\psi| \HamQ | \psi} 
&\approx \sum_{\mu = 1}^{N_H'} \: \HQCoeff{\mu} \: 
\sum_{j = 1}^{N_t} \: \lambda_{\mu, j} \: 
\text{Re} ( \braket{\psi | \: \UniGate{N_d}{\mu, j} 
( \bm{\beta}_{\mu, j}^G, \bm{\theta}_{\mu, j}^G, \bm{\varphi}_{\mu, j}^G ) \: | \psi} ) \nonumber
\\
&= \sum_{\mu = 1}^{N_H'} \: \HQCoeff{\mu} \: 
\sum_{j = 1}^{N_t} \: \lambda_{\mu, j} \: M_{\mu, j},
\end{align}
where $\{ M_{\mu, j} \}$ can be computed by a Hadamard test involving another ancilla qubit.
The controlled unitary in the Hadamard test includes two types of gates: Qubit-controlled qubit rotation and qubit-controlled ECD gate.
The latter is two-qubit one-qumode gate and can be decomposed in terms of two-qubit and qubit-qumode gates, as discussed in \Appx{\ref{app: controlled_ecd}}.
The full circuit for the generation of the trial state and computation of expectation values involving one qumode and two qubits is illustrated in \Fig{\ref{fig: full_circuit_ev}}.
The two summations in \Eq{\ref{eq: qubit_ham_ev}} can be efficiently computed on a classical device after getting the set of $\{ M_{\mu, j} \}$ from quantum device measurements. 
We name this approach to optimize \Eq{\ref{eq: vqe}} with ECD-rotation circuits as ECD-VQE. 

It should be noted that for some specific electronic structure systems, the \textit{global} ground state of the mapped qubit Hamiltonian $\HamQ$ may not have the desired electron number of the system of interest since after mapping the fermion to qubit mapping, the Hilbert space of qubits contains all of the fermion number sectors.
In that case, explicit particle number constraint can be imposed by modifying the cost function \cite{Ryabinkin2018constrained}
\begin{equation} \label{eq: vqe_constrained_number}
\min_{\psi, \lambda} C_N 
= \braket{\psi | \HamQ | \psi}
+ \lambda \: \Big[ \braket{\psi | \: ( \hat{N} - N ) \: | \psi} \Big]^2
\end{equation}
where $\lambda$ is a Lagrange multiplier, $N$ is the number of electrons and 
$ \hat{N} = \sum_{p = 1}^M \: \FC{p} \FA{p} $ is the fermionic total number operator, which can be mapped to the following qubit operator 
\begin{equation}
\hat{N}
\mapsto \sum_{p = 1}^M \frac{1}{2} \: ( \EYE_p + Z_p )
\end{equation}
using the Jordan-Wigner transformation.
The expectation value 
$ \braket{\psi | \hat{N} | \psi} $ can be computed following the discussion above for computing $\braket{\psi | \HamQ| \psi}$.
This constrained cost function approach of \Eq{\ref{eq: vqe_constrained_number}} can also be applicable to spin-symmetries by mapping the fermionic $\hat{S}_F^2$ to its corresponding qubit operator $\hat{S}_Q^2$ and optimizing the cost function below \cite{Ryabinkin2018constrained}
\begin{equation} \label{eq: vqe_constrained_spin}
\min_{\psi, \lambda} C_S 
= \braket{\psi | \HamQ | \psi}
+ \lambda \: \Big[ \braket{\psi | \: \hat{S}_Q^2 \: | \psi}
- S (S + 1) \Big]^2,
\end{equation}
where $S$ is the total spin quantum number.
However, optimization of \Eq{\ref{eq: vqe}} is sufficient for many electronic systems, including the dihydrogen molecule discussed here, which we discuss below.


\subsection{Ground state energy of dihydrogen molecule} \label{sec: h2_molecule}

\begin{figure}[b!]

\includegraphics[width=0.7\columnwidth]{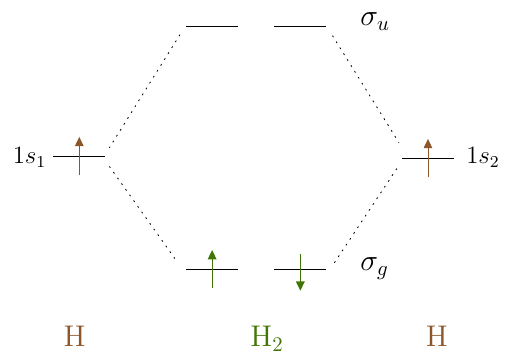}

\caption{
    A molecular orbital diagram corresponding to the H$_2$ molecule in a minimal basis. 
    The molecular orbitals, $\sigma_g$ and $\sigma_u$, are built from $1s$ atomic orbitals of the two hydrogen atoms. 
    In second quantization, the diagram represents the Slater determinant $\ket{0, 1}_F = \FC{0} \FC{1} \kVacP$, where the first and second spin-orbitals share the $\sigma_g$ spatial function.
}
\label{fig: mo_diagram_dihydrogen}
\end{figure}

As a specific example, we apply the ideas discussed above to simulate the electronic structure of the H$_2$ molecule in a minimal basis in this section. \cite{SzaboBook}
We choose one spatial function per hydrogen atom, which leads to the following molecular spatial orbitals  
\begin{subequations}
\begin{align}
\phi_g
&= \mathcal{N}_g \: ( 1s_1 + 1s_2 ),   
\\
\phi_u
&= \mathcal{N}_u \: ( 1s_1 - 1s_2 ),   
\end{align}
\end{subequations}
where $\mathcal{N}_g$ and $\mathcal{N}_u$ are the normalization factors based on the spatial functions chosen, and one popular choice is the STO-3G minimal basis \cite{Hehre1969} that approximates the Slater-type atomic functions with three real-valued Gaussian functions. \cite{SzaboBook}
Having two spatial orbitals means we have an electronic system of two electrons in four spin-orbitals, as shown by the molecular orbital diagram in \Fig{\ref{fig: mo_diagram_dihydrogen}}.
Let us define the four spin-orbitals as 
\begin{subequations}
\begin{align}
\ket{\chi_0}
&\equiv \ket{\phi_g, \alpha}, \quad
\ket{\chi_1}
\equiv \ket{\phi_g, \beta},
\\
\ket{\chi_2}
&\equiv \ket{\phi_u, \alpha}, \quad
\ket{\chi_3}
\equiv \ket{\phi_u, \beta},
\end{align}
\end{subequations}
where $\alpha$ and $\beta$ denote spin-orbitals with up and down electron spins, respectively.


The electronic structure Hamiltonian of the dihydrogen molecule in a minimal basis can be written as \cite{Whitfield2011} 
\begin{align} \label{eq: ham_dihydrogen} 
\HamF
&= h_{\text{nuc}} 
+ \HOne{0}{0} \: \FC{0} \FA{0} 
+ \HOne{1}{1} \: \FC{1} \FA{1} 
+ \HOne{2}{2} \: \FC{2} \FA{2} 
+ \HOne{3}{3} \: \FC{3} \FA{3} 
+ \HTwo{01}{10} \: \FC{0} \FC{1} \FA{1} \FA{0} 
+ \HTwo{23}{32} \: \FC{2} \FC{3} \FA{3} \FA{2} \nonumber
\\
&+ \HTwo{03}{30} \: \FC{0} \FC{3} \FA{3} \FA{0} 
+ \HTwo{12}{21} \: \FC{1} \FC{2} \FA{2} \FA{1} 
+ ( \HTwo{02}{20} -  \HTwo{02}{02} ) \: \FC{0} \FC{2} \FA{2} \FA{0} 
+ ( \HTwo{13}{31} -  \HTwo{13}{13} ) \: \FC{1} \FC{3} \FA{3} \FA{1} \nonumber
\\
&+ \HTwo{03}{12} \: ( \FC{0} \FC{3} \FA{1} \FA{2} + \HermConj ) 
+ \HTwo{01}{32} \: ( \FC{0} \FC{1} \FA{3} \FA{2} + \HermConj ),
\end{align}
which can then be expressed as the following four-qubit Hamiltonian using JWT\cite{Whitfield2011}
\begin{align} \label{eq: h2_qubit_ham}
\HamF \mapsto \HamQ 
&= \HQCoeff{1} 
+ \HQCoeff{2} \: ( Z_0 + Z_1 ) 
+ \HQCoeff{3} \: ( Z_2 + Z_3 ) 
+ \HQCoeff{4} \: Z_0 Z_1 
+ \HQCoeff{5} \: ( Z_0 Z_2 + Z_1 Z_3 ) 
+ \HQCoeff{6} \: ( Z_0 Z_3 + Z_1 Z_2 ) 
\nonumber
\\
&+ \HQCoeff{7} \: Z_2 Z_3 
+ \HQCoeff{8} \: ( 
X_0 Y_1 Y_2 X_3  
+ Y_0 X_1 X_2 Y_3  
- X_0 X_1 Y_2 Y_3  
- Y_0 Y_1 X_2 X_3 ),
\end{align}
where the scalars $\{ \HQCoeff{j} \}$ are defined in \Table{\ref{tab: h2_qubit_ham_coeff}} and tensor product is assumed. 
For example, the term $Z_1 Z_3$ represents the four-qubit operator 
$ \EYE \otimes Z \otimes \EYE \otimes Z $, and so on.
Following \Eq{\ref{eq: qubit_ham_grouped}}, we can also represent \Eq{\ref{eq: h2_qubit_ham}} as 
\begin{equation} \label{eq: h2_qubit_ham_grouped}
\HamQ 
= g_1 + \sum_{\mu = 2}^8 \: \HQCoeff{\mu} \: \QubitOp{4}{\mu},
\end{equation}
where $ \QubitOp{4}{2} = Z_0 + Z_1 $,  
$ \QubitOp{4}{3} = Z_2 + Z_3 $, and so on.
We have observed that after the optimization is achieved based on the loss function defined in \Eq{\ref{eq: opt_cost_fun}}, we can expand \Eq{\ref{eq: h2_qubit_ham_grouped}} as 
\begin{equation} \label{eq: h2_qubit_ham_grouped_expand}
\HamQ 
\approx g_1 + \sum_{\mu = 2}^8 \: \HQCoeff{\mu} \: 
\sum_{j = 1}^{15} \: \lambda_{\mu, j} \: 
\UniGate{10}{\mu, j} 
( \bm{\beta}_{\mu, j}^G, \bm{\theta}_{\mu, j}^G, \bm{\varphi}_{\mu, j}^G ),
\end{equation}
where each of the $\QubitOp{4}{\mu}$ operator is written as linear combination of $N_t = 15$ ECD-rotation unitaries, each with circuit depth $N_d = 10$.
The converged loss function for each of the $\QubitOp{4}{\mu}$ operators is shown in \Table{\ref{tab: h2_qubit_ham_terms_losses}}, where the Broyden--Fletcher--Goldfarb--Shanno (BFGS) method was used as the optimizer as implemented in SciPy. \cite{2020SciPy-NMeth}
The expectation value of $\HamQ$ for a qumode state generated by the ECD-rotation ansatz can now be computed by following the discussions in \Sec{\ref{sec: exp_val_ecd}} and \Fig{\ref{fig: full_circuit_ev}}.


\begin{table}[t]
\begin{tabular}{lll}
\hline
Coefficient && Definition
\\ \hline 
$ \HQCoeff{1} $ &&   
$ h_{\text{nuc}} 
+ \frac{1}{2} \: ( 
\HOne{0}{0} + \HOne{1}{1} 
+ \HOne{2}{2} + \HOne{3}{3} ) 
+ \frac{1}{4} \: \Big[
\HTwo{01}{10} + \HTwo{23}{32}    
+ \HTwo{03}{30} + \HTwo{12}{21}    
+ ( \HTwo{02}{20} - \HTwo{02}{02} )
+ ( \HTwo{13}{31} - \HTwo{13}{13} )
\Big] $   
\\ 
$ \HQCoeff{2} $ &&
$ - \frac{1}{2} \: \HOne{0}{0} 
- \frac{1}{4} \Big[
\HTwo{01}{10} 
+ \HTwo{03}{30} 
+ ( \HTwo{02}{20} - \HTwo{02}{02} )
\Big] $ 
\\ 
$ \HQCoeff{3} $ &&  
$ - \frac{1}{2} \: \HOne{2}{2} 
- \frac{1}{4} \Big[
\HTwo{23}{32}    
+ \HTwo{12}{21}    
+ ( \HTwo{02}{20} - \HTwo{02}{02} )
\Big] $
\\ 
$ \HQCoeff{4} $ && 
$ \frac{1}{4} \: \HTwo{01}{10} $  
\\
$ \HBCoeff{5} $ &&
$ \frac{1}{4} \: ( 
\HTwo{02}{20} - \HTwo{02}{02} ) $ 
\\
$ \HBCoeff{6} $ &&
$ \frac{1}{4} \: \HTwo{03}{30} $ 
\\
$ \HBCoeff{7} $ &&
$ \frac{1}{4} \: \HTwo{23}{32} $ 
\\
$ \HBCoeff{8} $ && 
$ \frac{1}{4} \: \HTwo{03}{12} $
\\ \hline
\end{tabular}

\caption{
    Definition of scalars in \Eq{\ref{eq: h2_qubit_ham}} using nuclear repulsion, one-electron and two-electron integral terms.
}
\label{tab: h2_qubit_ham_coeff}
\end{table}


\begin{table}[h!]
\begin{tabular}{ll}
\hline
Operator & Converged Loss 
\\ \hline
$ \QubitOp{4}{2} $ & $5 \times 10^{-12}$ 
\\
$ \QubitOp{4}{3} $ & $3 \times 10^{-11}$ 
\\
$ \QubitOp{4}{4} $ & $5 \times 10^{-12}$ 
\\
$ \QubitOp{4}{5} $ & $2 \times 10^{-11}$ 
\\
$ \QubitOp{4}{6} $ & $1 \times 10^{-11}$ 
\\
$ \QubitOp{4}{7} $ & $1 \times 10^{-11}$ 
\\
$ \QubitOp{4}{8} $ & $2 \times 10^{-12}$ 
\\ \hline
\end{tabular}

\caption{
    The converged losses for the Hamiltonian operator terms corresponding to \Eq{\ref{eq: h2_qubit_ham_grouped}} following the optimization of the loss function defined in \Eq{\ref{eq: opt_cost_fun}}.
    The results are for a linear combination of $N_t = 15$ unitaries, each of depth $N_d = 10$.
}
\label{tab: h2_qubit_ham_terms_losses}
\end{table}

\begin{figure}[t]

\includegraphics[width=0.7\columnwidth]{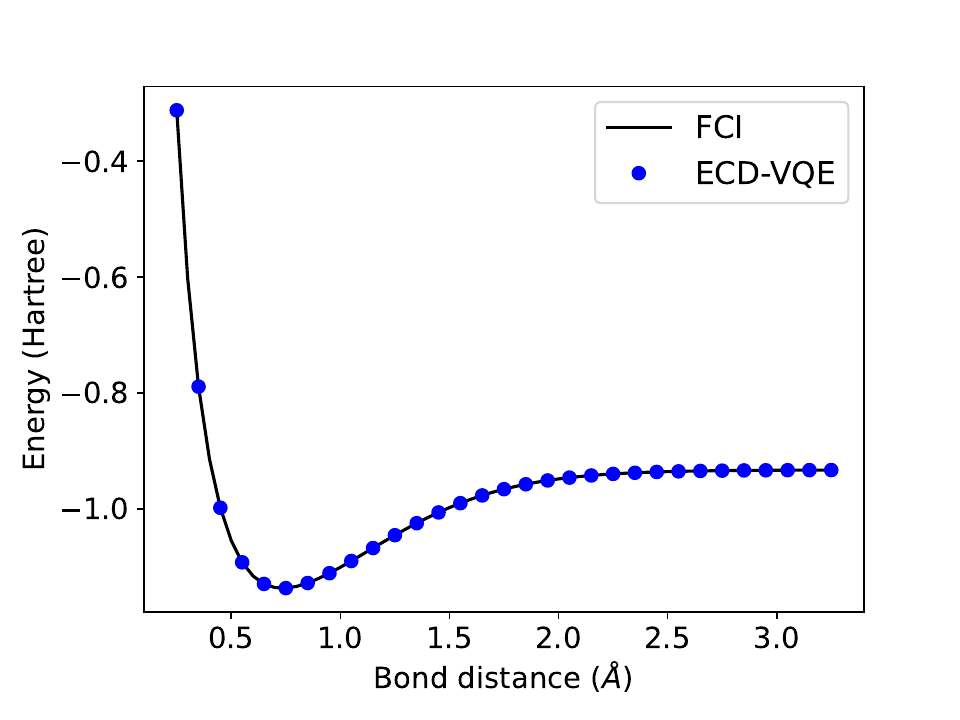}

\caption{
    Comparison of dihydrogen molecule ground state energies in STO-3G minimal basis using the ECD-VQE approach as discussed in \Sec{\ref{sec: mapping_ecd}} with the FCI energies.
    The ECD with qubit rotation ansatz circuit for the trial state preparation has a depth of $D = 9$.
}
\label{fig: vqe_en_ecd_rot}
\end{figure}
\begin{figure}[h!]

\includegraphics[width=0.7\columnwidth]{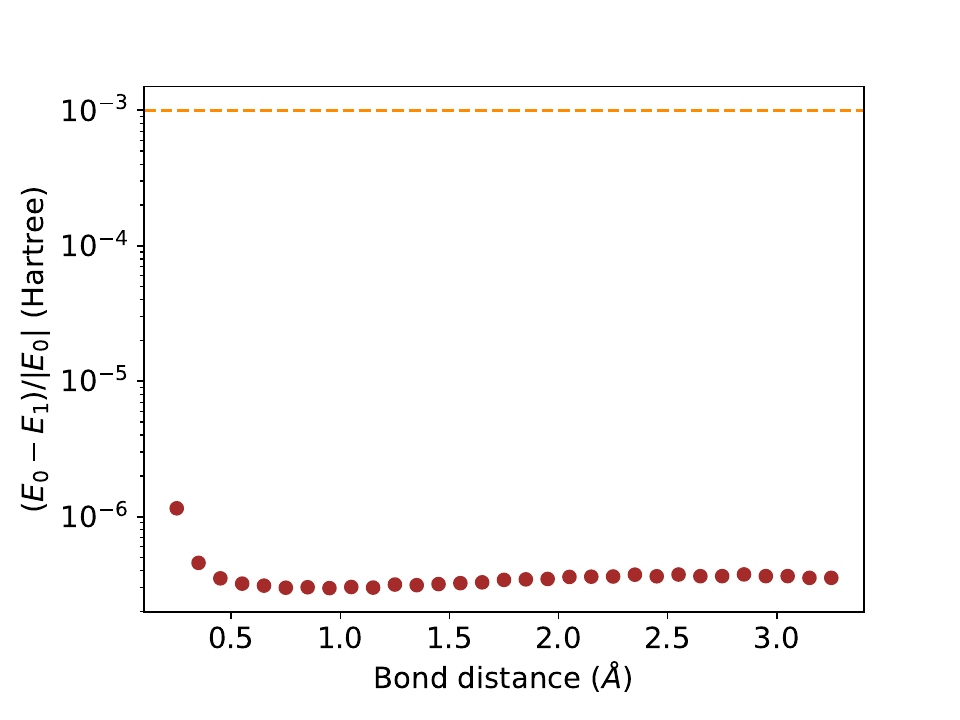}

\caption{
    The accuracy in energies computed using the linear decomposition involving ECD with qubit rotation unitaries. 
    Here, $E_0$ represents the energy computed with the original qubit Hamiltonian defined in \Eq{\ref{eq: h2_qubit_ham}}, whereas $E_1$ represents energy computed with its decomposition defined in \Eq{\ref{eq: h2_qubit_ham_grouped_expand}}.
    The qumode states for the expectation values are taken from the converged trial states corresponding to the ECD-VQE calculations shown in \Fig{\ref{fig: vqe_en_ecd_rot}}.
}
\label{fig: vqe_check_en}
\end{figure}

Computation of expectation values allowed the emulation of VQE for the dihydrogen molecule on a classical computer, where the ECD-VQE optimization as defined in \Eq{\ref{eq: vqe}} has been achieved using the BFGS method.
As shown in \Fig{\ref{fig: vqe_en_ecd_rot}}, the ECD-rotation ansatz circuit with a depth $D = 9$ accurately reproduces the ground state energies of the dihydrogen molecule in the STO-3G basis, where the exact ground state energies in this basis can be found by diagonalizing the Hamiltonian, also known as full configuration interaction (FCI). 
All calculations were done using QuTip \cite{Lambert2024qutip} and OpenFermion. \cite{Mcclean2022openfermion}
We compare the approximate trial energies computed using the decomposed Hamiltonian defined by \Eq{\ref{eq: h2_qubit_ham_grouped_expand}} with the exact trial energies in \Fig{\ref{fig: vqe_check_en}}. 
The energy errors in \Fig{\ref{fig: vqe_check_en}} show that it is possible that the approximate energies can sometimes go below the exact energies. 
However, these negative errors are numerically tolerable since the error ranges are far smaller than the chemical accuracy regime.


\section{Qumode mapping without ancilla qubit} \label{sec: mapping_snap} 


\begin{figure}[t]

\includegraphics[width=0.9\columnwidth]{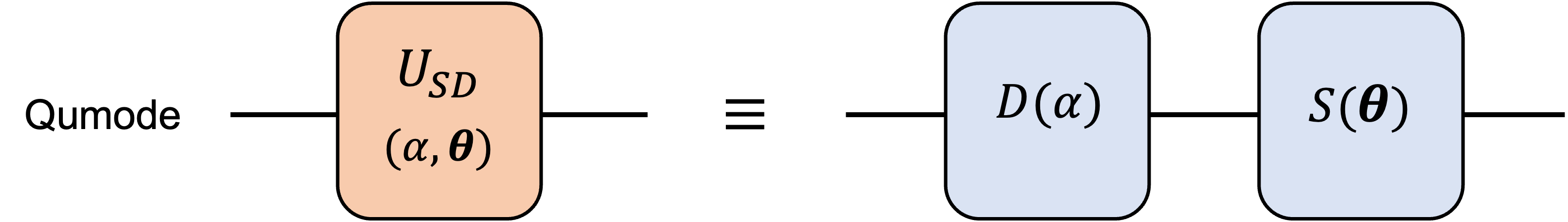}

\caption{
    Qumode gate consisting of one displacement gate as defined in \Eq{\ref{eq: displacement_operator}} and a SNAP gate as defined in \Eq{\ref{eq: snap}}.
}
\label{fig: single_snap_disp}
\end{figure}


Another universal bosonic unitary can be implemented by repeating blocks of the following parametrized gate \cite{Krastanov2015,Job2023efficient}
\begin{equation} \label{eq: snap_displacement}
\UniSNAP (\alpha, \bm{\theta})
= S (\bm{\theta}) D (\alpha), 
\end{equation}
where $D (\alpha)$ is the displacement gate and $S (\bm{\theta})$ is the selective number-dependent arbitrary phase (SNAP) gate \cite{Fosel2020}
\begin{equation} \label{eq: snap}
S (\bm{\theta})
= \sum_{n = 0}^{L - 1} \: e^{i \theta_n} \KetBra{n},
\end{equation}
where $\KetBra{n}$ is the Fock basis projection operator and $L$ is the Fock cutoff for the qumode.
We illustrate the SNAP-displacement gate in \Fig{\ref{fig: single_snap_disp}} and discuss more details on this gate set including its universality for a single qumode in \Appx{\ref{app: snap_disp}}.
Since the parametrization of the SNAP gate already includes a complex phase, the displacement coefficient in \Eq{\ref{eq: snap_displacement}} can be assumed to be real-valued.  

Following \Sec{\ref{sec: qubit_to_qumode_ecd}}, we can now approximate an arbitrary target unitary operator $V_T$ with our parametrized unitary 
\begin{equation} \label{eq: snap_disp_ansatz}
\UniGate{N_d}{} ( \bm{\alpha}^G, \bm{\theta}^G )
= \UniSNAP (\alpha_{N_d}, \bm{\theta}_{N_d}) \cdots 
\UniSNAP (\alpha_1, \bm{\theta}_1),
\end{equation}
where $N_d$ is the circuit depth. 
The corresponding optimization problem to find the parameters for a given target unitary $V_T$ now becomes
\begin{equation} \label{eq: opt_cost_fun_snap}
\min_{ \bm{\alpha}^G, \bm{\theta}^G } F 
= \frac{1}{L^2} \: \sum_{n, m = 0}^{L - 1} \: | 
\braket{n | V_T |m } 
- \braket{n | \UniGate{N_d}{} ( \bm{\alpha}^G, \bm{\theta}^G ) |m } |^2,
\end{equation}
where $\{ \ket{n} \}$ are the qumode Fock basis states and the Fock cutoff $L$ is given by the dimension of $V_T$.
We note that even though we can understand the parametrized unitary ansatz in \Eq{\ref{eq: snap_disp_ansatz}} as a sequence of qumode gates conceptually, the hardware implementation of these gates do require an ancilla qubit, as discussed in \Appx{\ref{app: snap_disp}}.

\subsection{Dihydrogen molecule}

We now apply this to the qubit Hamiltonian for the dihydrogen molecule in a minimal basis, as defined in \Eq{\ref{eq: h2_qubit_ham}}.
We have observed that it is possible to accurately approximate each of the Pauli words of the Hamiltonian in \Eq{\ref{eq: h2_qubit_ham}} with SNAP-displacement circuit of depth $N_d = 16$, as shown in 
\Table{\ref{tab: h2_qubit_ham_pauli_losses}}.
Thus, we can write the electronic structure Hamiltonian of the dihydrogen molecule as 
\begin{equation} \label{eq: h2_qubit_ham_expanded_snap}
\HamQ 
\approx g_1 + \sum_{j = 1}^{14} \: v_j \: 
\UniGate{16}{j} ( \bm{\alpha}_j^G, \bm{\theta}_j^G ),
\end{equation}
where the coefficients $\{ v_j \}$ can be easily deduced from \Eq{\ref{eq: h2_qubit_ham}}.


\begin{table}[t!]
\begin{tabular}{llll}
\hline
Operator & Converged Loss & Operator & Converged Loss
\\ \hline
$ Z \otimes \EYE \otimes \EYE \otimes \EYE $ 
& $8 \times 10^{-14}$ 
& $ \EYE \otimes Z \otimes Z \otimes \EYE $ 
& $5 \times 10^{-14}$ 
\\
$ \EYE \otimes Z \otimes \EYE \otimes \EYE $ 
& $1 \times 10^{-13}$ 
& $ \EYE \otimes Z \otimes \EYE \otimes Z $ 
& $2 \times 10^{-13}$
\\
$ \EYE \otimes \EYE \otimes Z \otimes \EYE $ 
& $3 \times 10^{-14}$ 
& $ \EYE \otimes \EYE \otimes Z \otimes Z $ 
& $3 \times 10^{-14}$ 
\\
$ \EYE \otimes \EYE \otimes \EYE \otimes Z $ 
& $9 \times 10^{-14}$ 
& $ X \otimes Y \otimes Y \otimes X $ 
& $5 \times 10^{-14}$ 
\\
$ Z \otimes Z \otimes \EYE \otimes \EYE $ 
& $8 \times 10^{-15}$ 
& $ Y \otimes X \otimes X \otimes Y $ 
& $1 \times 10^{-13}$
\\
$ Z \otimes \EYE \otimes Z \otimes \EYE $ 
& $3 \times 10^{-14}$ 
& $ X \otimes X \otimes Y \otimes Y $ 
& $1 \times 10^{-13}$
\\
$ Z \otimes \EYE \otimes \EYE \otimes Z $ 
& $4 \times 10^{-14}$ 
& $ Y \otimes Y \otimes X \otimes X $ 
& $6 \times 10^{-14}$
\\ \hline
\end{tabular}

\caption{
    The converged losses for the four-qubit Pauli words corresponding to \Eq{\ref{eq: h2_qubit_ham}} following the optimization of the loss function defined in \Eq{\ref{eq: opt_cost_fun_snap}}.
    The results are for the SNAP-displacement circuit as the parameterized circuit with depth $N_d = 16$. 
}
\label{tab: h2_qubit_ham_pauli_losses}
\end{table}

\begin{figure}[t!]

\includegraphics[width=0.9\columnwidth]{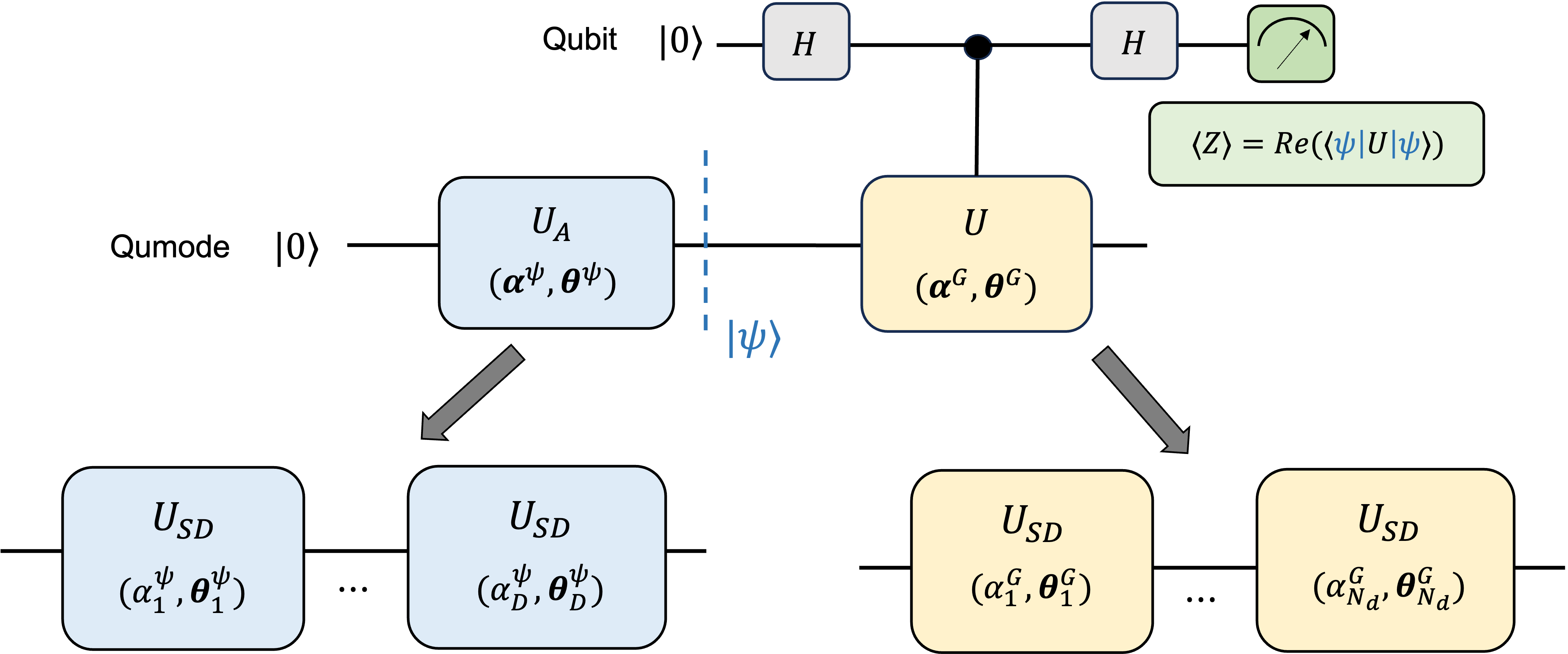}

\caption{
    Full circuit involving one qumode with one qubit for computing the expectation value of a qubit-qumode unitary $U$ for a qumode state $\ket{\psi}$.
    After generating a qumode state $\ket{\psi}$, the Hadamard test computes $\text{Re} ( \braket{\psi| U |\psi} )$.
}
\label{fig: full_circuit_ev_snap}
\end{figure}
\begin{figure}[b!]

\includegraphics[width=0.9\columnwidth]{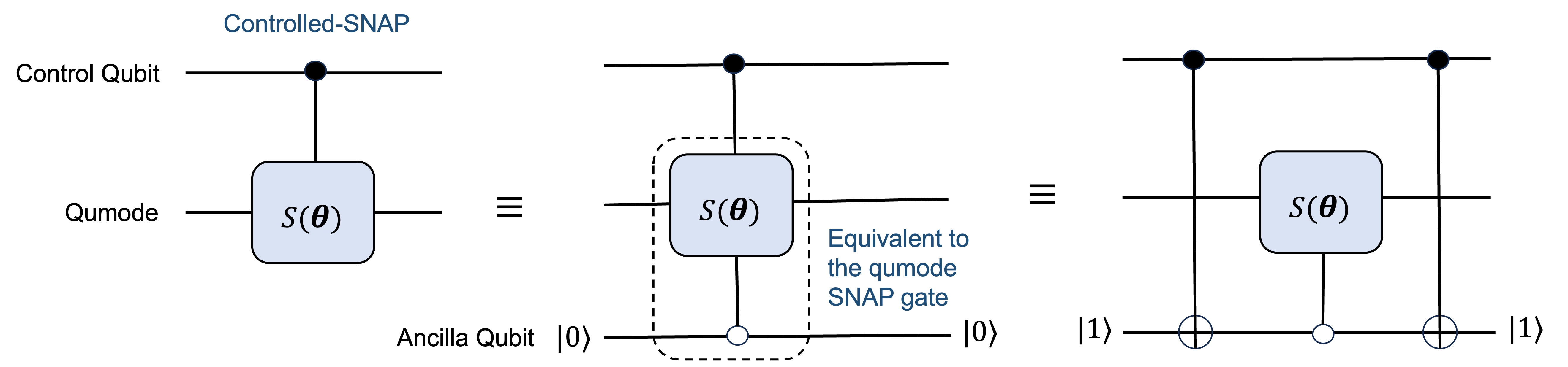}

\caption{
    Implementation of controlled-SNAP gate with the ancilla qubit dispersively coupled to the qumode explicitly shown. 
    The circuit is based on the fact the SNAP gate implementation needs the ancilla qubit to be returned to $\ket{0}$ state after the operation, as defined in \Eq{\ref{eq: snap_alt_qubit}}.
}
\label{fig: control_snap}
\end{figure}

The computation of expectation value $\braket{\psi| \HamQ | \psi}$ for a trial qumode state $\ket{\psi}$ can be achieved by a little modification of the circuit described in \Sec{\ref{sec: exp_val_ecd}} and \Fig{\ref{fig: full_circuit_ev}}. 
A major improvement of the SNAP-displacement approach over the ECD with qubit rotation approach can be understood by comparing \Eq{\ref{eq: h2_qubit_ham_grouped_expand}} and \Eq{\ref{eq: h2_qubit_ham_expanded_snap}}. 
Specifically, the Hadamard test of only 14 unitaries are needed in the case of SNAP-displacement approach compared to 120 unitaries in the case of ECD with qubit rotation for computing the trial energy of the dihydrogen molecule.
We illustrate the full circuit for the SNAP-displacement approach in \Fig{\ref{fig: full_circuit_ev_snap}}, where the qumode trial ansatz
$ \ket{\psi} = \UniAn{D} \ket{0} $
is also parametrized with a SNAP-displacement circuit ansatz
\begin{equation} \label{eq: snap_disp_ansatz_state}
\UniAn{D} ( \bm{\alpha}^\psi, \bm{\theta}^\psi )
= \UniSNAP (\alpha_{D}^\psi, \bm{\theta}_{D}^\psi) \cdots 
\UniSNAP (\alpha_1^\psi, \bm{\theta}_1^\psi), 
\end{equation}
where $D$ is the trial ansatz circuit depth and $\UniSNAP$ is defined in \Eq{\ref{eq: snap_displacement}}. 
We name this approach to optimize \Eq{\ref{eq: vqe}} with SNAP-displacement circuits as SNAP-VQE. 
As mentioned above, the auxiliary qubit for implementing the SNAP gate may be omitted in circuits conceptually. 
For the controlled SNAP-displacement ansatz, the two-qubit one-qumode controlled-SNAP gates can be implemented by combining two-qubit CNOT with qubit-qumode SNAP gates following \Eq{\ref{eq: snap_alt_qubit}}, as illustrated in \Fig{\ref{fig: control_snap}}. 
It should be noted that since the qumode state preparation is separate from the expectation value computation part in \Fig{\ref{fig: full_circuit_ev_snap}}, it can also be achieved with an ECD with qubit rotation ansatz or any other universal bosonic circuits in principle.


\begin{figure}[t!]
    \centering
    \begin{subfigure}[c]{0.47\textwidth}
    
        \centering
        \vskip-2ex
        \includegraphics[width=1.0\linewidth]{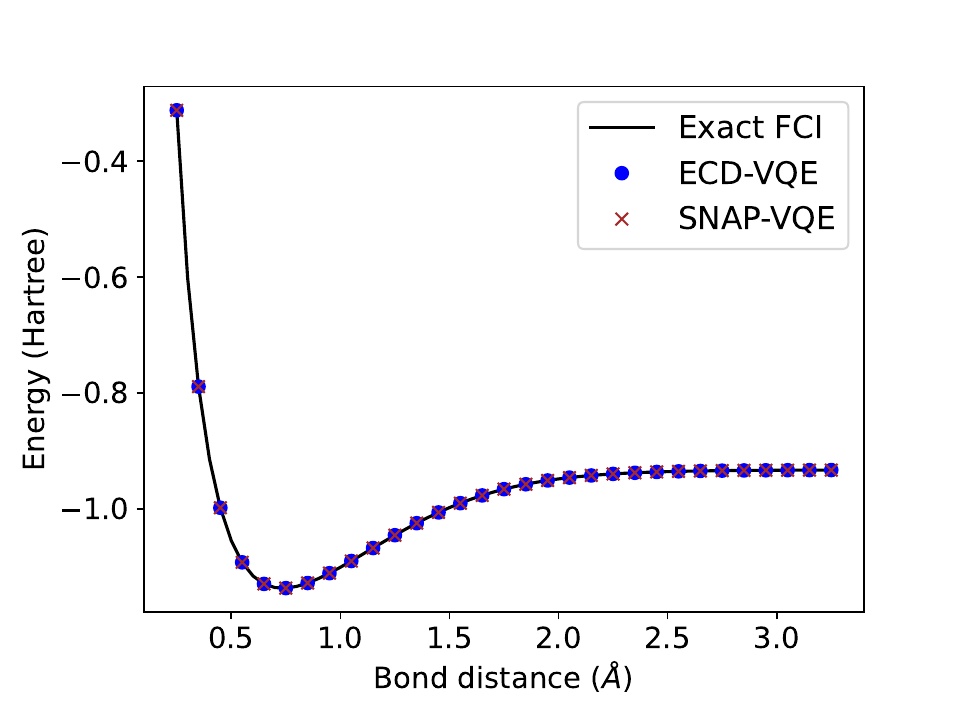}
        
    \end{subfigure}
    \hskip1ex
    \begin{subfigure}[c]{0.47\textwidth}
    
        \centering
        \includegraphics[width=0.94\linewidth]{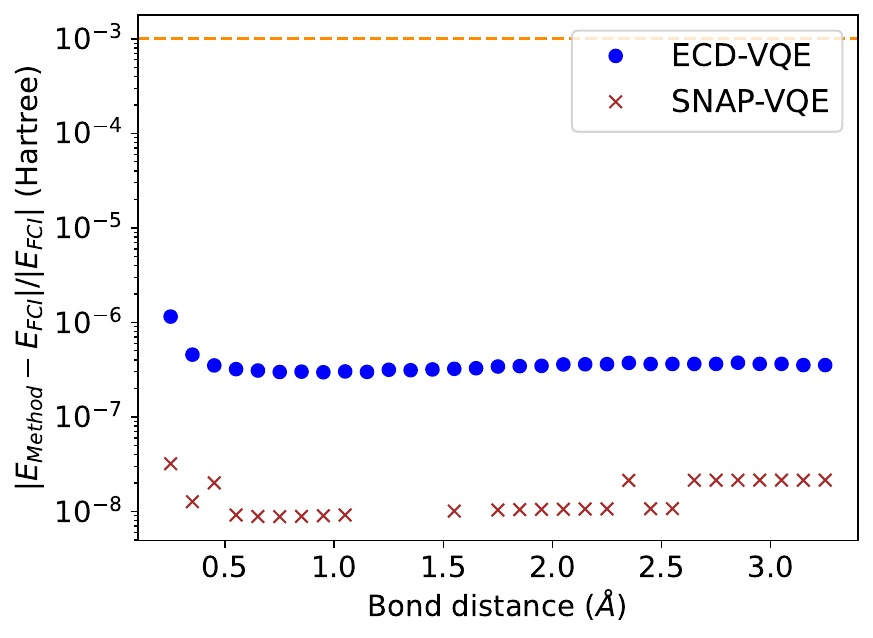}
        
    \end{subfigure}
    
    \caption{
        Comparison of dihydrogen molecule ground state energies in STO-3G minimal basis using ECD-VQE and SNAP-VQE approaches, as discussed in \Sec{\ref{sec: mapping_ecd}} and \Sec{\ref{sec: mapping_snap}}, respectively. 
        The ECD with qubit rotation ansatz circuit has a depth of $D = 9$ and the SNAP with displacement ansatz circuit has a depth of $D = 4$ for the trial state preparation parts.
        The orange horizontal line in the right panel represents the minimum energy error needed for chemical accuracy, usually in the miliHartree range.
     }
     \label{fig: vqe_en_ecd_snap}
\end{figure}

We compare the ECD-VQE and SNAP-VQE approaches for the dihydrogen molecule on a classical computer in \Fig{\ref{fig: vqe_en_ecd_rot}}, where the BFGS method has been applied for the classical optimization part as defined in \Eq{\ref{eq: vqe}}.
All calculations were done using QuTip and OpenFermion. 
The ECD-rotation ansatz circuit has a depth of $D = 9$ whereas the SNAP-displacement ansatz circuit has a depth of $D = 4$ for the trial state preparation part.   
It is clear from the right panel of 
\Fig{\ref{fig: vqe_en_ecd_snap}} 
that the SNAP-VQE is a better approach in terms of accuracy and circuit depth, although both the methods achieve energetic error well below the chemical accuracy regime. 
Overall, we have observed that the SNAP-displacement ansatz has greater variational flexibility than the ECD-rotation ansatz, thus can represent unitary circuits more compactly for both Pauli words and as a trial state. 
This is justified by more tunability of the SNAP parameters that can precisely affect each of the Fock basis states. 
An important distinction between ECD and SNAP gates is that the ECD can be implemented in the weakly dispersive regime between the qubit and qumode, \cite{Eickbusch2022} where a strong dispersive interaction is currently needed for implementing the latter. \cite{Krastanov2015}
We refer the readers to \Appx{\ref{app: comparison_ecd_snap}} for more details on the comparison of the hardware implementation of ECD and SNAP gates. 


\subsection{Generalization to multiple qumodes} \label{sec: multimode_snap}


\begin{figure}[t!]

\includegraphics[width=0.9\columnwidth]{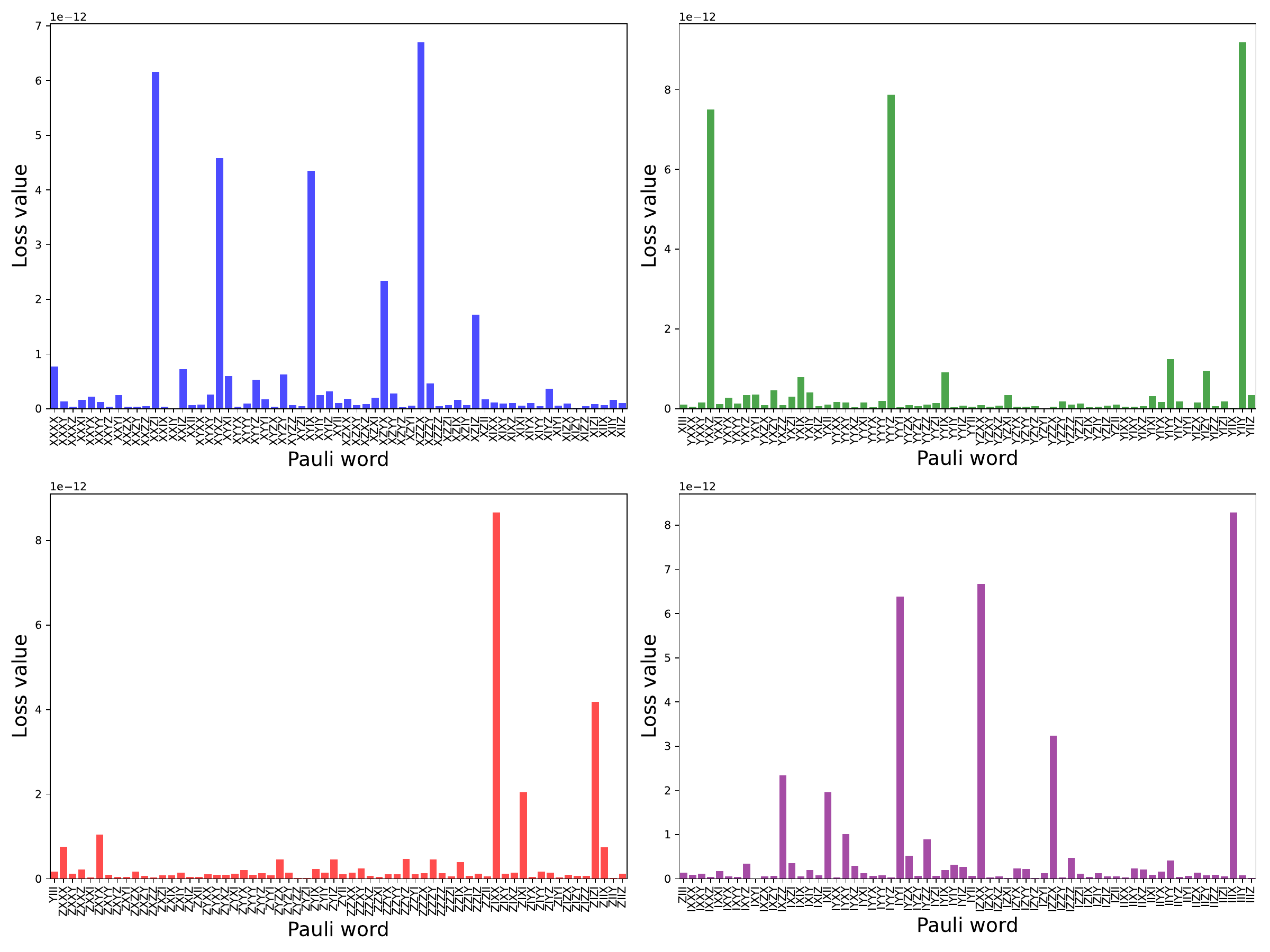}

\caption{
    The converged losses for all possible non-trivial Pauli words based on the optimization defined in \Eq{\ref{eq: opt_cost_fun_snap}}. 
    The results are for the SNAP-displacement universal circuit ansatz with depth $N_d = 16$. 
}
\label{fig: pauli_words_snap_losses}
\end{figure}


The SNAP-VQE approach allows a single unitary circuit with practical depth for computing expectation values using a Hadamard test. 
This allows the development of a modular approach to generalize the SNAP-VQE approach for an arbitrary number of qubits after mapping them to a few qumodes. 
We discuss how to extend the SNAP-VQE approach for multiple qumodes and the quantum hardware necessary to implement multimode trial states here. 

In principle, $N_Q$ qubits can be mapped to the unitary gates of a single qumode with a Fock cutoff of $L = 2^{N_Q} $. 
In practice, it is more appropriate to partition the tensor product space of $N_Q$ qubits, which still has the advantage of replacing many qubits with a few qumodes, while having more control over the hardware.
Let us assume we want to map $N_Q$ qubits to $N_R \: (< N_Q)$ qumodes with corresponding Fock cutoffs given by $\{ L_1, \cdots, L_{N_R} \}$. 
The $N_Q$-qubit Hamiltonian in \Eq{\ref{eq: qubit_ham}} can then be represented as 
\begin{equation} \label{eq: qubit_ham_multiple_qumodes}
\HamQ 
= \sum_{\mu = 1}^{N_H} \: \HQCoeff{\mu} \: 
\Pauli{N_{Q, 1}}{\mu} \otimes 
\cdots \otimes
\Pauli{N_{Q, N_R}}{\mu}
\mapsto \sum_{\mu = 1}^{N_H} \: 
\HQCoeff{\mu} \: 
\mathcal{U}_\mu^{(1)} \otimes 
\cdots \otimes
\mathcal{U}_\mu^{(N_R)},
\end{equation}
where 
$\Pauli{N_{Q, j}}{\mu}$ is a Pauli word of $N_{Q, j}$ qubits,  
$ N_Q = N_{Q, 1} + \cdots + N_{Q, N_R} $, and 
$\mathcal{U}_\mu^{(j)}$ is a SNAP-displacement ansatz with a Fock cutoff $ L_j = 2^{N_{Q, j}} $.
Let us also assume that any $N_{Q, j} \leq 4$, which means the Fock cutoff for each qumode can not be more than 16, consistent with the current qumode hardware capabilities. \cite{Wang2020} 
This gives us the blueprint to generalize SNAP-VQE to a qubit Hamiltonian with an arbitrary number of qubits, as mentioned below. 
\begin{itemize}

\item The optimization problem of \Eq{\ref{eq: opt_cost_fun_snap}} is solved for all possible Pauli words of up to 4 qubits. 
This classical optimization involves matrices up to $16 \times 16$ dimensions. 
The optimal parameters are then stored as a library. 
    
\item Any qubit Hamiltonian $H_Q$ with $N_Q$ number of qubits is partitioned into subsystems, each containing no more than 4 qubits.  

\item The mapping of \Eq{\ref{eq: qubit_ham_multiple_qumodes}} is known based on the parameter library mentioned above, and a maximum of four-fold resource reduction can be achieved in terms of the qubit to qumode mapping. 

\end{itemize}
We show the optimal loss function values defined in \Eq{\ref{eq: opt_cost_fun_snap}} for all possible four-qubit Pauli words in \Fig{\ref{fig: pauli_words_snap_losses}}. 
It is clear from \Fig{\ref{fig: pauli_words_snap_losses}} that all the converged values are at least less than $10^{- 11}$, which represents the error bound for the Hamiltonian mapping.


\begin{figure}[t!]

\includegraphics[width=0.9\columnwidth]{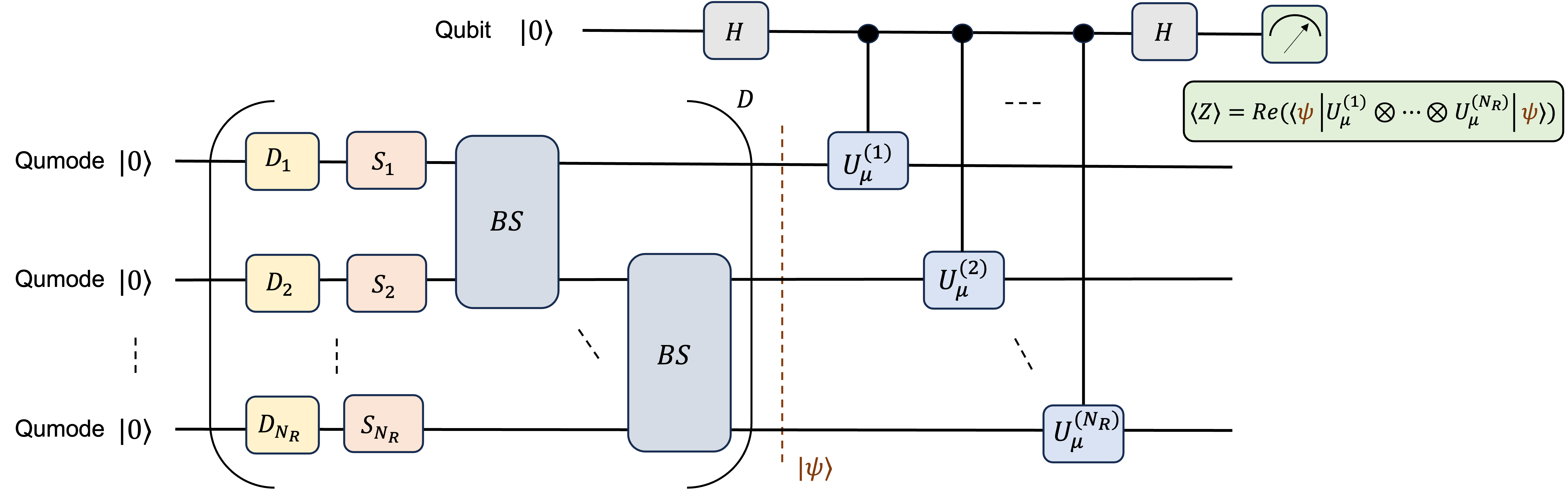}

\caption{
    Full circuit involving multiple qumodes with one qubit for computing the expectation value as defined in \Eq{\ref{eq: ev_multiple_qumodes}} based on the trial state defined in \Eq{\ref{eq: bs_snap_disp_ansatz_state}}.
    The circuit contains an additional ancilla qubit not shown here which is coupled to all qumodes.
}
\label{fig: full_circuit_ev_snap_multimodes}
\end{figure}


\begin{figure}[t!]

\includegraphics[width=0.4\columnwidth]{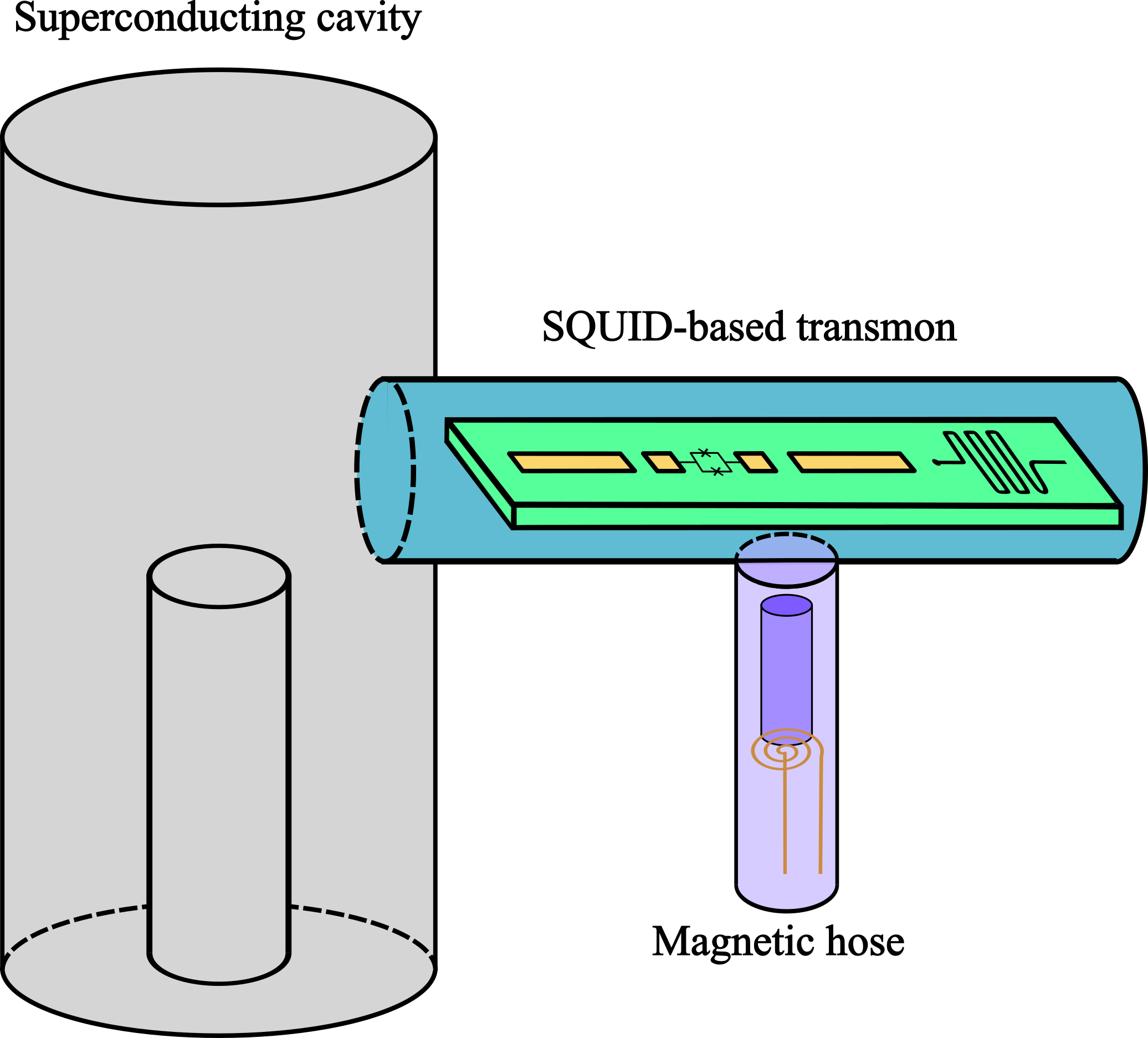}

\caption{
    Qubit-qumode hardware that can efficiently switch between different interaction regimes, as described in \Reference{\citenum{Valadares2024demand}}.
    The microwave resonator is coupled to a superconducting quantum interference device (SQUID)-based transmon, which is connected to a magnetic hose made from aluminium and mu-metal layers.   
}
\label{fig: cavity_transmon_squid}
\end{figure}


The Hamiltonian mapping to multiple qumodes discussed above also allows us to generalize the circuit from the one qumode scenario for computing the expectation value to the multimode case. 
The hardware needed involves one transmon qubit connected to multiple qumodes. \cite{Diringer2024}
Given a multi-qumode trial state $\ket{\psi}$, we can compute the expectation value using a Hadamard test approach 
\begin{equation} \label{eq: ev_multiple_qumodes}
\braket{\psi| \HamQ | \psi} 
= \sum_{\mu = 1}^{N_H} \: \HQCoeff{\mu}
\braket{\psi| \: 
\mathcal{U}_\mu^{(1)} \otimes 
\cdots \otimes
\mathcal{U}_\mu^{(N_R)} \: | \psi}
\end{equation}
using a sequence of qubit-controlled SNAP displacement unitaries $\{ \mathcal{U}_\mu \}$ followed by measuring the ancilla qubit.
The multiple-qumode trial ansatz state can be implemented by augmenting the SNAP-displacement ansatz with a two-qumode beam-splitter gate 
\begin{equation}
BS_{j, k} (\beta, \phi)
= e^{ i \frac{\beta}{2} \: \big( 
e^{i \phi} \: \BC{j} \BA{k}
+ \HermConj \big) }
\end{equation}
since universality for one qumode combined with beam-splitter implies multimode universality. \cite{You2024Crosstalk,Zhang2023energy}  
Specifically, the multimode trial state can then be written as 
\begin{subequations} \label{eq: bs_snap_disp_ansatz_state}
\begin{align}
\ket{\psi}
&= \UniAn{D} ( \mathbf{v} ) \ket{0_1, \cdots, 0_{N_R}}, 
\\
\UniAn{D} ( \mathbf{v} ) 
&= U_{BSD} ( \bm{\theta}^{(D)}, \bm{\phi}^{(D)}, \bm{\beta}^{(D)}, \bm{\phi}^{(D)} ) 
\: \cdots \: 
U_{BSD} ( \bm{\theta}^{(1)}, \bm{\phi}^{(1)}, \bm{\beta}^{(1)}, \bm{\phi}^{(1)} ), 
\\
U_{BSD} 
&= \Big[ \prod_{1 \leq j < k \leq N_R} 
BS_{j, k} (\beta_{j, k}, \phi_{j, k}) 
\Big] \: \Big[ 
\prod_{j = 1}^{N_R} 
S_{j} (\bm{\theta}_j) D_{j} (\alpha_j) \Big], 
\end{align}
\end{subequations}
where the subscripts $j, k$ represent qumodes,
$\mathbf{v}$ represents all the parameters, and $D$ is the trial ansatz circuit depth. 

We will now discuss the implementation of the universal ansatz on near-term qubit-qumode devices. 
As mentioned above, an ancilla transmon qubit is assumed for implementing all the qumode gates such as SNAP, whereas the controlled-SNAP gates for each of the qumodes can be implemented following \Fig{\ref{fig: control_snap}}.
Indeed, a tunable beam-splitter between two qumodes in the cQED formalism can be realized by coupling the resonator qumodes to a bichromatically driven superconducting transmon, \cite{Gao2018,Zhang2019}
although it can also be implemented by connecting the resonators with a superconducting coupler. \cite{Lu2023high,Chapman2023StrongWeakDispersive} 
The full circuit is illustrated in \Fig{\ref{fig: full_circuit_ev_snap_multimodes}}, where it is assumed that one transmon qubit is helping implement all one-qumode SNAP or displacement gates and two-qumode beam-splitter gates where all the other qumodes are assumed to be unaffected.  
In reality, this approach can lead to \textit{crosstalk} among the qumodes in the strongly dispersive regime where the SNAP-displacement ansatz is implemented. \cite{You2024Crosstalk}
A straightforward way to avoid the crosstalk would be to have separate ancilla transmon qubits for each of the qumodes. 
Even though this approach will need as many qubits as the qumodes, the qubits are only used for implementing qumode gates. 
There has been a pioneering development in \Reference{\citenum{Valadares2024demand}} which avoids crosstalk in qumodes by realizing efficient switching of the qubit-qumode interaction regimes without impacting the resonators.
The authors in \Reference{\citenum{Valadares2024demand}} achieve this by combining a standard resonator coupled to a superconducting quantum interference device (SQUID)-based transmon with a cleverly designed magnetic hose. \cite{Gargiulo2021fast}
The magnetic hose is a cylinder consisting of concentric aluminum and mu-metal layers and is placed perpendicular to the SQUID coax line, as illustrated in \Fig{\ref{fig: cavity_transmon_squid}}.


\begin{figure}[t!]
    \centering
    \begin{subfigure}[c]{0.45\textwidth}
    
        \centering
        \vskip-2ex
        \includegraphics[width=1.0\linewidth]{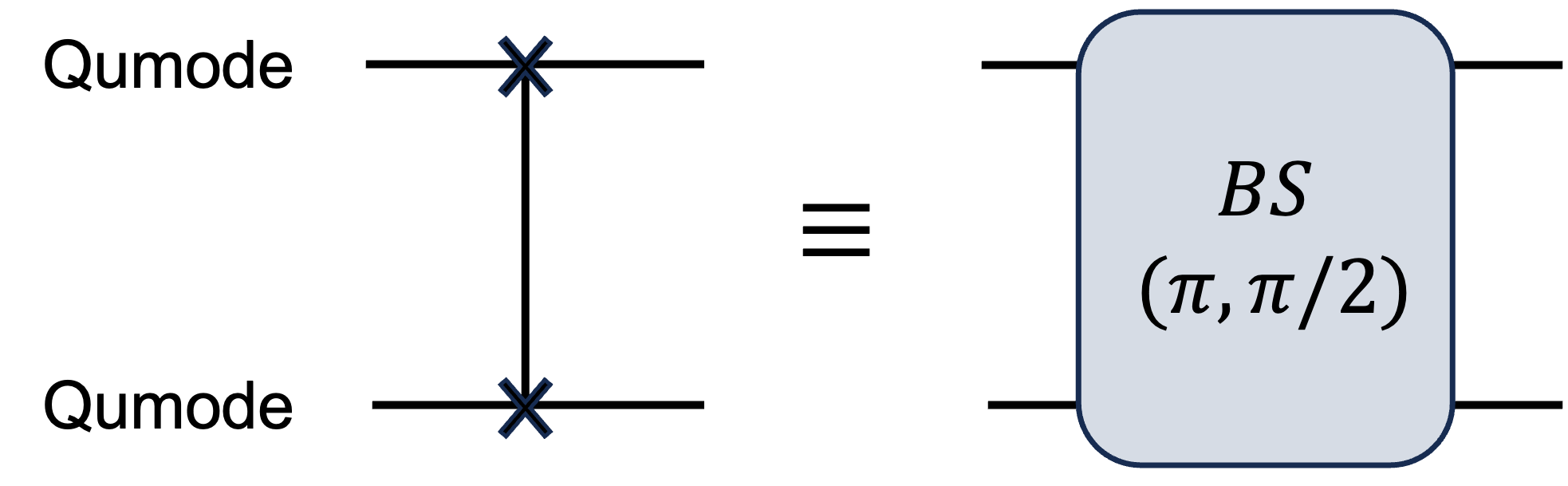}
        
    \end{subfigure}
    \hskip7ex
    \begin{subfigure}[c]{0.4\textwidth}
    
        \centering
        \vskip-2ex
        \includegraphics[width=1.0\linewidth]{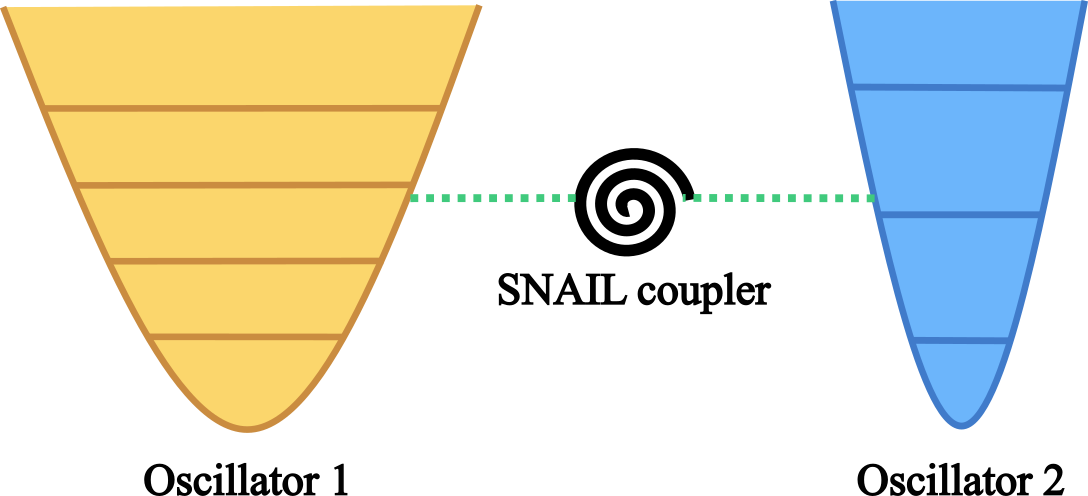}
        
    \end{subfigure}
    
    \caption{
        Left: Illustration of how a two-qumode SWAP gate can be implemented by a beam-splitter. 
        Right: Illustration of realization of beam-splitter of two microwave resonator qumodes without an ancilla transmon by coupling them with a superconducting nonlinear asymmetric inductive element (SNAIL) device.
     }
     \label{fig: bs_snail_coupler}
\end{figure}


\begin{figure}[t!]

\includegraphics[width=0.9\columnwidth]{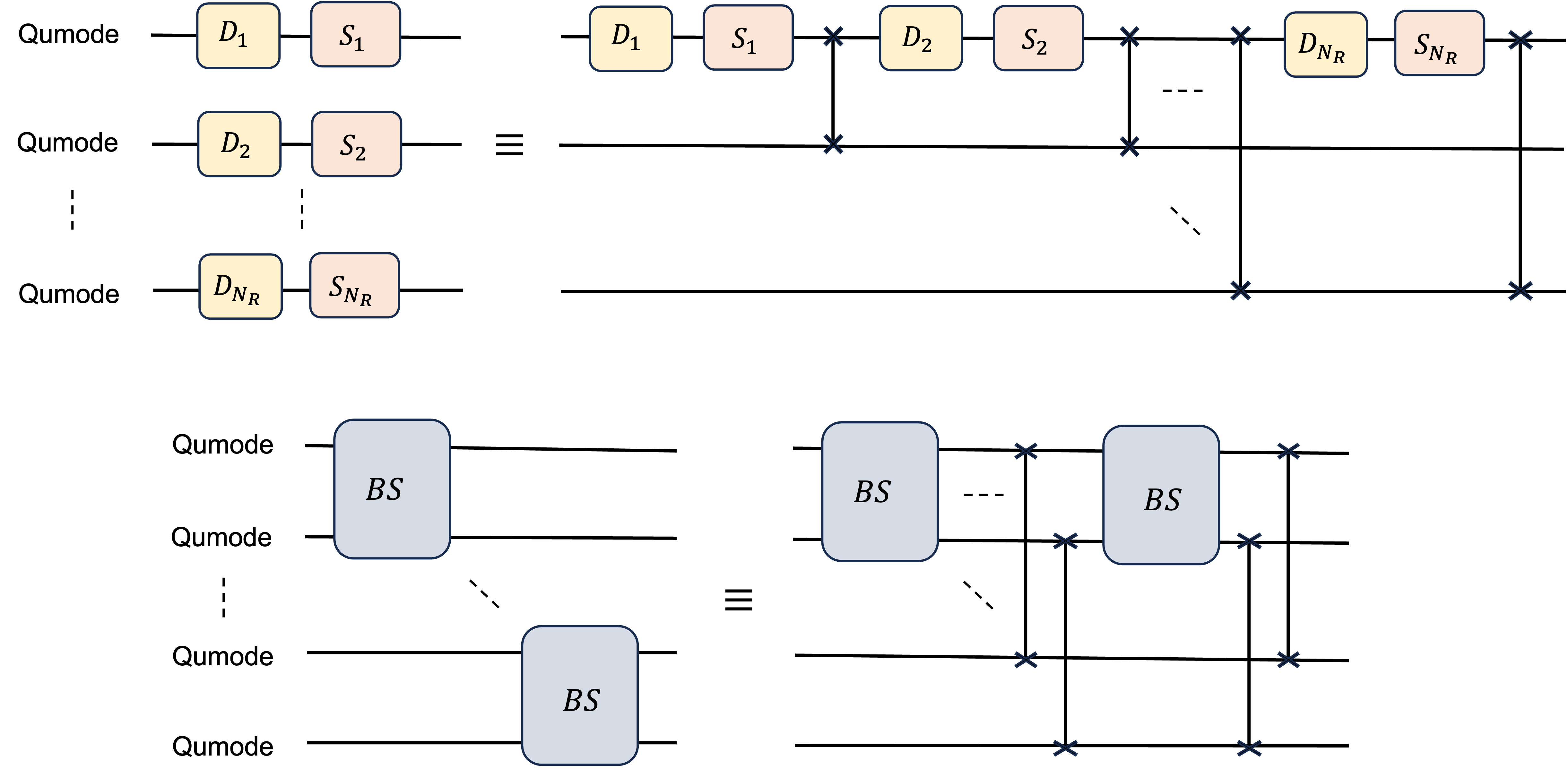}

\caption{
    A crosstalk-resistant implementation of multi-qumode trial ansatz circuit originally illustrated in \Fig{\ref{fig: full_circuit_ev_snap_multimodes}}. 
    The qumode SWAP gates make sure that all the other gates are implemented using the first two qumodes which are coupled to an ancilla transmon qubit not shown here. 
    The qumode SWAP gates between the first two and the rest of the qumodes can be realized by a superconducting coupler, as illustrated in \Fig{\ref{fig: bs_snail_coupler}}.
}
\label{fig: mm_ansatz_implement}
\end{figure}


\begin{figure}[b!]

\includegraphics[width=0.9\columnwidth]{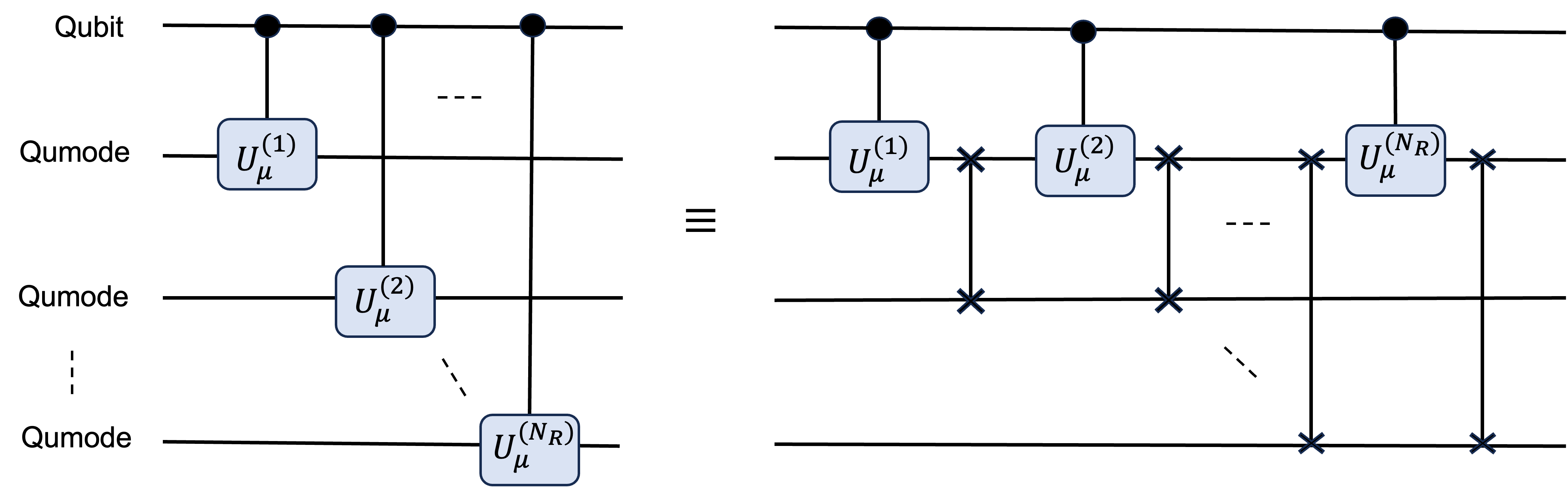}

\caption{
    A crosstalk-resistant implementation of part of the circuit corresponding to the multi-qumode Hadamard test originally discussed in \Fig{\ref{fig: full_circuit_ev_snap_multimodes}}. 
    The qumode SWAP gates make sure that all the other gates are implemented using the first qumode which is coupled to an ancilla transmon qubit not shown here. 
    The qumode SWAP gates between the first two and the rest of the qumodes can be realized by a superconducting coupler, as illustrated in \Fig{\ref{fig: bs_snail_coupler}}.
}
\label{fig: mm_control_implement}
\end{figure}


We also discuss an alternative approach below, where the ancilla transmon qubit is connected only to two \textit{primary} qumodes whereas the other qumode gates are implemented using qumode-SWAP gates.
The two-qumode SWAP gates can be implemented as $BS (\pi, \pi/2)$, \cite{Fujii2003exchange,Liu2024qumodequbitreview} 
which means that beam-splitter interactions are needed between the primary and the other qumodes for this approach.
As mentioned above, the beam-splitter between two resonators without an ancilla transmon qubit can be achieved by implementing a superconducting coupler between them, such as a superconducting nonlinear asymmetric inductive element (SNAIL) \cite{Frattini2017} as demonstrated in \Reference{\citenum{Chapman2023StrongWeakDispersive}}.
We illustrate the qumode SWAP and beam-splitter gates in \Fig{\ref{fig: bs_snail_coupler}}.
Then the multimode trial ansatz as defined in \Eq{\ref{eq: bs_snap_disp_ansatz_state}} can now be implemented using SNAP and displacement gates on the primary qumodes combined with SWAP gates with other qumodes, as illustrated in \Fig{\ref{fig: mm_ansatz_implement}}. 
Similarly, the controlled unitaries for the Hadamard test can also be implemented on the primary qumodes with SWAP gates, as illustrated in \Fig{\ref{fig: mm_control_implement}}. 
This approach is also resource-adaptive since more than two primary qumodes can be chosen by introducing additional ancilla transmon qubits based on the connectivity of the couplers connecting the primary qumodes to other qumodes. 


\begin{figure}[t!]
    \centering
    \begin{subfigure}[c]{0.47\textwidth}
    
        \centering
        \vskip-2ex
        \includegraphics[width=1.0\linewidth]{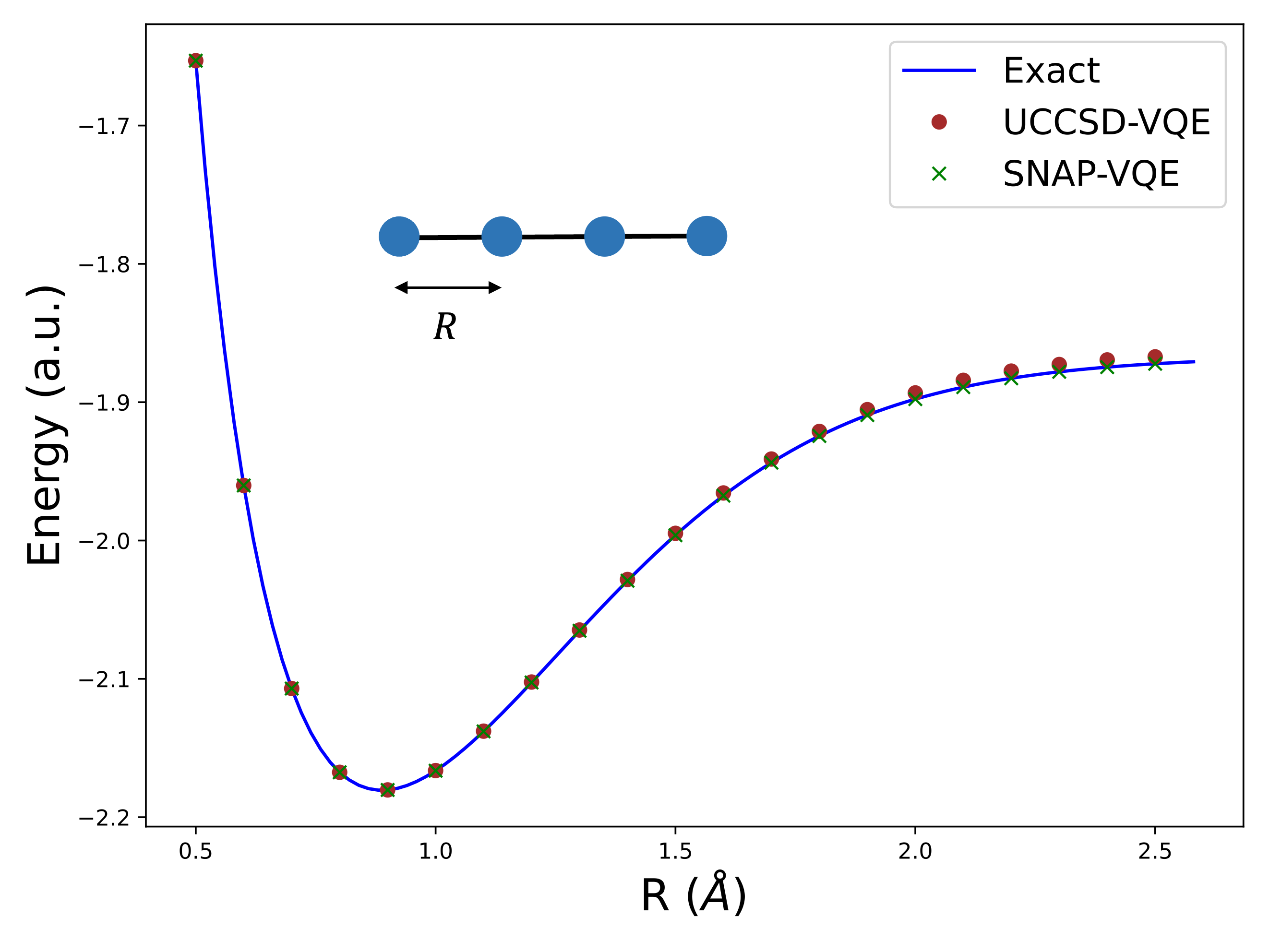}
        
    \end{subfigure}
    \hskip1ex
    \begin{subfigure}[c]{0.47\textwidth}
    
        \centering
        \vskip-2ex
        \includegraphics[width=1.0\linewidth]{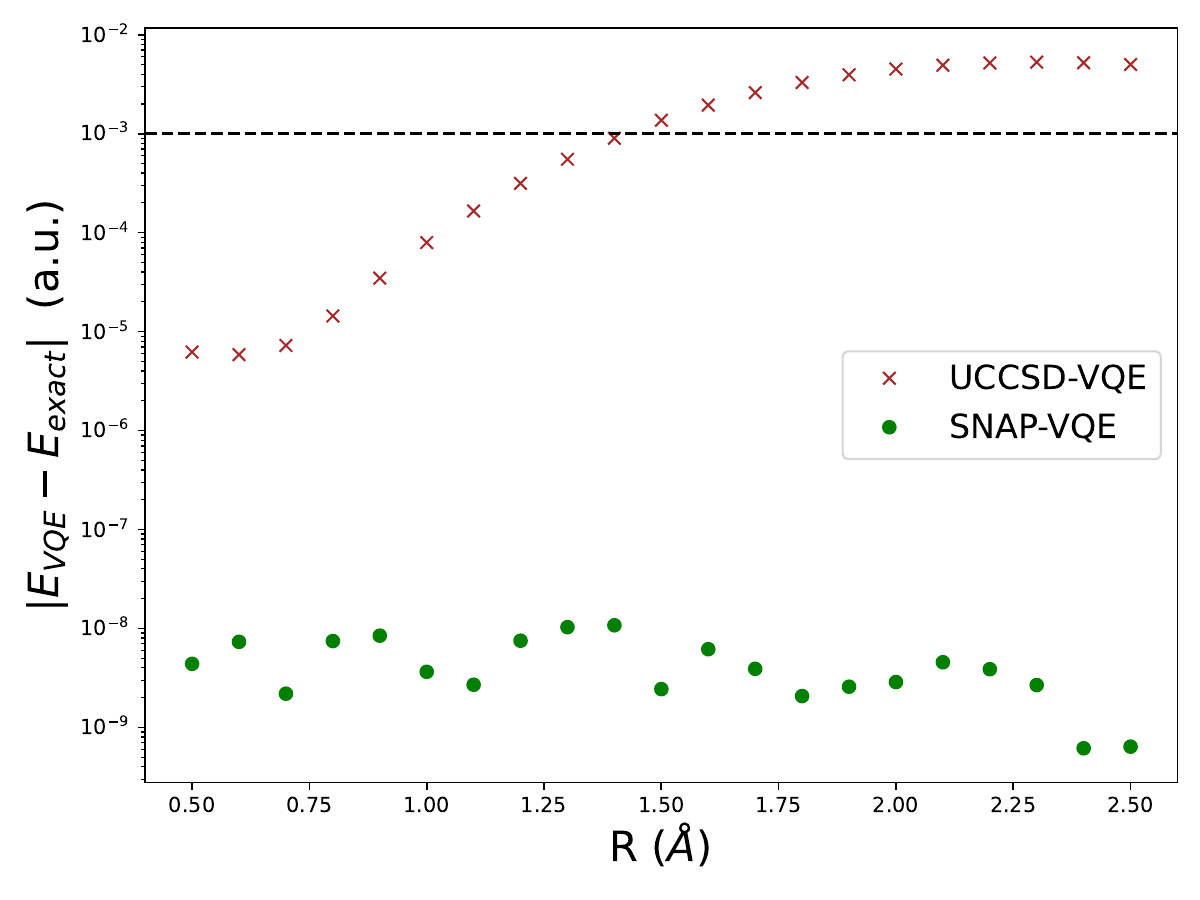}
        
    \end{subfigure}
    
    \caption{
        Comparison of linear H$_4$ molecule ground state energies in STO-3G minimal basis using qubit-based UCCSD-VQE and qumode-based SNAP-VQE approach discussed in \Sec{\ref{sec: multimode_snap}}.
        All the H--H bond distances are assumed to be the same and plotted on the horizontal axis. 
        The multimode ansatz circuit for the SNAP-VQE has a depth of $D = 20$ for the trial state preparation part.
        The black horizontal line in the right panel represents the minimum energy error needed for chemical accuracy, usually in the miliHartree range.
     }
     \label{fig: h4_vqe_en}
\end{figure}


We apply our multimode approach to the ground states for the dissociation of linear H$_4$ molecule, where all the H-H bonds are assumed the same.
Although this is a simple system with four electrons and eight spin-orbitals in a minimal basis, it is known to be a challenging problem when the molecule dissociates giving rise to strongly correlated electronic systems. \cite{Vaquero2024physically}
Indeed, the classical gold-standard electronic structure methods such as traditional coupled cluster completely fail to describe the dissociation curves of hydrogen chains, making it a family of benchmark problems for strong electron correlation. \cite{Motta2017Simons}
After the JWT mapping, the qubit Hamiltonian for the H$_4$ molecule in STO-3G basis can be represented in the form of \Eq{\ref{eq: qubit_ham}} with $N_H = 185$ Pauli word terms. 
Each of the resulting eight-qubit Pauli words is then mapped to two qumodes. 
Thus, each of the qumode represented four qubits with a Fock cutoff $L = 16$. 
We compare our multimode SNAP-VQE for the H$_4$ molecule with qubit-based VQE based on the unitary coupled cluster with singles and doubles (UCCSD) in \Fig{\ref{fig: h4_vqe_en}}. \cite{Peruzzo2014}
The SNAP-VQE was implemented using QuTip and OpenFermion with the BFGS optimizer as implemented using TensorFlow, \cite{tensorflow2015-whitepaper}
whereas UCCSD-VQE was implemented using Qiskit with a limited memory variant of the BFGS optimizer. \cite{Javadi2024quantum}
The trial states for SNAP-VQE with $D = 20$ blocks provide highly accurate ground state energies compared to the exact energies even when the molecule dissociates. 
The depth for the qubit-based UCCSD scales as $\ComCom{N^2 N_Q^3}$, where $N$ is the number of electrons and $N_Q$ is the number of qubits. 
It is clear from the right panel of \Fig{\ref{fig: h4_vqe_en}} that in the strong correlation regime for H$_4$, the SNAP-VQE approach outperforms the qubit-based UCCSD-VQE energy errors which stay barely close to the chemical accuracy range.  


\section{Discussions} \label{sec: final}  

We have introduced a general scheme for mapping the molecular electronic structure Hamiltonian in terms of unitary operations native to bosonic quantum devices such as in the cQED architecture involving microwave resonators coupled with transmon qubits with the help of fermion to qubit mapping such as the Jordan--Wigner transformation. 
Our work opens the door for simulating molecular electronic structure, and by extension, any many-fermion or many-qubit system, on bosonic quantum devices that use qumodes as the building blocks of quantum information.

After mapping the fermionic Hamiltonian to a qumode Hamiltonian, one can consider the electronic structure of interest as a bosonic problem and apply a bosonic ansatz as the trial state for variationally finding the ground state using a classical-quantum hybrid approach.
This is related to the qubit coupled cluster approach, where a hardware-efficient qubit ansatz is used after mapping the electronic structure Hamiltonian to a qubit Hamiltonian. \cite{Ryabinkin2018}
We have shown how to compute the expectation values using a hybrid quantum device involving one qumode coupled with up to two ancilla qubits.
We have also discussed two different hybrid qubit-qumode universal ansatze that efficiently reproduce the exact energies for the potential energy surface of the H$_2$ molecule.
The SNAP-VQE approach is shown to be more robust than the ECD-VQE from the perspective of mapping the qubit Hamiltonian. 
It is possible to numerically map a Pauli word to SNAP-displacement gates with manageable circuit depth, whereas the ECD-rotation needs a linear combination of unitary approach to make the circuit depth practical. 

The robustness of the SNAP-VQE approach inspired us to generalize the qubit to qumode mapping to multiple qumodes. 
We have introduced a modular approach to partition an arbitrary number of qubits into a few subsystems, each mapped to the Fock basis of a qumode.  
We have shown that the expectation values can be computed by a sequence of qubit-controlled one-qumode gates followed by measurements in the ancilla qubit. 
This allowed us to introduce a multi-qumode universal circuit as the trial state ansatz. 
The universality for the multi-qumode ansatz is achieved by combining the SNAP-displacement gates with two-qumode beam-splitter interactions.
We have proposed three hardware-efficient approaches to tackle crosstalk constraints between multiple qumodes: using one ancilla transmon qubit per qumode, adapting recent progress on SQUID-based transmons, and employing qumode-SWAP gates to minimize the qubit-qumode connectivity required for high-fidelity operations.

We have applied our multi-qumode SNAP-VQE approach to the linear H$_4$, a benchmark molecular electronic system that shows strong electron correlation when all three H-H bonds are stretched. 
We have numerically demonstrated that the SNAP-VQE approach can provide highly accurate ground state energies for the linear H$_4$ molecule with two qumodes coupled to one qubit instead of eight qubits in a qubit-centric approach, with a manageable number of qumode gates.  
We concluded by comparing our method with the benchmark qubit-based UCCSD-VQE method with a noiseless simulator, which suggests that our approach has the potential to outperform the traditional qubit-based variational approaches for molecular electronic structure in terms of fewer quantum resources and circuit depth.
An important aspect of the qumode gates we have explored here would be understanding their optimization landscape for VQE,\cite{Zhang2023energy} which we leave for future development. 
We remark that the circuits for the two examples presented in our work (H$_2$ molecule and linear H$_4$ chain) are shallow such that the intrinsic noisiness of quantum computers, especially photon loss in the qumode resonators, is less significant. 
We leave the exploration of the effects of noise on our qumode approaches to future work. 


\begin{acknowledgement}

We acknowledge support from the NSF for the Center for Quantum Dynamics on Modular Quantum Devices (CQD-MQD) under grant number CHE-2124511. 
R.D. thanks Brandon Allen and Francesco Calcagno for useful discussions. 
N.P.V. acknowledges support from the Lafayette College Bergh Family Fellows Program.

\end{acknowledgement}


\begin{suppinfo}

Additional theoretical details including overview of the electronic structure problem, 
justification and derivation of the DMS operator mapping, bosonic Hamiltonian for the H$_2$ molecule alongside illustrations of the parametric dependence of its coefficients (Figure S1), heatmap of the Hamiltonian matrix elements (Figure S2), cQED-based subspace tomography for photon transfer expectation values, hybrid variational approach with the universal bosonic ansatz for two qumodes and one ancilla qubit (Figure S3), and comparison of ground state energies of H$_2$ molecule with various approaches (Figure S4). 

\end{suppinfo}


\section*{Code and Data Availability}

The Python code and data for the ansatz optimization, state preparation simulations and figure generation can be found at \href{https://github.com/CQDMQD/qumode_est_paper}{https://github.com/CQDMQD/rishabdchem/qumode\_est\_paper}.


\appendix


\section{Direct mapping from fermions to bosons} \label{app: direct_mapping}

We will assume real-valued molecular orbitals from now on which leads to the following relations between electron integral tensor elements 
\begin{subequations} \label{eq: elec_integral_symmetries}
\begin{align}
\HOne{p}{q}
&= \HOne{q}{p},    
\\
\HTwo{pq}{rs}
&= \HTwo{pr}{qs}
= \HTwo{sq}{rp}
= \HTwo{sr}{qp},
\end{align}    
\end{subequations}
in addition to $ \HTwo{pq}{rs} = \HTwo{qp}{sr} $ due to the indistinguishability of electrons. 
It is then possible to write $\HElec$ in an alternate form  
\begin{align} \label{eq: molecular_ham_compact}
\HElec  
&= \Big[ \frac{1}{2} \: \sum_{p} \HOne{p}{p} \: \FC{p} \FA{p}    
+ \sum_{p > q} \Big( 
\HOne{p}{q} \: \FC{p} \FA{q} 
+ \frac{1}{2} \: \HATwo{pq}{pq} \: \FC{p} \FC{q} \FA{p} \FA{q} \Big) \nonumber
\\
&+ \sum_{p > q > r} \Big(  
\HATwo{pq}{pr} \: \FC{p} \FC{q} \FA{p} \FA{r} 
+ \HATwo{pq}{qr} \: \FC{p} \FC{q} \FA{q} \FA{r} 
+ \HATwo{pr}{qr} \: \FC{p} \FC{r} \FA{q} \FA{r} \Big) \nonumber
\\
&+ \sum_{p > q > r > s} \Big( 
\HATwo{pq}{rs} \: \FC{p} \FC{q} \FA{r} \FA{s} 
+ \HATwo{pr}{qs} \: \FC{p} \FC{r} \FA{q} \FA{s} 
+ \HATwo{ps}{qr} \: \FC{p} \FC{s} \FA{q} \FA{r} \Big) 
\Big] + \HermConj,
\end{align} 
where we have defined  
$ \HATwo{pq}{rs} 
\equiv \HTwo{pq}{rs} - \HTwo{pq}{sr} $ such that
\begin{equation} \label{eq: as_integral_symmetries}
\HATwo{pq}{rs} 
= \HATwo{qp}{rs} 
= - \HATwo{pq}{sr} 
= \HATwo{rs}{pq}, 
\end{equation}
and $\HermConj$ represents the Hermitian conjugate of its preceding operator term.

Each term of the electronic Hamiltonian in \Eq{\ref{eq: molecular_ham_compact}} has creation and annihilation operators in pairs, which reflects the fact that $\HElec$ is number-conserving. 
Let us define the bilinear fermionic operators \cite{Jordan1935} 
\begin{equation} \label{eq: bilinear_op}
\BiF{p}{q} 
\equiv \FC{p} \FA{q}
= ( \BiF{q}{p} )^\dagger,
\end{equation}
which is equivalent to the number operator when $p = q$ and generalized singles excitation otherwise. \cite{Nakatsuji2000}
A set of $\{ \BiF{p}{q} \}$ can be successively applied to transform between any two Slater determinants with the same number of electrons. 
The bilinear fermionic operators follow a simple commutation relation 
\begin{equation} \label{eq: bilinear_commutation}
[ \BiF{p}{q}, \BiF{r}{s} ] 
= \delta_{qr} \: \BiF{p}{s} 
- \delta_{ps} \: \BiF{r}{q},
\end{equation}
and generate the $u(M)$ Lie algebra, \cite{Fukutome1981} 
where $M$ is the number of spin orbitals.
Thus, we can rewrite the electronic Hamiltonian as  
\begin{align} \label{eq: molecular_ham_bilinear}
\HElec 
&= \Big[ \frac{1}{2} \: \sum_{p} \HOne{p}{p} \: \BiF{p}{p} 
+ \sum_{p > q} \Big( 
\HOne{p}{q} \: \BiF{p}{q}
+ \frac{1}{2} \: \HATwo{pq}{qp} \: \BiF{p}{p} \BiF{q}{q} \Big) \nonumber
\\
&+ \sum_{p > q > r} \Big(  
\HATwo{pq}{rp} \: \BiF{p}{p} \BiF{q}{r} 
+ \HATwo{pq}{qr} \: \BiF{q}{q} \BiF{p}{r} 
+ \HATwo{pr}{rq} \: \BiF{r}{r} \BiF{p}{q} \Big) \nonumber
\\
&+ \sum_{p > q > r > s} \Big( 
\HATwo{pq}{sr} \: \BiF{p}{r} \BiF{q}{s} 
+ \HATwo{pr}{sq} \: \BiF{p}{q} \BiF{r}{s}
+ \HATwo{ps}{rq} \: \BiF{p}{q} (\BiF{r}{s} )^\dagger \Big) 
\Big] + \HermConj, 
\end{align}
where we have applied the adjoint relation for the bilinear fermionic operators and the relations in \Eq{\ref{eq: as_integral_symmetries}}.
Thus, we have written \Eq{\ref{eq: molecular_ham_bilinear}} in such a way that the knowledge about the bilinear fermionic operators $\{ \BiF{p}{q} \}$ with $p \geq q$ is sufficient to represent $\HElec$.

Our goal is to map \Eq{\ref{eq: free_fermion_state}} to a bosonic state and \Eq{\ref{eq: molecular_ham}} to a bosonic Hamiltonian, so that the molecular electronic structure problem can be tackled with bosonic quantum computers.  
The key results for the direct mapping are given below. 
\begin{itemize}
    
\item A system with $N$ fermions can be mapped to a system of $N$ quantum harmonic oscillators (QHOs) or bosonic modes with a maximum of $M - N + 1$ oscillator levels for each. 

\item An exact injective state mapping exists between Slater determinants and Fock states of QHOs. 

\item An exact mapping exists between $\{ \BiF{p}{q} \}$ and Fock state projection operators of QHOs. 

\end{itemize}

\subsection{State mapping}


\begin{figure}[t]
    \centering
    \includegraphics[width=0.9\textwidth]{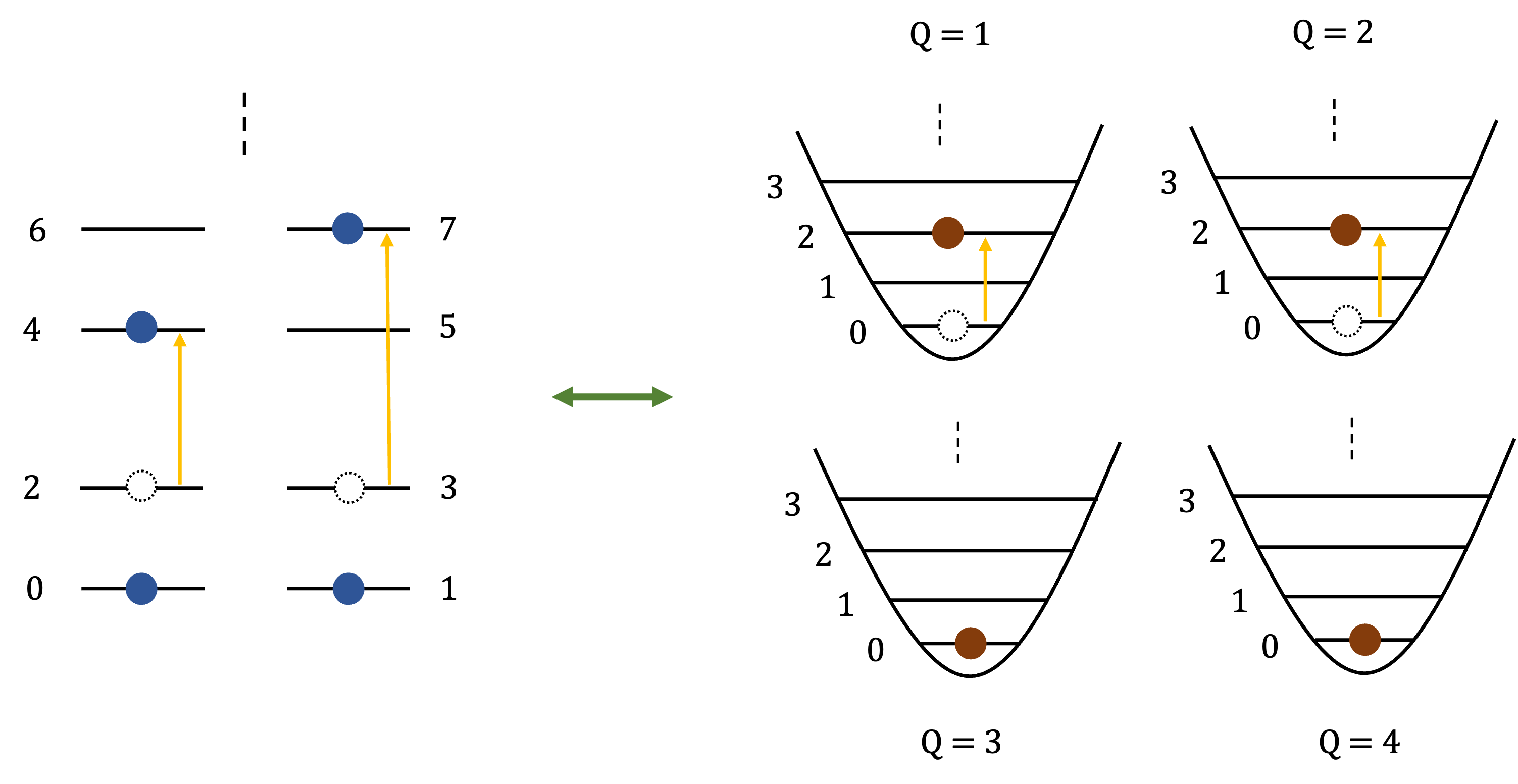}

\caption{
    State diagrams corresponding to the state mapping as defined in \Eq{\ref{eq: state_mapping}} and \Eq{\ref{eq: fb_index_mapping}}.
    Here, a system with $N = 4$ electrons is mapped to a system of four quantum harmonic oscillators.
    In the initial state, the Slater determinant $\ket{0, 1, 2, 3}_F$ corresponds to four electrons occupying the lowest four spin-orbitals, which is mapped to the oscillator vacuum state $\ket{0, 0, 0, 0}_B$.
    When some of the occupied spin-orbitals are now excited to get the Slater determinant $\ket{0, 1, 4, 7}_F$, it gets mapped to the Fock state $\ket{2, 2, 0, 0}_B$. 
    The occupied spin-orbitals for electrons and excitation levels for oscillators are represented by blue and brown circles, respectively.
}
\label{fig: state_map}
\end{figure}


Elementary bosonic operators follow the canonical commutation relation (CCR) 
\begin{subequations} \label{eq: ccr}
\begin{align}
[ \BC{p}, \BC{q} ] 
&= \BC{p} \BC{q} - \BC{q} \BC{p}  
= 0, \quad  
\\
[ \BA{p}, \BC{q} ]
&= \BA{p} \BC{q} - \BC{q} \BA{p}  
= \delta_{pq},
\end{align}
\end{subequations}
where $\BC{p}$ and $\BA{p}$ are the bosonic creation and annihilation operators.
These operators are defined such that their action on the Fock state $\{ \ket{q} | \: 0 \leq q \leq \infty \}$ of a single qumode is 
\begin{subequations} \label{eq: action_boson_ops}
\begin{align}
\BC{} \ket{q}_B 
&\equiv \sqrt{q + 1} \ket{q + 1}_B, 
\\  
\BA{} \ket{q}_B 
&\equiv \sqrt{q} \ket{q - 1}_B, \quad q > 0
\\
\BA{} \ket{0}_B 
&\equiv 0, 
\end{align}
\end{subequations}
which can be trivially generalized to multimodal bosonic systems by taking tensor products of single-mode Fock states.
Similar to the fermionic case, the bosonic mode indices in \Eq{\ref{eq: ccr}} represent an orthogonal one-particle basis.
Each of the bosonic or QHO mode can have inifinite levels or occupancies since there is no nilpotency in the CCR.
Thus, we define a Fock state of $N$ QHOs as
\begin{equation} \label{eq: free_boson_state}
\kIndB{q}{N}
\equiv \frac{ ( \BC{1} )^{q_1} \cdots ( \BC{N} )^{q_N} }{ \sqrt{q_1! \cdots q_N! }} \kVacB,    
\end{equation}
where $\kVacB$ is the ground state of the $N$ oscillators. 
It should be noted that we only require a bosonic Fock basis of \Eq{\ref{eq: free_boson_state}} for this paper, and our approach is agnostic of the properties of the underlying oscillators such as their anharmonicity.

The indices in \Eq{\ref{eq: free_boson_state}} represent occupied levels of each mode, in contrast to \Eq{\ref{eq: free_fermion_state}} where the occupied modes themselves are indexed.
This distinction between \Eq{\ref{eq: free_fermion_state}} and \Eq{\ref{eq: free_boson_state}} is simply the result of Pauli exclusion principle and how we have chosen the index ordering in \Eq{\ref{eq: free_fermion_state}}.
For example, the bosonic states 
$ \ket{2, 2}_B, \ket{2, 3}_B, \ket{3, 2}_B $, and $ \ket{3, 3}_B $ are all legitimate bosonic states involving the second and third levels of two bosonic modes, whereas $ \ket{2, 3}_F $ is the only legitimate fermionic state involving the second and third spin-orbitals. 
Note that 
$ \ket{2, 3}_F 
= \FC{2} \FC{3} \kVacP  
= - \FC{3} \FC{2} \kVacP $ 
respects the permutation of its underlying operators and the resulting sign change, which is similarly true for any Slater determinant defined in \Eq{\ref{eq: free_fermion_state}}.  

An injective state mapping exists between Slater determinants of $N$ fermions defined in \Eq{\ref{eq: free_fermion_state}} and state of $N$ QHOs defined in \Eq{\ref{eq: free_boson_state}} \cite{Ohta1998,Suryanarayana2006} 
\begin{equation} \label{eq: state_mapping}
\kIndF{p}{N}
\leftrightarrow \kIndB{q}{N},  
\end{equation}
where the relation between the two sets of indices are
\begin{subequations} \label{eq: fb_index_mapping}
\begin{align}
q_j 
&= p_1, \quad \text{if} \quad j = N,
\\    
&= p_{N - j + 1} - p_{N - j} - 1, \quad \text{otherwise}.
\end{align}
\end{subequations}
We refer the reader to the \SuppInfo
for a justification of the state mapping.
Clearly, the Fermi vacuum 
$\kVacF$ maps to the Fock ground state 
$\kVacB$ 
following \Eq{\ref{eq: fb_index_mapping}}.
Then the physical interpretation of \Eq{\ref{eq: state_mapping}} is that the holes created from the Fermi vacuum and their impact on it are regarded as bosonic excitations, as in photoelectron spectroscopy.
A schematic for an example state mapping of a system with $N = 4$ electrons is shown in \Fig{\ref{fig: state_map}}. 

It is thus possible to apply the state mapping of \Eq{\ref{eq: state_mapping}} to map the full configuration interaction (FCI) state for an $N$-fermion system as
\begin{equation} \label{eq: fci_boson_mapping}
\ket{\Psi}
= \sum_{1 \leq p_1 < \cdots < p_N \leq M} C_{p_1 \cdots p_N} \kIndF{p}{N}
\mapsto \sum_{1 \leq p_1 < \cdots < p_N \leq M} C_{p_1 \cdots p_N} \kIndB{q}{N},
\end{equation}
where the scalars $\{ C_{p_1 \cdots p_N} \}$ are the FCI coefficients and the $\{ q_j \}$ indices are defined in \Eq{\ref{eq: fb_index_mapping}}.
Since any $N$-fermion state can be represented as a special case of FCI, \Eq{\ref{eq: fci_boson_mapping}} allows mapping any state corresponding to a fermionic system with a fixed particle number to a bosonic state with the number of modes same as the number of fermions.
Based on \Eq{\ref{eq: fb_index_mapping}}, it is easy to see that the highest integer corresponding to the indices $\{ q_j \}$ in \Eq{\ref{eq: fci_boson_mapping}} is
$ L  = M - N $.
Thus, the state mapping naturally truncates the dimension of the Fock basis, i.e., number of qumode levels, based on the $M$ number of spin-orbitals for a given electronic system, which makes the relevant bosonic Hilbert space isomorphic to the Hilbert space of $N$ qudits \cite{Wang2020Qudits} of $M - N + 1$ dimension. 

As a specific example, let us discuss the state mapping of \Eq{\ref{eq: state_mapping}} for a system with $N = 2$ with arbitrary $M > 2$ number of spin-orbitals.
Mapping between an arbitrary Slater determinant 
\begin{equation} \label{eq: free_fermion_two_els} 
\ket{p, q}_F 
\equiv \FC{p} \FC{q} \kVacP,    
\end{equation}
and the mapped state of two qumodes 
\begin{equation} \label{eq: free_boson_two_els}
\ket{j, k}_B 
\equiv \frac{1}{\sqrt{j! \: k!}} \: (\BC{1})^j \: (\BC{2})^k \ket{0, 0}_B   
\end{equation}
is given by the following relations 
\begin{equation} \label{eq: state_mapping_two_els}
p = k, \quad
j = q - p - 1, \quad 
q = j + k + 1.    
\end{equation} 
For example, if $M = 4$, then the transformations are  
\begin{subequations} \label{eq: state_map_dihydrogen}
\begin{align}
\ket{0, 1}_F 
&\leftrightarrow \ket{0, 0}_B, \quad 
\ket{0, 2}_F 
\leftrightarrow \ket{1, 0}_B, \quad
\ket{0, 3}_F
\leftrightarrow \ket{2, 0}_B, 
\\
\ket{1, 2}_F
&\leftrightarrow \ket{0, 1}_B, \quad 
\ket{1, 3}_F
\leftrightarrow \ket{1, 1}_B, \quad 
\ket{2, 3}_F
\leftrightarrow \ket{0, 2}_B.
\end{align}    
\end{subequations}
We have so far focused on mapping a Slater determinant into a multimodal bosonic state, but as evident from \Eq{\ref{eq: state_mapping}}, the reverse is also true. 
For example, we write the bosonic states that did not appear in \Eq{\ref{eq: state_map_dihydrogen}} but still correspond to two harmonic oscillator modes with three levels below
\begin{equation}
\ket{1, 2}_B 
\leftrightarrow \ket{2, 4}_F, \quad 
\ket{2, 1}_B 
\leftrightarrow \ket{1, 4}_F, \quad 
\ket{2, 2}_B 
\leftrightarrow \ket{2, 5}_F, 
\end{equation}
which are mapped to a Slater determinants of a $N = 2$ system that have $M > 4$ spin-orbitals. 


\subsection{Operator mapping} \label{sec: op_map}  

The Dhar--Mandal--Suryanarayana (DMS) transformation maps $\{ \BiF{p}{q} \}$ operators into Fock state projection operators of QHOs. \cite{Dhar2006} 
The DMS transformation was derived from the state mapping of \Eq{\ref{eq: state_mapping}} in \Reference{\citenum{Dhar2006}}. 
We simply state the resulting expressions of the DMS transformation here and refer the reader to 
the \SuppInfo for more insight into its derivation. 

Let us define the bosonic Fock space projection operator corresponding to a given set of $k$ harmonic oscillator modes as 
\begin{equation} \label{eq: multimode_proj_op}
\BigP{r_1, \cdots, r_k}  
\equiv \KetBra{r_1, \cdots, r_k}, 
\end{equation}
and similarly define a related operator as 
\begin{equation} \label{eq: conditional_fock_projection}
\BigC{k, a}
= \sum_{ \substack{r_1 + \cdots + r_k \\
= \: a }} \BigP{r_1, \cdots, r_k},   
\end{equation}
where each index $\{ r_j \: | \: 0 \leq r_j \leq L \}$ has a specific range based on the highest physical mode level that needs to be accessed following the state mapping. 
The expectation value of the operator in \Eq{\ref{eq: conditional_fock_projection}} can be computed from the same set of photon number measurements.
Let us also denote the identity operator acting on the first $k$ harmonic oscillator modes as 
\begin{equation} \label{eq: multimode_identity_op}
\BigEYE{k}  
\equiv \EYE_1 \: \otimes \: \cdots \: \otimes \: \EYE_k.   
\end{equation}
Then the DMS mapping for any number operator $\BiF{p}{p}$ is 
\begin{equation} \label{eq: dms_map_diag}
\BiF{p}{p}
\mapsto \BigEYE{N - 1} \otimes \KetBra{p} 
+ \sum_{k = 1}^{N - 1} \: \BigEYE{k - 1} \otimes \BigC{N - k + 1, p - N + k},
\end{equation}
where $0 \leq p \leq M - 1$.
Thus, there is $N$ number of operator terms in \Eq{\ref{eq: dms_map_diag}} of the form defined in \Eq{\ref{eq: conditional_fock_projection}}. 
Operator terms involving more than one number operators can be similarly expressed and simplified due to the projection operator in \Eq{\ref{eq: dms_map_diag}}.

Let us now define the \textit{normalized} bosonic creation and annihilation operators
\begin{subequations} \label{eq: normalized_boson_ops}
\begin{align}
\NBC{} \ket{q}_B 
&\equiv \ket{q + 1}_B, 
\\  
\NBA{} \ket{q}_B 
&\equiv \ket{q - 1}_B, \quad q > 0
\\
\NBA{} \ket{0}_B 
&\equiv 0, 
\end{align}
\end{subequations}
which can easily be extended for multimodal systems.
The DMS mapping expression for the $p > q$ case consists of Fock projection operators as in \Eq{\ref{eq: dms_map_diag}} with the $\{ \NBC{k}, \NBA{k} \}$ operators. 
We show an example of the generalized singles excitation mapping with $q = p + 1$ below 
\begin{align}
\BiF{q + 1}{q}
&\mapsto \NBC{1} \: \BigC{N, q - N + 1}     
+ \sum_{k = 1}^{N - 1} \: 
\BigEYE{k - 1} \otimes 
\sum_a \: \NBA{k} \: \BigP{q + a} 
\otimes \NBC{k + 1} \: \BigC{N - k, q - N + k + 1}, 
\end{align}
and state the general expression for the mapping of $\{ \BiF{p}{q} \}$ operators in the Supplementary Information.

As a specific example, let us discuss the DMS operator mapping for the specific case of $N = 2$ with an arbitrary $M > 2$ number of spin-orbitals. 
The number operators can be mapped as 
\begin{equation} \label{eq: dms_map_diag_two_els}
\BiF{p}{p} 
\mapsto \EYE \otimes \ket{p} \bra{p} 
+ \sum_{j + k \: = \: p - 1} \ket{j, k} \bra{j, k},
\end{equation}
where $p = 0, 1, \cdots, M-1$. 
The off-diagonal fermionic bilinear operators can be mapped as
\begin{align} \label{eq: dms_map_offdiag_two_els}
\BiF{q + p}{q} 
&\mapsto (\NBC{1})^p 
\sum_{j + k \: = \: q - 1} \ket{j, k} \bra{j, k} 
+ \NBA{1}^p \: (\NBC{2})^p \: \sum_j \: \ket{j + p, q} \bra{j + p, q} \nonumber
\\ 
&- \sum_{j = 0}^{p - 2} \: (\NBC{1})^{p - 2 - j} \: \NBA{1}^j \: (\NBC{2})^{j + 1} \ket{j, q} \bra{j, q},
\end{align}
where $q = 0, 1, \cdots, M-1$ and $p = 1, 2, \cdots, M-q-1$.
It is also possible to have an alternate representation of the DMS mapping for the $N = 2$ case by applying \Eq{\ref{eq: normalized_boson_ops}}
\begin{subequations} \label{eq: dms_map_alt_two_els}
\begin{align}
\BiF{p}{p} 
&\mapsto \EYE \otimes \ket{p} \bra{p} 
+ \sum_{j + k \: = \: p - 1} \ket{j, k} \bra{j, k}, 
\\ 
\BiF{q + p}{q} 
&\mapsto \sum_{j + k = q - 1} \ket{j + p, k} \bra{j, k} 
+ \sum_j \: \ket{j, q + p} \bra{j + p, q} \nonumber
\\ 
&- \sum_{j = 0}^{p - 2} \ket{p - 2 - j, q + j + 1} \bra{j, q}.
\end{align}
\end{subequations}
We mention two examples of mapping the number operators below 
\begin{subequations} 
\begin{align}
\BiF{0}{0} 
&\mapsto \EYE \otimes \ket{0} \bra{0},
\\ 
\BiF{1}{1}
&\mapsto \EYE \otimes \ket{1} \bra{1}
+ \ket{0, 0} \bra{0, 0}.  
\end{align}
\end{subequations}
Similarly, the off-diagonal bilinear fermionic operators can be mapped, with two examples given below 
\begin{subequations}
\begin{align}
\BiF{1}{0} 
&\mapsto \sum_{j = 1}^L \: \ket{j - 1, 1} \bra{j, 0},
\\ 
\BiF{2}{0} 
&\mapsto \Big( \sum_{j = 1}^{L - 1} \: \ket{j - 1, 2} \bra{j + 1, 0} \Big)
- \ket{0, 1} \bra{0, 0},
\end{align}
\end{subequations}
where we truncated the expansion based on the highest relevant level of the bosonic modes.
We refer the reader to the \SuppInfo for applying the direct mapping to the electronic structure Hamiltonian of the dihydrogen molecule in a minimal basis.


\section{Matrix representation of bosonic operators} \label{app: bosonic_matrices}

We review the finite matrix representation of bosonic operators in the Fock basis, where the matrix dimensions are $L \times L$ with $L$ being the Fock cutoff chosen for the qumode. 
For a bosonic operator $\mathcal{O}$, the matrix elements are given by 
$ \mathcal{O}_{n, m} 
= \braket{n | \mathcal{O} | m } $, 
where $\{ \ket{n} \}$ are the Fock basis states.
The matrices for the bosonic creation and annihilation operators are
\begin{subequations} \label{eq: bosonic_elementary_matrices}
\begin{align}
\BA{} 
&= \begin{pmatrix}
0 & 0 & 0 & \cdots & 0 
\\
\sqrt{1} & 0 & 0 & \cdots & 0 
\\
0 & \sqrt{2} & 0 & \cdots & 0 
\\
\vdots & \vdots & \vdots & \ddots & \vdots \\
0 & 0 & 0 & \cdots & \sqrt{L-1}
\end{pmatrix},
\\
\BC{}
&= \begin{pmatrix}
0 & \sqrt{1} & 0 & \cdots & 0 \\
0 & 0 & \sqrt{2} & \cdots & 0 \\
0 & 0 & 0 & \ddots & 0 \\
\vdots & \vdots & \vdots & \ddots & \sqrt{L-1} \\
0 & 0 & 0 & \cdots & 0
\end{pmatrix}.
\end{align}    
\end{subequations}
The matrix form for the bosonic number operator is simply 
\begin{equation} \label{eq: bosonic_num_matrix}
\BN{}
= \begin{pmatrix}
0 & 0 & 0 & \cdots & 0 \\
0 & 1 & 0 & \cdots & 0 \\
0 & 0 & 2 & \cdots & 0 \\
\vdots & \vdots & \vdots & \ddots & \vdots \\
0 & 0 & 0 & \cdots & L-1
\end{pmatrix}.
\end{equation}
Thus, any qumode operator can now be represented with $L \times L$ matrices by using matrix multiplications involving \Eq{\ref{eq: bosonic_elementary_matrices}} and \Eq{\ref{eq: bosonic_num_matrix}}.


\section{ECD with qubit rotation} \label{app: ecd_rotation}

\subsection{Alternate forms} \label{app: ecd_forms}

The position and momentum operators of the qumode are
\begin{subequations}
\begin{align}
\BX{}
&= \frac{1}{\sqrt{2}} \: ( \BC{} + \BA{} ),
\\
\BP{}
&= \frac{i}{\sqrt{2}} \: ( \BC{} - \BA{} ),
\end{align}
\end{subequations}
where we assumed atomic units.
The displacement operator can now be written as 
\begin{equation}
D (\beta) 
= e^{ \beta \BC{} - \beta^* \BA{} }
= e^{ i \sqrt{2} \: \big[ 
\text{Im} (\beta) \: \BX{} 
- \text{Re} (\beta) \: \BP{} \big] }.
\end{equation} 
The conditional displacement operator is defined as 
\begin{equation} \label{eq: cond_disp_exp_form}
CD (\beta)
= e^{ i \: Z \: \otimes \: (\beta \BC{} - \beta^* \BA{}) }
= e^{ i \: Z \: \otimes \: \sqrt{2} \: \big[ 
\text{Im} (\beta) \: \BX{} 
- \text{Re} (\beta) \: \BP{} \big] }
= e^{ i \: Z \: \otimes \: \mathcal{A} (\beta) }, 
\end{equation}
where 
$ \mathcal{A} (\beta) = - \mathcal{A} (- \beta) $. 
We now Taylor expand \Eq{\ref{eq: cond_disp_exp_form}} to get 
\begin{align} \label{eq: cond_disp_expansion}
CD (\beta)
&= \EYE \otimes \Big[ \EYE 
+ \frac{i^2}{2!} \: \mathcal{A}^2 (\beta) 
+ \cdots \Big]
+ Z \otimes \Big[ i \: \mathcal{A} (\beta)
+ \frac{i^3}{3!} \: \mathcal{A}^3 (\beta) 
+ \cdots \Big] \nonumber
\\
&= \big( \KetBra{0} + \KetBra{1} \big) 
\otimes \Big[ \EYE 
+ \frac{i^2}{2!} \: \mathcal{A}^2 (\beta) 
+ \cdots \Big] \nonumber
\\ 
&+ \big( \KetBra{0} - \KetBra{1} \big) 
\otimes \Big[ i \: \mathcal{A} (\beta)
+ \frac{i^3}{3!} \: \mathcal{A}^3 (\beta) 
+ \cdots \Big],
\end{align}
where we have used the relation $Z^2 = \EYE$. 
We now rearrange \Eq{\ref{eq: cond_disp_expansion}} to arrive at an alternate form of the conditional displacement operator 
\begin{equation} \label{eq: cond_disp_lin_form}
CD (\beta)
= \KetBra{0} \: \otimes \: D (\beta) 
+ \KetBra{1} \: \otimes \: D (- \beta). 
\end{equation}
Thus the ECD operator is related to the conditional displacement as 
\begin{equation} \label{eq: ecd_connect_cd}
ECD (\beta) 
= ( \sigma_x \otimes \EYE ) \: CD (\beta / 2)
= \ket{1} \bra{0} \: \otimes \: D (\beta / 2) 
+ \ket{0} \bra{1} \: \otimes \: D (- \beta / 2). 
\end{equation}
The ECD with qubit rotation operator 
$ \UniECD (\beta, \theta, \varphi)
= ECD (\beta) \: 
\Big[ \EYE \otimes R (\theta, \varphi) \Big] $
can be written in a block matrix form as \cite{Eickbusch2022,Zhang2019}
\begin{equation} \label{eq: ecd_rotation_matrix}
\UniECD (\beta, \theta, \varphi)
= \begin{bmatrix}
e^{i \: (\varphi - \pi/2)} \: \sin ( \theta/2 ) \: 
\mathbf{D} (-\beta/2) &
\cos ( \theta/2 ) \: \mathbf{D} (-\beta/2)
\\
\cos ( \theta/2 ) \: \mathbf{D} (\beta/2) &
- e^{- i \: (\varphi - \pi/2)} \: \sin ( \theta/2 ) \: 
\mathbf{D} (\beta/2)
\end{bmatrix},
\end{equation}
where the qubit rotation 
$ R (\theta, \varphi)
= \exp \big[ - i \: (\theta/2) \: 
( \cos \varphi \: \sigma_x
+ \sin \varphi \: \sigma_y ) \big] $
is generated by the $\sigma_x$ and $\sigma_y$ Pauli matrices.
Here, $\mathbf{D} (\beta)$ in \Eq{\ref{eq: ecd_rotation_matrix}} represents the $L \times L$ matrix representation of the displacement operator following \Appx{\ref{app: bosonic_matrices}}, where $L$ is Fock cutoff chosen for the qumode. 

\subsection{Universality} \label{app: ecd_universality}

A Hamiltonian $H$ is called a generator of the corresponding unitary $U = e^{i H}$. 
The set of ECD gates and qubit rotations has shown to be universal since linear combinations of repeated nested commutators of the elementary set generators cover the full Lie algebra corresponding to the combined space of a qubit-qumode system. \cite{Eickbusch2022,Zhang2019}
Let us review the justification here. 
Given a set of generating Hamiltonians $A$ and $B$, the following relations are true \cite{Lloyd1999,Braunstein2005}
\begin{subequations} \label{eq: generator_decomposition}
\begin{align}
e^{ \delta t^2 \: [A, B]}
&= e^{ - i \delta t \: A } \: e^{ - i \delta t \: B }
e^{ i \delta t \: A } \: e^{ i \delta t \: B } 
+ \ComCom{\delta t^3},
\\
e^{ i \delta t \: ( A + B)}
&= e^{ i \delta t / 2 \: A } \: 
e^{ i \delta t / 2 \: B }
e^{ i \delta t / 2 \: B } \: 
e^{ i \delta t \: A} 
+ \ComCom{\delta t^3},
\end{align}    
\end{subequations}
which can be applied to generate the unitary corresponding to the Hamiltonians $ -i \: [A, B] $ and $A + B$ in the 
$ \delta t \rightarrow 0 $ limit. 
By the repeated application of \Eq{\ref{eq: generator_decomposition}}, it is possible to generate unitaries corresponding to the linear combination of the nested commutators of the original set of generators. 
Thus, universality for a qubit-qumode system means the ability to implement any unitary transformation that can be generated from an arbitrary linear combination of Hamiltonians of the form 
$ \sigma_i \: \BX{}^j \BP{}^k $, where $j, k$ are non-negative integers and $\sigma_i \in \{ \EYE_2, \sigma_x, \sigma_y, \sigma_z \}$ is one of the Pauli matrices. 

As discussed in \Sec{\ref{app: ecd_forms}}, the generators of ECD with qubit rotation are 
$ \{ \sigma_z \BX{}, \sigma_z \BP{}, \sigma_x, \sigma_y \} $.
We will now show how to expand this generator set using commutators to achieve universality. 
First, we apply the following commutators 
\begin{equation}
[ \SZ{} \BX{}, \SX{}]
= 2i \: \SY{} \BX{}, \quad
[ \SZ{} \BX{}, \SY{}]
= - 2i \: \SX{} \BX{}, \quad
[ \SZ{} \BP{}, \SX{}]
= 2i \: \SY{} \BP{}, \quad
[ \SZ{} \BP{}, \SY{}]
= - 2i \: \SX{} \BP{}
\end{equation}
to include the generators $\{ \sigma_a \BX{}, \sigma_a \BP{} \}$ with $a \in \{x, y, z \}$.
Repeated applications of the following commutators
\begin{subequations}
\begin{align}
[ \SX{} \BX{}, \SY{} \BX{}]
&= 2i \: \SZ{} \BX{}^2, \quad 
[ \SY{} \BX{}, \SZ{} \BX{}]
= 2i \: \SX{} \BX{}^2, \quad
[ \SZ{} \BX{}, \SX{} \BX{}]
= 2i \: \SY{} \BX{}^2,
\\
[ \SX{} \BP{}, \SY{} \BP{}]
&= 2i \: \SZ{} \BP{}^2, \quad 
[ \SY{} \BP{}, \SZ{} \BX{}]
= 2i \: \SX{} \BP{}^2, \quad
[ \SZ{} \BP{}, \SX{} \BX{}]
= 2i \: \SY{} \BP{}^2, 
\\ 
[ \SX{} \BX{}, \SY{} \BX{}^2]
&= 2i \: \SZ{} \BX{}^3, \quad 
[ \SY{} \BX{}, \SZ{} \BX{}^2]
= 2i \: \SX{} \BX{}^3, \quad
[ \SZ{} \BX{}, \SX{} \BX{}^2]
= 2i \: \SY{} \BX{}^3,
\\
[ \SX{} \BP{}, \SY{} \BP{}^2]
&= 2i \: \SZ{} \BP{}^3, \quad 
[ \SY{} \BP{}, \SZ{} \BP{}^2]
= 2i \: \SX{} \BP{}^3, \quad
[ \SZ{} \BP{}, \SX{} \BP{}^2]
= 2i \: \SY{} \BP{}^3,
\\
&\vdots \nonumber
\end{align} 
\end{subequations}
include generators $ \sigma_a \BX{}^j $ with $j \geq 2$.
Similarly, by repeated application of the following commutators 
\begin{subequations}
\begin{align}
[ \SX{} \BX{}^{j + 1}, \SY{} \BP{}]
&= 2i \: \SZ{} \BP{} \: \BX{}^{j + 1} 
+ (j + 1) \: i \: \SX{} \SY{} \BX{}^j, 
\\
[ \SY{} \BX{}^{j + 1}, \SZ{} \BP{}]
&= 2i \: \SX{} \BP{} \: \BX{}^{j + 1} 
+ (j + 1) \: i \: \SY{} \SZ{} \BX{}^j
\\
[ \SZ{} \BX{}^{j + 1}, \SX{} \BP{}]
&= 2i \: \SY{} \BP{} \: \BX{}^{j + 1} 
+ (j + 1) \: i \: \SZ{} \SX{} \BX{}^j, 
\\ 
[ \SX{} \BP{} \: \BX{}^{j + 1}, \SY{} \BP{}]
&= 2i \: \SZ{} \BP{}^2 \BX{}^{j + 1} 
+ (j + 1) \: i \: \SX{} \SY{} \BP{} \: \BX{}^j,
\\
[ \SY{} \BX{}^{j + 1} \BP{}, \SZ{} \BP{}]
&= 2i \: \SX{} \BP{}^2 \BX{}^{j + 1} 
+ (j + 1) \: i \: \SY{} \SZ{} \BP{} \: \BX{}^j, 
\\
[ \SZ{} \BX{}^{j + 1} \BP{}, \SX{} \BP{}]
&= 2i \: \SY{} \BP{}^2 \BX{}^{j + 1} 
+ (j + 1) \: i \: \SZ{} \SX{} \BP{} \: \BX{}^j, 
\\ 
&\vdots \nonumber
\end{align}    
\end{subequations}
we have covered all polynomial terms $ \sigma_c \BX{}^j \BP{}^k $ with 
$ c \in \{ x, y, z \} $, which has sufficient for universality for the composite qubit-qumode system, where we have used the relation 
$ [ \BX{}^{j + 1},  \BP{}] 
= ( j + 1) \: i \: \BX{}^j $.
The universality for only the qumode can be shown by using the commutator 
\begin{equation}
[ \sigma_a \BP{}^j \BX{}^{k + 1}, 
\sigma_a \BP{}]
= i \: (k + 1) \: \BP{}^j \BX{}^k,    
\end{equation}
which eliminates the Pauli operators.
It should be noted that this proof does not specify how many blocks of ECD gate with qubit rotation can reproduce an arbitrary qubit-qumode unitary or an arbitrary qumode unitary. 


\section{SNAP-displacement ansatz} \label{app: snap_disp}

The SNAP gate defined in \Eq{\ref{eq: snap}} allows the application of different phases on each Fock basis state of a qumode and can also be equivalently defined as 
\begin{equation} \label{eq: snap_alt}
S (\bm{\theta})
= \exp \Big( i \sum_{n = 0}^{L - 1} \: \theta_n \KetBra{n} \Big),
\end{equation}
where $L$ is the Fock cutoff chosen for the qumode.
Although conceptually understood as a qumode operator, realistically the SNAP gate is implemented via strongly dispersive qubit-cavity interactions in which the ancillary qubit is rotated whenever the cavity has $n$ photons, consecutively for each $n$ between $0$ and $L-1$ \cite{Heeres2015}. 
Since the qubit remains in $\ket{0}$ after the SNAP operation, \cite{Krastanov2015} it can also be written as 
\begin{equation} \label{eq: snap_alt_qubit}
S (\bm{\theta})
= \KetBra{0} \otimes \exp \Big( i \sum_{n = 0}^{L - 1} \: \theta_n \KetBra{n} \Big)
+ \KetBra{1} \otimes \EYE.
\end{equation}
Equivalently, it can also be represented by the following qubit-qumode operator \cite{Liu2024qumodequbitreview}
\begin{equation} 
S (\bm{\theta})
= \exp \Big( i Z \otimes \sum_{n = 0}^{L - 1} \: \theta_n \KetBra{n} \Big).
\end{equation}
We note that the exact energies for the potential energy surface for the H$_2$ can be obtained by the Hadamard test as shown in Figure \ref{fig: full_circuit_ev_snap}, where each layer of the controlled-(SNAP-displacement) ansatz can be decomposed as a controlled-SNAP and controlled-displacement. 
Similar to \Appx{\ref{app: ecd_universality}}, to prove the universality of the SNAP with displacement ansatz, as defined in \Eq{\ref{eq: snap_disp_ansatz}}, we focus on its initial generator set. 
The generators for the displacement operator $D (\alpha)$ with real-valued $\alpha$ is $ \BP{} $ and the generator for SNAP operator is
\begin{equation}
Q_n 
= \sum_{n' = 0}^n \KetBra{n'}. 
\end{equation}
The commutator of the initial generator set
\begin{equation}
J_n 
= i \: [\BP{}, Q_n]
= \sqrt{n + 1} \: \big( \ket{n} \bra{n + 1} + \HermConj \big)
\end{equation}
can selectively couple the basis states $\ket{n}$ and $\ket{n - 1}$.
Thus, for any integer $L > 0$, the operators 
$ \{ J_n \}_{n=0}^{L-1} $ and 
$ \{ Q_n \}_{n=0}^{L-1} $ are sufficient to
generate the Lie algebra $u(L)$ over the truncated Fock space, which implies universal oscillator control. \cite{Krastanov2015}


\section{Hardware implementation of universal bosonic gates} \label{app: hardware_universal_ansatz}

\subsection{Decomposition of controlled-ECD operation} \label{app: controlled_ecd}

We provide here an explicit, hardware-efficient compilation of the controlled-ECD (cECD) operation, defined as a qubit $(a)$ controlling the native dispersive interaction between a bosonic mode $(b)$ and its auxiliary qubit: \begin{equation}
    c_aECD_b (\beta) 
    = \KetBra{0_a} \otimes \EYE_b 
    + \KetBra{1_a} \otimes 
    \underbrace{\big[\ket{1_b} \bra{0_b} \otimes D_b(\beta/2) + 
    \ket{0_b} \bra{1_b} \otimes D_b (-\beta/2) 
    \big] }_{ECD_b(\beta) \text{ as defined in \Eq{\ref{eq: ecd}} 
    }}.
\end{equation}
Here, we assume native access to qubit-qubit CNOT operations and conditional displacement (CD) gates. Apart from previous demonstration in the strong dispersive limit \cite{Leghtas2013CD}, \Eq{\ref{eq: ecd_connect_cd}} suggests that the CD gate is also implementable in the weakly dispersive regime with one native ECD operation and one bit-flip\begin{equation} \label{eq: CD_compile_ECD}
CD_b(\beta/2) 
= (X_b \otimes \EYE_b) \: ECD_b(\beta).
\end{equation} 
We now show analytically that the compiled circuit as shown in Figure \ref{fig: compiled-cECD} holds. Indeed,
\begin{align}
    &CD_b\left(\frac{-\beta}{4}\right) (c_aX_b) CD_b\left(\frac{\beta}{4}\right) \nonumber 
    \\
    &= \left[\mathbb{I}_a\otimes \left(\KetBra{0_b} \otimes D(-\beta/4) + \KetBra{1_b} \otimes D(\beta/4)\right)\right] (c_aX_b) \nonumber
    \\
    & \times \left[\mathbb{I}_a\otimes \left(\KetBra{0_b} \otimes D(\beta/4) + \KetBra{1_b} \otimes D(-\beta/4)\right)\right] \nonumber
    \\ 
    &= \left[\mathbb{I}_a\otimes \left(\KetBra{0_b} \otimes D(-\beta/4) + \KetBra{1_b} \otimes D(\beta/4)\right)\right]  \nonumber
    \\
    & \times \big[ \KetBra{0_a}\otimes \left(\KetBra{0_b} \otimes D(\beta/4) + \KetBra{1_b} \otimes D(-\beta/4)\right) + \nonumber
    \\
    & \KetBra{1_a}\otimes \left(\ket{1_b} \bra{0_b} \otimes D(\beta/4) + \ket{0_b} \bra{1_b} \otimes D(-\beta/4)\right)\big] \nonumber
    \\
    &= \KetBra{0_a} \otimes \mathbb{I}_b + \KetBra{1_a} \otimes \big[ \ket{1_b} \bra{0_b} \otimes D(\beta/2) + \ket{0_b} \bra{1_b} \otimes D(-\beta/2) \big] \nonumber
    \\
    &= c_aECD_b(\beta).
\end{align} 


\begin{figure}[h!]
            \centering
            \mbox{
                \Qcircuit @C=0.6em @R=1.5em {
                \ket{\phi_a} \quad \quad \quad \quad & \qw & \ctrl{1} & \qw &
                    & \quad & \qw & \qw & \ctrl{1} & \qw & \qw & \qw \\
                \ket{\phi_b} \quad \quad \quad \quad & \qw & \multigate{1}{ECD(\beta)} & \qw & \raisebox{-1.75em}{=} 
                    & \quad & \multigate{1}{CD(\beta/4)} & \qw & \targ & \qw & \multigate{1}{CD(-\beta/4)} & \qw \\
                \ket{\psi_b} \quad \quad \quad \quad & \qw & \ghost{ECD(\beta)} & \qw &
                    & \quad & \ghost{CD(\beta/4)} & \qw & \qw & \qw & \ghost{CD(-\beta/4)} & \qw 
                }
            }
            
            \caption{
                Compiled circuit for controlled-ECD operation using native gates.
                $\ket{\psi_b}$ and its corresponding \textit{wire} represents a qumode whereas the other wires represent ancilla qubits.
            }
            \label{fig: compiled-cECD}
\end{figure}


\subsection{Comparison of ECD-rotation and SNAP-displacement gates} \label{app: comparison_ecd_snap}

Both the ECD with qubit rotation and SNAP with displacement gates are implemented at the hardware by tuning the dynamics of an oscillator qumode with an ancilla qubit. 
The Hamiltonian of the qubit-qumode system for the SNAP gate is \cite{Krastanov2015}
\begin{equation}
\hat{H} 
= \hat{H}_0 + \hat{H}_1 + \hat{H}_2.
\end{equation}
Here, $\hat{H}_0$ represents a dispersively coupled qubit and cavity oscillator 
\begin{equation}
\hat{H}_0
= \omega_q \KetBra{e}
+ \omega_c \: \BN{}
- \chi \KetBra{e} \BN{},
\end{equation}
where $\omega_q$ is the transition frequency between the qubit states $\ket{g}$ and $\ket{e}$, $\omega_c$ is the oscillator frequency, $ \BN{} = \BC{} \BA{} $ is the number operator of the qumode, and $\chi$ is the dispersive coupling. 
The Hamiltonian $\hat{H}_1$ represents the time-dependence of the oscillator 
\begin{equation}
\hat{H}_1
= \epsilon (t) \: e^{i \omega_c t} \: \BC{} + \HermConj,
\end{equation}
with the oscillator drive denoted by $\epsilon (t)$. 
The Hamiltonian $\hat{H}_2$ represents the time-dependence of the qubit 
\begin{equation}
\hat{H}_1
= \Omega (t) \: e^{i \omega_q t} \: \ket{e} \bra{g} 
+ \HermConj,
\end{equation}
with the qubit drive denoted by $\Omega (t)$. 
The control scheme for the SNAP gate requires that 
$ | \Omega (t) | \ll \chi $, i.e.,  the qubit drive is weak compared with the dispersive coupling. \cite{Krastanov2015}
In contrast, the ECD gate is implemented in the weak dispersive regime, where $\chi \leq \text{max} ( \Gamma_1, \Gamma_2, \kappa )$, where $\Gamma_1, \Gamma_2$ are qubit decoherence and relaxation rates, and $\kappa$ is the oscillator relaxation rate. \cite{Eickbusch2022}
In other words, the SNAP operation is only natively available in the strong dispersive region, whereas the ECD operation only operates in the weak dispersive region and involves unselective ancilla control which allows higher resiliency against crosstalk. \cite{You2024Crosstalk}
A mixed hardware architecture comprising both SNAP and ECD gates could potentially be enabled by a programmable, fast beam-splitter with a three-wave mixing coupler, \cite{Chapman2023StrongWeakDispersive} which remains an exciting future direction.


\clearpage
\include{arxiv_supp_info}


\bibliography{QBM}


\end{document}

%% file: arxiv_supp_info.tex
\begin{center}
\bf{
{\large Supporting Information for: \\ 
Simulating electronic structure on bosonic quantum computers}}
\end{center}


\section*{Overview of the electronic structure problem} \label{app: es_problem_background_si}

Here, we provide further context and motivation for the electronic structure problem.
The molecular Hamiltonian in atomic units can be written as \cite{SzaboBook}
\begin{equation} \label{eq: molecular_full_ham}
\mathcal{H} 
= - \frac{1}{2} \: \sum_i \: \nabla_i^2   
- \frac{1}{2} \: \sum_A \: \frac{\nabla_A^2}{M_A} 
- \sum_i \sum_A \: \frac{Z_A}{r_{iA}} 
+ \sum_i \sum_{j > i} \: \frac{1}{r_{ij}} 
+ \sum_A \sum_{B > A} \: \frac{Z_A Z_B}{R_{AB}}, 
\end{equation}
where $i, j$ are electron indices, 
$A, B$ are nuclear indices,
$\nabla_i^2$ and $\nabla_A^2$ are Laplacian operators representing differentiation with respect to the coordinates of the $i\textsuperscript{th}$ electron and $A\textsuperscript{th}$ nucleus, 
$M_A$ and $Z_A$ are the mass and atomic number of nucleus $A$,  
$ r_{iA} = | \mathbf{r}_i - \mathbf{R}_A | $ is the distance between $i\textsuperscript{th}$ electron and $A\textsuperscript{th}$ nucleus,  
$ r_{ij} = | \mathbf{r}_i - \mathbf{r}_j | $ is the distance between $i\textsuperscript{th}$ and $j\textsuperscript{th}$ electrons, and 
$ R_{AB} = | \mathbf{R}_A - \mathbf{R}_B | $ is the distance between $A\textsuperscript{th}$ and $B\textsuperscript{th}$ nuclei.
The operator terms in \Eq{\ref{eq: molecular_full_ham}} represent the kinetic energy of electrons, kinetic energy of nuclei, Coulombic attraction between electrons and nuclei, repulsion between electrons, and repulsion between nuclei, respectively.

The Born--Oppenheimer approximation assumes the molecular electrons are moving in the field of fixed nuclei since they are much lighter. \cite{SzaboBook}
This allows one to neglect the nuclear kinetic energy term in \Eq{\ref{eq: molecular_full_ham}} and consider the nuclear-nuclear repulsion term to be constant.
Thus, the remaining terms of \Eq{\ref{eq: molecular_full_ham}} constitute the molecular electronic structure Hamiltonian 
\begin{equation} \label{eq: molecular_ham_fq}
\mathcal{H}_{\text{elec}} 
= - \frac{1}{2} \: \sum_i \: \nabla_i^2   
- \sum_i \sum_A \: \frac{Z_A}{r_{iA}} 
+ \sum_i \sum_{j > i} \: \frac{1}{r_{ij}}.
\end{equation}
Our goal is to solve the time-independent Schr{\"o}dinger equation for the molecular electronic structure 
\begin{equation} \label{eq: el_tise} 
\mathcal{H}_{\text{elec}} \Psi_\mu (\mathbf{r}) 
= E_\mu \Psi_\mu (\mathbf{r})
\end{equation}
where $\{ \Psi_\mu \}$ are the electronic wavefunctions with corresponding energies $\{ E_\mu \}$ for a given molecular nuclear coordinates with $\{ \mathbf{r}_j \}$ being the set of electronic coordinates.
As an example, $\Psi_0$ and $E_0$ are the ground electronic wavefunction and its energy.
Finding the $\{ \Psi_n \}$ wavefunctions on a classical computer is a notoriously hard problem because of the combinatorial growth of the dimensionality with increasing number of electrons $N$ in the molecule. 
This is where quantum computing promises to be impactful. 

An electronic wavefunction $\Psi (\mathbf{r})$ depends on a set of $N$ electron coordinates $\{ \mathbf{r}_j \}$.
However, one should also include the electron spin into the picture, and denote the wavefunction as $\Psi (\mathbf{x})$ instead, where \textbf{x} represents the combined spatial and spin coordinates of the electrons.
Spin does not fundamentally arise in the non-relativistic premise of electronic structure theory. 
Nevertheless, spin must be included as a bookkeeping tool to respect the fermionic antisymmetry of electrons 
\begin{equation}
\Psi (\cdots, \mathbf{x}_j, \cdots, \mathbf{x}_k, \cdots) 
= - \Psi (\cdots, \mathbf{x}_k, \cdots, \mathbf{x}_j, \cdots), 
\end{equation}
even in approximate wavefunctions.
A good starting point for approximately solving the electronic structure is the Hartree--Fock (HF) method, \cite{SzaboBook} which transform the many-electron problem of \Eq{\ref{eq: el_tise}} to an effective one-electron problem in the mean-field created by the other electrons.
The HF method provides $M$ number ($M > N$) of orthonormal one-electron functions $\{ \chi_p (\mathbf{x}) \}$, called the molecular spin-orbitals. 
We are assuming $M$ to be an even integer since there is an underlying $M / 2$ number of spatial functions $\{ \phi_p (\mathbf{r}) \}$ which can associate with either up-spin $\alpha (\omega)$ or down-spin $\beta (\omega)$ functions 
\begin{equation}
\chi_{2p \uparrow} (\mathbf{x})
\equiv \phi_p (\mathbf{r}) \: \alpha (\omega), \quad    
\chi_{2p \downarrow} 
\equiv \phi_p (\mathbf{x}) \: \beta (\omega).
\end{equation}
Thus, $N$ electrons in $M$ molecular spin-orbitals give rise to ${M \choose N}$ number of many-electron basis states, each of which is an antisymmetrized product state 
\begin{equation} \label{eq: slater_determinant}
\ket{p_1, \cdots, p_N}_F 
\equiv \frac{1}{\sqrt{N!}} \:
\begin{vmatrix}
\chi_{p_1} (\textbf{x}_1) & \cdots & \chi_{p_N} (\textbf{x}_1)
\\
\chi_{p_1} (\textbf{x}_2) & \cdots & \chi_{p_N} (\textbf{x}_2)
\\
\cdots & \ddots & \cdots 
\\
\chi_{p_1} (\textbf{x}_N) & \cdots & \chi_{p_N} (\textbf{x}_N)
\end{vmatrix},    
\end{equation}
where $0 \leq p_1 < \cdots < p_N \leq M - 1$.
The wavefunction in \Eq{\ref{eq: slater_determinant}} is the so-called Slater determinant and based on the ordering of $\{ p_j \}$ indices, approximates the exact ground and excited electronic states.  
For example, the Slater determinant 
$\kVacF$ is the electronic ground state wavefunction under the HF approximation. 

\section*{Justification of the state mapping} \label{app: state_mapping_derivation_si}

The state mapping described in the \MainText is justified if we can prove that acting with bosonic operators $\{ \BC{j}, \BA{j} \}$ on the mapped fermionic states still preserve their commutation relations. 
We follow the derivation done in \Reference{\citenum{Dhar2006}} here, although a similar justification was first given by \Reference{\citenum{Ohta1998}}.

The first step to deduce the action of bosonic creation operators on a Slater determinant is to apply the state mapping to get
\begin{equation}
\BC{j} \kIndF{p}{N} 
= \BC{j} \kIndB{q}{N} 
= \sqrt{q_j + 1} \ket{q_1, \cdots, q_j + 1, \cdots, q_N}_B,
\end{equation}
which can be again mapped back to
\begin{equation} \label{eq: bc_sd_gen_si}
\BC{j} \kIndF{p}{N} 
= \sqrt{p_{N - j + 1} - p_{N - j}} 
\ket{p_1, \cdots, p_{N - k}, p_{N - k + 1} + 1, \cdots, p_N + 1}_F,
\end{equation}
where $j < N$. 
The same expression for the special case of $j = N$ is given by 
\begin{equation}
\BC{N} \kIndF{p}{N} 
= \sqrt{p_1 + 1} \ket{p_1 + 1, \cdots, p_N + 1}_F,
\end{equation}
The action of bosonic annihilation operators on a Slater determinant can be similarly derived as 
\begin{equation} \label{eq: ba_sd_gen_si}
\BA{j} \kIndF{p}{N} 
= \sqrt{p_{N - j + 1} - p_{N - j} - 1}
\ket{p_1, \cdots, p_{N - k}, p_{N - k + 1} - 1, \cdots, p_N - 1}_F,
\end{equation}
when $j < N$ and 
\begin{equation}
\BA{N} \kIndF{p}{N} 
= \sqrt{p_1} \ket{p_1 - 1, \cdots, p_N - 1}_F.
\end{equation}
Let us now combine \Eq{\ref{eq: bc_sd_gen_si}} and \Eq{\ref{eq: ba_sd_gen_si}} to arrive at 
\begin{align} \label{eq: ba_bc_sd_gen_si}
\BA{j} \BC{k} \kIndF{p}{N} 
&= \sqrt{ ( p_{N - k + 1} - p_{N - k} ) \: ( p_{N - j + 1} - p_{N - j} - 1 ) } \nonumber
\\
&\times \ket{ p_1, \cdots, p_{N - k}, p_{N - k + 1} + 1, \cdots, p_{N - j} + 1, p_{N - j + 1}, \cdots, p_N }_F,
\end{align}
where $j < k$. 
Similarly, by reversing the order of the bosonic operators, we get 
\begin{align} \label{eq: bc_ba_sd_gen_si}
\BC{k} \BA{j} \kIndF{p}{N} 
&= \sqrt{ ( p_{N - k + 1} - p_{N - k} ) \: ( p_{N - j + 1} - p_{N - j} - 1 ) } \nonumber
\\
&\times \ket{ p_1, \cdots, p_{N - k}, p_{N - k + 1} + 1, \cdots, p_{N - j} + 1, p_{N - j + 1}, \cdots, p_N }_F,
\end{align}
where $j < k$. 
Thus, the right hand sides of \Eq{\ref{eq: ba_bc_sd_gen_si}} and \Eq{\ref{eq: bc_ba_sd_gen_si}} are the same, which proves 
$ [ \BA{j}, \BC{k} ] = 0 $ for $j < k$. 
The $j > k$ case can be similarly derived as above. 
Let us now consider the $j = k$ case
\begin{equation} \label{eq: ba_bc_sd_diag_si}
\BA{k} \BC{k} \kIndF{p}{N} 
= ( p_{N - k + 1} - p_{N - k} ) \kIndF{p}{N}.
\end{equation}
Similarly, by reversing the order, we get 
\begin{equation} \label{eq: bc_ba_sd_diag_si}
\BC{k} \BA{k} \kIndF{p}{N} 
= ( p_{N - k + 1} - p_{N - k} - 1 ) \kIndF{p}{N}.
\end{equation}
Thus, \Eq{\ref{eq: ba_bc_sd_diag_si}} and \Eq{\ref{eq: bc_ba_sd_diag_si}} shows that 
$[ \BA{k}, \BC{k} ] = 1$, which proves that the state mapping preserves the relation 
$ [ \BA{j}, \BC{k} ] = \delta_{jk} $. 


\section*{Derivation of the operator mapping} \label{app: op_mapping_derivation_si}

The derivation of the mapping $\{ \BiF{p}{q} \}$ operators is shown in \Reference{\citenum{Dhar2006}}, which we call Dhar--Mandal--Suryanarayana (DMS) mapping or simply direct mapping. 
We gain insight into the derivation here by discussing all the steps for the specific cases of $N = 1$ and $N = 2$.

Let us first discuss the operator mapping with an $N = 1$ system. 
The state mapping is then simply 
$ \ket{j}_F \leftrightarrow \ket{j}_B $, where the one-particle states are defined as 
\begin{equation}
\ket{j}_F 
\equiv \FC{j} \ket{-}_F, \quad 
\ket{j}_B 
\equiv \frac{ \BC{j} }{\sqrt{j!}} \ket{0}_B. 
\end{equation}
Let us figure out how the bilinear fermionic operators act on the state $\ket{j}_B$ by using the state mapping. 
The states $\{ \ket{j}_F \}$ are eigenstate of the number operator $\BiF{p}{p}$, and combining this with the state mapping leads to 
\begin{equation}
\BiF{p}{p} \ket{j}_F 
= \delta_{p j} \ket{j}_B 
= ( \ket{j} \bra{p} ) \ket{j}_B 
= ( \ket{p} \bra{p} ) \ket{j}_B, 
\end{equation}
where $\ket{p} \bra{p}$ is the projection operator in the bosonic Fock basis with the subscripts ``B'' suppressed to avoid symbolic clutter.
Since $\ket{j}_B$ can now be mapped back to $\ket{j}_F$, the number operator is thus mapped as 
\begin{equation}
\BiF{p}{p}  
\mapsto \ket{p} \bra{p},  
\end{equation}
$p = 0, 1, \cdots, M - 1$.
Similarly, acting with the off-diagonal bilinear fermionic operator on $\ket{j}_F$ leads to 
\begin{equation} \label{eq: deriv_offdiag_dms_one_el_si}
\BiF{q}{q + p} \ket{j}_F 
= \delta_{q j} \: \FC{j + p} \ket{-}_F 
= \delta_{q j} \ket{j + p}_F, 
\end{equation}
which maps to the state $\ket{j + p}_B$. 
We can now write 
\begin{equation}
\BiF{q + p}{q} \ket{j}_F 
= ( \NBC{} )^p \: \delta_{q, j} \ket{j}_B 
= ( \NBC{} )^p \: ( \ket{q} \bra{q} ) \ket{j}_B. 
\end{equation}
Since $\ket{j}_B$ can now be mapped back to $\ket{j}_F$, the $\BiF{q + p}{q}$ operator can be mapped as 
\begin{equation}
\BiF{q + p}{q}  
\mapsto ( \NBC{} )^p \: \ket{q} \bra{q},
\end{equation} 
where $ q = 1, \cdots, M - 1 $ and 
$ p = 1, \cdots, M - p - 1 $.
Because of its adjoint relation 
$ ( \BiF{p}{q}  )^\dagger = \BiF{q}{p} $, mapping bilinear fermionic operators $\{ \BiF{p}{q} \}$ for $p \geq q$ is sufficient. 

Let us now discuss the $N = 2$ case, for which the state mapping is given by, 
$ \ket{p, q}_F 
\leftrightarrow \ket{j, k}_B $.
Let us start our derivation by writing down how the number operator $\BiF{r}{r}$ acts on an arbitrary Slater determinant  
\begin{equation}
\BiF{r}{r} \ket{p, q}_F
= ( \delta_{p, r} + \delta_{q, r} ) \ket{p, q}_F,   
\end{equation}
and after applying the state mapping back and forth, we arrive at
\begin{align}
\BiF{r}{r} \ket{p, q}_F 
&\mapsto ( \delta_{k, r} + \delta_{j + k, r - 1} ) \ket{j, k}_B \nonumber
\\
&= \Big[ \delta_{k, r} 
+ \sum_{a + b = r - 1} \delta_{j, a} \: \delta_{k, b} \Big] \ket{j, k}_B \nonumber
\\
&= \Big[ \EYE \otimes \ket{r} \bra{r} 
+ \sum_{a + b = r - 1} \ket{a, b} \bra{a, b} 
\Big] \ket{j, k}_B \nonumber
\\
&\mapsto \Big[ \EYE \otimes \ket{r} \bra{r} 
+ \sum_{a + b = r - 1} \ket{a, b} \bra{a, b} 
\Big] \ket{p, q}_F,
\end{align}
where $\EYE$ is the identity operator acting on the first mode. 
We extended the projection operator trick in the derivation of the $N = 1$ system here, i.e., the goal is to find a Kronecker delta involving one of the indices ($j$ and $k$) corresponding to the two modes.
We can now redefine the dummy indices above and express the mapping for the number operators as 
\begin{equation} \label{eq: num_op_dms_two_els_si}
\BiF{p}{p} 
\mapsto \EYE \otimes \ket{p} \bra{p} 
+ \sum_{a + b = p - 1} \ket{a, b} \bra{a, b}. 
\end{equation}
Let us now map the off-diagonal operators. 
We first act $\BiF{s + r}{s}$ on an arbitrary Slater determinant 
\begin{equation}
\BiF{s + r}{s} \ket{p, q}_F 
= \delta_{p, s} \: \FC{p + r} \FC{q} \ket{-}_F 
+ \delta_{q, s} \: \FC{p} \FC{q + r} \ket{-}_F,
\end{equation}
which can be further written as 
\begin{align} \label{eq: blf_off_diag_two_els_derivation_si}
\BiF{s + r}{s} \ket{p, q}_F 
&= \delta_{p, s} \: \Big( 
\sum_{a = p + 1}^{q - 1} \: \delta_{p + r, a} \ket{p + r, q}_F 
- \sum_{a = q + 1}^\infty \: \delta_{p + r, a} \ket{q, p + r}_F \Big) \nonumber
\\
&+ \delta_{q, s} \ket{p, q + r}_F.
\end{align}
Similar to the derivation for the $N = 1$ case, we now apply the state mapping and the normalized bosonic operators. 
The third term of \Eq{\ref{eq: blf_off_diag_two_els_derivation_si}} then reduces to  
\begin{equation}
\delta_{q, s} \: \ket{p, q + r}_F  
\mapsto \delta_{j + k, s - 1} \ket{j + r, k}_B 
= ( \NBC{1} )^r  \: \delta_{j + k, s - 1} \ket{j, k}_B,
\end{equation}
Similarly the first term of \Eq{\ref{eq: blf_off_diag_two_els_derivation_si}} can be rewritten as 
\begin{align}
\delta_{p, s} \sum_{a = p + 1}^{q - 1} \: \delta_{p + r, a} \ket{p + r, q}_F 
&\mapsto \delta_{p, s} \: 
\sum_{a = p + 1}^{q - 1} \: \delta_{p + r, a} \ket{j - r, k + r}_B \nonumber
\\
&= \NBA{1}^r \: ( \NBC{2} )^r \: \sum_{a = 0}^\infty \: \delta_{j, r + a} \: \delta_{k, s} \ket{j, k}_B,
\end{align} 
whereas the second term of \Eq{\ref{eq: blf_off_diag_two_els_derivation_si}} turns to  
\begin{align} 
\delta_{p, s} \sum_{a = q + 1}^\infty \: \delta_{p + r, a} \ket{q, p + r}_F 
&\mapsto \delta_{p, s} \sum_{a = 0}^\infty  \: \delta_{p + r, a + q + 1} 
\ket{r - 2 - j, j + k + 1}_B \nonumber   
\\
&= \sum_{a = 0}^{r - 2} \: 
( \NBC{1} )^{r - 2 - a} \: \NBA{1}^a \: 
( \NBC{2} )^{a + 1} \:
\delta_{j, a} \: \delta_{k, s} \ket{j, k}_B. 
\end{align}
Applying the projection operator relation and the state mapping back to the Slater determinants, the action of the 
$\BiF{s + r}{s}$ on an arbitrary Slater determinant can now be written as 
\begin{align}
\BiF{s + r}{s} \ket{p, q}_F 
&= \Big[ \NBA{1}^r \: ( \NBC{2} )^r \: \sum_{a = 0}^\infty \: \ket{r + a, s} \bra{r + a, s} 
- \sum_{a = 0}^{r - 2} \: 
( \NBC{1} )^{r - 2 - a} \: \NBA{1}^a \: 
( \NBC{2} )^{a + 1} \ket{a, s} \bra{a, s} \nonumber  
\\
&+ ( \NBC{1} )^r \sum_{a + b = s - 1} \ket{a, b} \bra{a, b}
\Big] \ket{p, q}_F. 
\end{align}
We can now redefine the dummy indices above and express the mapping for the off-diagonal operators as 
\begin{align} \label{eq: off_diag_op_dms_two_els_si}
\BiF{q + p}{q} 
&\mapsto \NBA{1}^p \: ( \NBC{2} )^p \: \sum_{a = 0}^\infty \: \ket{p + a, q} \bra{p + a, q} 
- \sum_{a = 0}^{p - 2} \: 
( \NBC{1} )^{p - 2 - a} \: \NBA{1}^a \: 
( \NBC{2} )^{a + 1} \ket{a, q} \bra{a, q} \nonumber  
\\
&+ ( \NBC{1} )^p \sum_{a + b = q - 1} \ket{a, b} \bra{a, b}. 
\end{align}
Thus, we have shown how to derive the DMS operator mapping for the $N = 1$ and $N = 2$ cases.


\section*{General DMS mapping expression} \label{app: gen_dms_map_si}

The expression for the DMS mapping of $\{ \BiF{p}{q} \}$ with $p > q$ is given by \cite{Dhar2006}
\begin{align} \label{eq: dms_map_offdiag_si}
\BiF{q + p}{q}
&\mapsto \sum_{ \substack{r_1 + \cdots + r_N \\ \: = \: q - N + 1 }} 
( \NBC{1} )^p \: \BigP{r_1, \cdots, r_N}
+ \sum_{a = 0}^\infty \: 
\sum_{ \substack{r_2 + \cdots + r_N \\ \: = \: q - N + 2 }}  
\NBA{1}^p ( \NBC{2} )^p \: \BigP{p + a, r_2, \cdots, r_N}
\nonumber
\\
&- \sum_{\mu = 0}^{p - 2} \: 
\sum_{ \substack{r_2 + \cdots + r_N \\ \: = \: q - N + 2 }}  
( \NBC{1} )^{p - 2 - \mu} \NBA{1}^\mu 
( \NBC{2} )^{\mu + 1} \: 
\BigP{\mu, r_2, \cdots, r_N} 
\nonumber 
\\
&+ \sum_{k = 2}^{N - 1} \: \Big[
\BigEYE{k - 1} \otimes  
\sum_{a = 0}^\infty \: 
\sum_{ \substack{r_{k + 1} + \cdots + r_N \\ \: = \: q - N + k + 1 }}  
\NBA{k}^p ( \NBC{k + 1} )^p \: 
\BigP{p + a, r_{k + 1}, \cdots, r_N} 
\nonumber
\\
&- \BigEYE{k - 2} \otimes  
\sum_{a = 0}^\infty \: 
\sum_{\mu = 0}^{p - 2} \: 
\sum_{r_{k - 1} = a + 1}^{p + a - 1} \:  
\sum_{ \substack{r_{k + 1} + \cdots + r_N \\ \: = \: q - N + k + 1 }}  
\BigT{1}{p, k, \mu} \:  
\BigP{r_{k - 1}, \mu, r_{k + 1}, \cdots, r_N}
\nonumber
\\
&+ ( - 1 )^k \: \Big( \sum_{r_1 = 0}^\infty \cdots \sum_{r_k = 0}^\infty 
- \sum_{a = 0}^\infty \: 
\sum_{ \substack{r_1 + \cdots + r_k \\ \: = \: p + a - k }} \Big) \:  
\sum_{ \substack{r_{k + 1} + \cdots + r_N \\ \: = \: q - N + k + 1 }} 
\BigT{2}{p, k, \mu_1, \cdots, \mu_k} \:  
\BigP{r_1, \cdots, r_N}
\nonumber 
\\
&+ \sum_{j = 2}^{k - 1} \: ( - 1 )^j \:
\BigEYE{k - j - 1} \otimes 
\sum_{a = 0}^\infty \: \Big(
\sum_{ \substack{r_{k - j} + \cdots + r_k \\ \: = \: p + a - j }} 
\sum_{ \substack{r_{k + 1} + \cdots + r_N \\ \: = \: q - N + k + 1 }} 
\BigT{3}{p, k, j, r_{k - j + 1}, \cdots, r_k} \: 
\BigP{r_{k - j}, \cdots, r_N}
\nonumber 
\\
&- \sum_{ \substack{r_{k - j + 1} + \cdots + r_k \\ \: = \: p + a - j }} 
\sum_{ \substack{r_{k + 1} + \cdots + r_N \\ \: = \: q - N + k + 1 }} 
\BigT{3}{p, k, j, r_{k - j + 1}, \cdots, r_k} \: 
\ket{0} \bra{0} \otimes \BigP{r_{k - j + 1}, \cdots, r_N} 
\Big) \Big], 
\end{align}
where $q = 0, \cdots, M - 2$ and $p = 1, \cdots, M - q - 1$ and the operators $\{ \BigT{\mu}{} \}$ are defined in \Table{\ref{tab: dms_intermediate_ops_si}}. 
All the summations in \Eq{\ref{eq: dms_map_offdiag_si}} will naturally truncate following the highest Fock state allowed for a qumode based on the state mapping described in the \MainText.
There are $\ComCom{N^2}$ number of terms in \Eq{\ref{eq: dms_map_offdiag_si}} that need to be taken care of in case of computing the expectation value of the operator $\BiF{q + p}{q}$. 
The expression for $\{ \BiF{p}{q} \}$ with $p < q$ can be found by taking the Hermitian conjugate of \Eq{\ref{eq: dms_map_offdiag_si}}, while the mapping for $p = q$ is given in the \MainText. 


\begin{table*}[h!]
\begin{tabular}{llllll}
\hline
Operator &&&&&  Definition
\\ \hline \\
$ \BigT{1}{p, k, \mu} $ &&&&&   
$ \NBA{k - 1}^{p - 1 - \mu} \: 
( \NBC{k} )^{p - 2 - \mu} \: 
\NBA{k}^{\mu} \: 
( \NBC{k + 1} )^{\mu + 1} $   
\\ \\ 
$ \BigT{2}{p, k, r_1, \cdots, r_k} $ &&&&&   
$ ( \NBC{1} )^{ p - 1 - k - ( r_1 + \cdots + r_k ) } 
 \: 
\NBA{1}^{r_1} \: 
( \NBC{2} )^{r_1} \: \cdots 
\NBA{k - 1}^{r_{k - 1}} \: 
( \NBC{k} )^{r_{k - 1}} \: 
\NBA{k}^{r_k} \: 
( \NBC{k + 1} )^{r_k + 1} $
\\ \\
$ \BigT{3}{p, k, j, \mu_{k - j + 1}, \cdots, \mu_k} $ &&&&&   
$ \NBA{k - j}^{ p - j - ( \mu_{k - j + 1} + \cdots + \mu_k ) } \: 
( \NBC{k - j + 1} )^{ p - j - 1 - ( \mu_{k - j + 1} + \cdots + \mu_k ) } $ 
\\
&&&&&   
$ \times \: \NBA{k - j + 1}^{ \mu_{k - j + 1} } \: 
( \NBC{k - j + 2} )^{ \mu_{k - j + 1} } \:
\cdots \: 
\NBA{k - 1}^{ \mu_{k - 1} } \: 
( \NBC{k} )^{ \mu_{k - 1} } \: 
\NBA{k}^{ \mu_k } \: 
( \NBC{k + 1} )^{ \mu_k + 1 } $
\\ \\
\hline 
\end{tabular}

\caption{
    Definitions for the intermediate operator terms used in \Eq{\ref{eq: dms_map_offdiag_si}}.
}
\label{tab: dms_intermediate_ops_si}
\end{table*}


\section*{Bosonic Hamiltonian for the dihydrogen molecule} \label{app: bosonic_ham_dihydrogen_si}

The Hamiltonian of the dihydrogen molecule in a minimal basis can be written as \cite{Whitfield2011} 
\begin{align} 
\HamF
&= \HOne{0}{0} \: \FC{0} \FA{0} 
+ \HOne{1}{1} \: \FC{1} \FA{1} 
+ \HOne{2}{2} \: \FC{2} \FA{2} 
+ \HOne{3}{3} \: \FC{3} \FA{3} 
+ \HTwo{01}{10} \: \FC{0} \FC{1} \FA{1} \FA{0} 
+ \HTwo{23}{32} \: \FC{2} \FC{3} \FA{3} \FA{2} \nonumber
\\
&+ \HTwo{03}{30} \: \FC{0} \FC{3} \FA{3} \FA{0} 
+ \HTwo{12}{21} \: \FC{1} \FC{2} \FA{2} \FA{1} 
+ ( \HTwo{02}{20} -  \HTwo{02}{02} ) \: \FC{0} \FC{2} \FA{2} \FA{0} 
+ ( \HTwo{13}{31} -  \HTwo{13}{13} ) \: \FC{1} \FC{3} \FA{3} \FA{1} \nonumber
\\
&+ \HTwo{03}{12} \: ( \FC{0} \FC{3} \FA{1} \FA{2} + \HermConj ) 
+ \HTwo{01}{32} \: ( \FC{0} \FC{1} \FA{3} \FA{2} + \HermConj ),
\end{align}
which can be written in terms of the bilinear fermionic operators 
\begin{align} \label{eq: ham_dihydrogen_bilinear_si} 
\HamF
&= ( \HOne{0}{0} 
+ \HTwo{01}{10} 
+ \HTwo{03}{30} 
+ \HTwo{02}{20} 
- \HTwo{02}{02} ) \: \BiF{0}{0} 
+ ( \HOne{1}{1}
+ \HTwo{12}{21} 
+ \HTwo{13}{31} 
- \HTwo{13}{13} ) \: \BiF{1}{1} 
+ ( \HOne{2}{2} + \HTwo{23}{32} ) \: \BiF{2}{2} 
+ \HOne{3}{3} \: \BiF{3}{3} \nonumber 
\\
&- \HTwo{01}{10} \: \BiF{0}{1} \BiF{1}{0}  
- \HTwo{23}{32} \: \BiF{2}{3} \BiF{3}{2} 
- \HTwo{03}{30} \: \BiF{0}{3} \BiF{3}{0} 
- \HTwo{12}{21} \: \BiF{1}{2} \BiF{2}{1} 
- ( \HTwo{02}{20} - \HTwo{02}{02} ) \: \BiF{0}{2} \BiF{2}{0} 
- ( \HTwo{13}{31} - \HTwo{13}{13} ) \: \BiF{1}{3} \BiF{3}{1} \nonumber 
\\ 
&- \HTwo{03}{12} \: ( \BiF{0}{1} \BiF{3}{2} 
+ \HermConj ) 
- \HTwo{01}{32} \: ( \BiF{0}{3} \BiF{1}{2} 
+ \HermConj ), 
\end{align}
which means that we need to map the lone operators 
$ \BiF{p}{p} $ with $p = 0, 1, 2, 3$, 
the \textit{symmetric} operator couples 
$ \BiF{p}{q} \BiF{q}{p} $ with $p = 0, 1, 2$ and $q = p + 1$, and the \textit{transition} operator couples 
$ \BiF{0}{1} \BiF{3}{2} $ and  
$ \BiF{0}{3} \BiF{1}{2} $.

Let us map each of the operator terms of the Hamiltonian in \Eq{\ref{eq: ham_dihydrogen_bilinear_si}}. 
The maps corresponding to the operator terms with single bilinear fermionic operators are 
\begin{subequations} \label{eq: num_op_dms_examples_si}
\begin{align}
\BiF{0}{0} 
&\mapsto \EYE \otimes \ket{0} \bra{0},
\\ 
\BiF{1}{1}
&\mapsto \EYE \otimes \ket{1} \bra{1}
+ \ket{0, 0} \bra{0, 0},  
\\
\BiF{2}{2}
&\mapsto \EYE \otimes \ket{2} \bra{2} 
+ \ket{0, 1} \bra{0, 1}  
+ \ket{1, 0} \bra{1, 0},
\\
\BiF{3}{3}
&\mapsto \ket{0, 2} \bra{0, 2}  
+ \ket{1, 1} \bra{1, 1} 
+ \ket{2, 0} \bra{2, 0},
\end{align}
\end{subequations}
where we have truncated the mapping expression based on the relevant bosonic subspace for our problem.
The expressions after mapping the symmetric operator terms are  
\begin{subequations} \label{eq: sym_op_dms_dihydrogen_si} 
\begin{align}
\BiF{0}{1} \BiF{1}{0}
&\mapsto \ket{1, 0} \bra{1, 0}
+ \ket{2, 0} \bra{2, 0},
\\
\BiF{0}{3} \BiF{3}{0}
&\mapsto \ket{0, 0} \bra{0, 0} 
+ \ket{1, 0} \bra{1, 0}, 
\\
\BiF{1}{2} \BiF{2}{1}
&\mapsto \ket{0, 0} \bra{0, 0} 
+ \ket{1, 1} \bra{1, 1} 
+ \ket{2, 1} \bra{2, 1}, 
\\ 
\BiF{0}{2} \BiF{2}{0}
&\mapsto \ket{0, 0} \bra{0, 0} 
+ \ket{2, 0} \bra{2, 0}, 
\\
\BiF{1}{3} \BiF{3}{1}
&\mapsto \ket{0, 0} \bra{0, 0}
+ \ket{0, 1} \bra{0, 1} 
+ \ket{2, 1} \bra{2, 1}, 
\\ 
\BiF{2}{3} \BiF{3}{2}
&\mapsto \ket{0, 1} \bra{0, 1} 
+ \ket{1, 0} \bra{1, 0} 
+ \ket{1, 2} \bra{1, 2} 
+ \ket{2, 2} \bra{2, 2}.
\end{align}
\end{subequations}
The rest of the operator terms are similarly mapped as 
\begin{subequations} \label{eq: transition_op_dms_dihydrogen_si} 
\begin{align}
\BiF{0}{1} \BiF{3}{2}
&\mapsto \ket{2, 0} \bra{0, 1}, 
\\
\BiF{0}{3} \BiF{1}{2}
&\mapsto - \ket{0, 0} \bra{0, 2}. 
\end{align}
\end{subequations}
We now combine \Eq{\ref{eq: num_op_dms_examples_si}}, \Eq{\ref{eq: sym_op_dms_dihydrogen_si}}, and \Eq{\ref{eq: transition_op_dms_dihydrogen_si}} to arrive at 
\begin{align} 
\HamB
&= \Big( \HOne{0}{0} 
+ \HOne{1}{1} 
+ \HTwo{01}{10} \Big) \ket{0, 0} \bra{0, 0} 
+ \Big( \HOne{1}{1} 
+ \HOne{2}{2} 
+ \HTwo{12}{21} \Big) \ket{0, 1} \bra{0, 1} \nonumber 
\\
&+ \Big( \HOne{0}{0} 
+ \HOne{2}{2} 
+ \HTwo{02}{20} 
- \HTwo{02}{02} \Big) \ket{1, 0} \bra{1, 0} 
+ \Big( \HOne{1}{1} 
+ \HOne{3}{3} 
+ \HTwo{13}{31} 
- \HTwo{13}{13} \Big) \ket{1, 1} \bra{1, 1} \nonumber  
\\
&+ \Big( \HOne{2}{2} 
+ \HOne{3}{3} 
+ \HTwo{23}{32} \Big) \ket{0, 2} \bra{0, 2} 
+ \Big( \HOne{0}{0} 
+ \HOne{3}{3} 
+ \HTwo{03}{30} \Big) \ket{2, 0} \bra{2, 0} \nonumber  
\\
&+ \HTwo{01}{32} \: \big( \ket{0, 0} \bra{0, 2} + \HermConj \big) 
- \HTwo{03}{12} \: \big( \ket{2, 0} \bra{0, 1} + \HermConj \big). 
\end{align}  
We can simplify even more by taking advantage of the symmetries of the four spin-orbitals of the H$_2$ molecule in a minimal basis, which leads to the following relations \cite{Whitfield2011} 
\begin{subequations}
\begin{align}
\HOne{0}{0}
&= \HOne{1}{1},    
\\
\HOne{2}{2}
&= \HOne{3}{3},  
\\
\HTwo{02}{20}
&= \HTwo{13}{31}
= \HTwo{12}{21}
= \HTwo{03}{30},
\\
\HTwo{02}{02}
&= \HTwo{03}{12}
= \HTwo{01}{32}
= \HTwo{13}{13}.
\end{align}    
\end{subequations}
Thus, we can finally map the Hamiltonian in \Eq{\ref{eq: ham_dihydrogen_bilinear_si}} to the bosonic form below 
\begin{align} 
\HamB
&= \Big( \HOne{0}{0} 
+ \HOne{1}{1} 
+ \HTwo{01}{10} \Big) \ket{0, 0} \bra{0, 0} 
+ \Big( \HOne{0}{0} 
+ \HOne{2}{2} 
+ \HTwo{02}{20} \Big) \ket{0, 1} \bra{0, 1} \nonumber 
\\
&+ \Big( \HOne{0}{0} 
+ \HOne{2}{2} 
+ \HTwo{02}{20} 
- \HTwo{02}{02} \Big) \ket{1, 0} \bra{1, 0} 
+ \Big( \HOne{0}{0} 
+ \HOne{2}{2} 
+ \HTwo{02}{20} 
- \HTwo{02}{02} \Big) \ket{1, 1} \bra{1, 1} \nonumber  
\\
&+ \Big( 2 \HOne{2}{2} 
+ \HTwo{23}{32} \Big) \ket{0, 2} \bra{0, 2} 
+ \Big( \HOne{0}{0} 
+ \HOne{2}{2} 
+ \HTwo{02}{20} \Big) \ket{2, 0} \bra{2, 0} \nonumber  
\\
&+ \HTwo{02}{02} \: \big( \ket{0, 0} \bra{0, 2} + \HermConj \big) 
- \HTwo{02}{02} \: \big( \ket{2, 0} \bra{0, 1} + \HermConj \big). 
\end{align} 
The final form of the Hamiltonian mapping of \Eq{\ref{eq: ham_dihydrogen_bilinear_si}} becomes
\begin{align} \label{eq: ham_dihydrogen_bosonic_si}
\HamF
\mapsto \HamB
&= \HBCoeff{1} \ket{0, 0} \bra{0, 0} 
+ \HBCoeff{2} \ket{0, 2} \bra{0, 2} 
+ \HBCoeff{3} \big( \ket{0, 1} \bra{0, 1} 
+ \ket{2, 0} \bra{2, 0} \big) \nonumber
\\
&+ \HBCoeff{4} \big( \ket{1, 0} \bra{1, 0} 
+ \ket{1, 1} \bra{1, 1} \big) 
+ \HBCoeff{5} \: \big( \ket{0, 0} \bra{0, 2} + \HermConj \big) \nonumber
\\ 
&- \HBCoeff{5} \: \big( \ket{2, 0} \bra{0, 1} + \HermConj \big),
\end{align}  
where the scalars $\{ \HBCoeff{p} \}$ are defined in \Table{\ref{tab: ham_coeff_expression_si}}.


\begin{table}[h!]
\begin{tabular}{llllll}
\hline
Coefficient &&&&&  Definition
\\ \hline 
$ \HBCoeff{1} $ &&&&&   
$ \HOne{0}{0} 
+ \HOne{1}{1}
+ \HTwo{01}{10} $   
\\ 
$ \HBCoeff{2} $ &&&&&  
$ 2 \HOne{2}{2} 
+ \HTwo{23}{32} $ 
\\ 
$ \HBCoeff{3} $ &&&&&   
$ \HOne{0}{0} 
+ \HOne{2}{2}
+ \HTwo{02}{20} $
\\ 
$ \HBCoeff{4} $ &&&&& 
$ \HOne{0}{0} 
+ \HOne{2}{2}
+ \HTwo{02}{20} 
- \HTwo{02}{02} $  
\\
$ \HBCoeff{5} $ &&&&& 
$ \HTwo{02}{02} $ 
\\ \hline
\end{tabular}

\caption{
   The bosonic Hamiltonian coefficients of the dihydrogen molecule in a minimal basis defined in \Eq{\ref{eq: ham_dihydrogen_bosonic_si}} in terms of the one-electron and two-electron integrals. 
}
\label{tab: ham_coeff_expression_si}
\end{table}

\begin{figure}[h!]

\includegraphics[width=0.75\columnwidth]{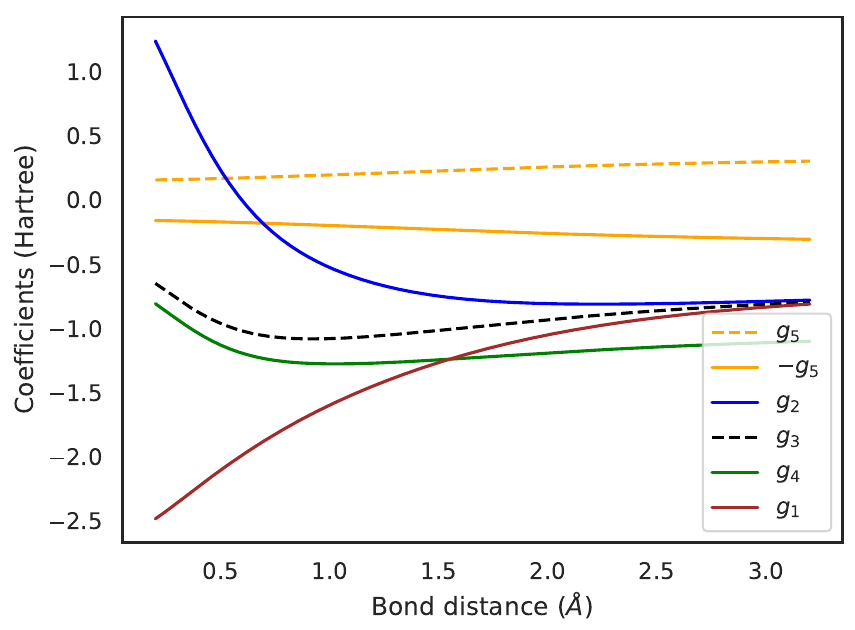}

\caption{
    The parametric dependence of the mapped bosonic Hamiltonian coefficients for the dihydrogen molecule in the STO-3G minimal basis, as defined in \Eq{\ref{eq: ham_dihydrogen_bosonic_si}}, on the H--H bond distance.
    The Hamiltonian coefficients are defined in \Table{\ref{tab: ham_coeff_expression_si}}.
}
\label{fig: bosonic_coeffs_si}
\end{figure}
\begin{figure}[h!]

\includegraphics[width=0.7\columnwidth]{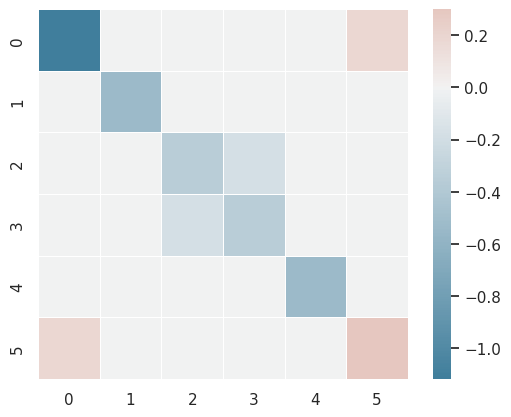}

\caption{
   Heatmap of the Hamiltonian matrix elements for the dihydrogen molecule in the STO-3G minimal basis after the direct mapping. 
   The matrix elements are computed for the fixed H--H bond distance of 0.7414 \AA \ and match exactly with their analogous fermionic Hamiltonian matrix elements.  
}
\label{fig: fci_mat_si}
\end{figure}

The dependence of the bosonic Hamiltonian coefficients of \Eq{\ref{eq: ham_dihydrogen_bosonic_si}} on the H--H bond distance is shown in \Fig{\ref{fig: bosonic_coeffs_si}}. 
The simplification of the bosonic Hamiltonian originates from the fact that the projection operators should not correspond to a basis state that is outside of the physical Hilbert space for the dihydrogen molecule.
Thus, the bosonic Hamiltonian in \Eq{\ref{eq: ham_dihydrogen_bosonic_si}} can also be understood as a Hamiltonian of two qutrits, i.e., qudits with three dimensions.
The mapped bosonic Hamiltonian has six physical Fock basis states and the corresponding matrix heatmap is shown in \Fig{\ref{fig: fci_mat_si}} for the H--H bond distance of 0.7414 \AA, which matches exactly with its corresponding fermionic FCI matrix elements.


Let us now discuss the expectation values of the Hamiltonian in \Eq{\ref{eq: ham_dihydrogen_bosonic_si}} given a bosonic state. 
There are two classes of operator terms possible for a bosonic Hamiltonian expressed in terms of tensor products of projection operators, namely, the photon counting operator such as $\KetBra{1, 0}$ and photon transfer operators such as $\ket{2, 0} \bra{0, 1} + \HermConj$.
The expectation values for the photon counting operators are easy to interpret. 
For example, the expectation value of $\KetBra{1, 0}$ for a given trial state 
\begin{equation}
\braket{\psi | ( \KetBra{1, 0} ) | \psi} 
= | \braket{1, 0 | \psi} |^2  
\end{equation}
is equivalent to the probability of measuring one photon in the first and zero photons in the second qumode.
The photon counting can be measured by optical detectors in the case of photonic quantum computing, \cite{Divochiy2008,Kardynal2008} or using cavity-transmon parity measurements in the case of cQED approach. \cite{Wang2020}

Computing the expectation value of photon transfer operators can be done via at least two ways. 
A conceptually straightforward approach involves generalization of the qubit-based Pauli-$X$ operator expectation value for a pairs of qudit states.
Let us elaborate with a specific example below  
\begin{equation}
\bar{X} 
\equiv \ket{2, 0} \bra{0, 1} + \HermConj
= \bar{H} \bar{Z} \bar{H},
\end{equation} 
where the generalizations of Pauli-$Z$ and Hadamard operators defined for the relevant qudit subspace are 
\begin{subequations}
\begin{align}
\bar{Z}
&\equiv \ket{2, 0} \bra{2, 0} - \ket{0, 1} \bra{0, 1}, 
\\
\bar{H}
&\equiv \frac{1}{\sqrt{2}} \: ( \ket{2, 0} + \ket{0, 1} ) \bra{2, 0} 
+ \frac{1}{\sqrt{2}} \: ( \ket{2, 0} - \ket{0, 1} ) \bra{0, 1}. 
\end{align}    
\end{subequations}
Since $\bar{H}$ is a unitary operator, expectation value of $\bar{X}$ reduces to photon counting in a rotated basis 
\begin{equation}
\braket{\psi | \bar{X} | \psi}
= \braket{\psi' | \bar{Z} | \psi'}
= | \braket{2, 0 | \psi'} |^2
- | \braket{0, 1 | \psi'} |^2,
\end{equation} 
where $\ket{\psi'} = \bar{H} \ket{\psi}$. 
The expectation values for the other Hamiltonian terms of \Eq{\ref{eq: ham_dihydrogen_bosonic_si}} can be similarly expressed. 
The operator $\bar{H}$ can be implemented with a photonic setup, \cite{Xu2024DensityMatrix} whereas operators like $\bar{H}$ in cQED approach can be implemented by driving cascaded three-wave or four-wave mixing transitions using a dispersively coupled ancilla qubit, such as 
$ \ket{2,0,g} 
\leftrightarrow \ket{0,0,e}
\leftrightarrow \ket{0,1,g} $, where 
$\ket{g}$ and $\ket{e}$ represent the ground and excited states of the ancilla. \cite{Hofheinz2008,Gao2018,Zhang2019}
A potentially more scalable approach for computing the expectation value of photon transfer operators between arbitrary multimode Fock states is the recently introduced subspace tomography in cQED, \cite{Gertler2023}
which does not rely on the $\bar{H}$ operators and instead uses phase space displacement operations that can be implemented efficiently. 


\section*{Subspace tomography for computing photon transfer expectation values} \label{app: tomography}

The cQED-based subspace tomography approach described in \Reference{\citenum{Gertler2023}} can be implemented with the help of an ancilla transmon qutrit and can be divided into two broad parts.
Let us denote the three levels of the ancilla to be $\ket{g}$, $\ket{e}$, and $\ket{f}$.  
The first part uses a unitary operator coupled to the states $\ket{e, g}$ that transforms the full density matrix of a qumode state into a chosen subspace density matrix coupled to the $\ket{e}$ state.  
The second part applies phase displacement operator(s) followed by a photon-number state projection and measure the corresponding probability in the $\ket{f}$ state. 
We discuss the resulting expressions below. 

Let us first understand the above protocol for one qumode, whose state can be represented in the Fock basis as 
\begin{equation}
\ket{\Psi}
= \sum_{n = 0}^\infty C_n \ket{n},    
\end{equation}
where $\{ C_n \}$ are the complex-valued Fock basis coefficients. 
Let us assume we want to compute the expectation value of the following photon transfer operator 
$ \ket{j} \bra{k} + \HermConj $
\begin{equation}
T_{j, k}
= \braket{\Psi | \big( \ket{j} \bra{k} + \HermConj \big) | \Psi}  
= C_j C_k^* + C_k C_j^*,   
\end{equation}
where the operator pairing for the off-diagonal parts ensure Hermiticity.
The subspace tomography approach will choose to handle the corresponding subspace density matrix 
\begin{equation}
\rho_{j, k} 
= \Big( \sum_{n = j, k} C_n \ket{n} \Big) \: 
\Big( \sum_{n = j, k} C_n^* \bra{n} \Big) 
= \sum_{n, m \: \in \: \mathcal{S}}
C_n C_m^* \ket{n} \bra{m},  
\end{equation}
where $\mathcal{S}$ represent the subspace chosen. 
Let us now apply the phase space displacement operator 
$ D (\alpha) 
= e^{ \alpha \BC{} - \alpha^* \BA{} } $,
which creates all possible photon excitation and deexcitation from the Fock state it acts on
\begin{equation}
D (\alpha) \ket{n}
= \sum_{j = 0}^\infty \: \DCoeff{n, j} \ket{j},     
\end{equation}
where $\{ \DCoeff{n, j} \}$ are the known and easily tunable linear coefficients associated with the displacement operator. 
The displacement operator transforms $\rho_{j, k}$ as 
\begin{align} \label{eq: disp_dm}
R_{j, k}^{(1)} 
&\equiv D (\alpha) \: \rho_{j, k} \: D^\dagger (\alpha) \nonumber 
\\
&= \sum_{n, m \: \in \: \mathcal{S}} 
C_n C_m^* \: \Big[ D (\alpha) \ket{n} \bra{m} D^\dagger (\alpha) \Big] \nonumber 
\\
&= \sum_{n, m \: \in \: \mathcal{S}} C_n C_m^* 
\sum_{j, k \: \in \: \mathbb{N}} 
\DCoeff{n, j} \: ( \DCoeff{m, k} )^* \: 
\ket{j} \bra{k}.    
\end{align}
The final observable can now be represented as 
\begin{equation} \label{eq: proj_disp_dm}
R_{j, k, p}^{(2)} 
\equiv \text{Tr} ( \KetBra{p} R_{j, k}^{(1)} \KetBra{p} ) 
= \sum_{n, m \: \in \: \mathcal{S}} 
C_n C_m^* \: \DCoeff{n, p} \: ( \DCoeff{m, p} )^*.   
\end{equation}
Assuming all $\{ \DCoeff{n, p} \}$ coefficients to be real-valued, \Eq{\ref{eq: proj_disp_dm}} can be rewritten as 
\begin{align}
R_{j, k, p}^{(2)}  
&= \sum_{n, m \: \in \: \mathcal{S}} 
\DCoeff{n, p} \: \DCoeff{m, p} \: 
C_n C_m^*    
\nonumber
\\
&= \DCoeff{j, p} \: \DCoeff{j, p} \: 
C_j C_j^*   
+ \DCoeff{k, p} \: \DCoeff{k, p} \: 
C_k C_k^*   
+ \DCoeff{j, p} \: \DCoeff{k, p} \: 
( C_j C_k^* + C_k C_j^* ) 
\nonumber
\\
&= \DCoeff{j, p}^2 \: 
| \braket{j | \Psi} |^2 
+ \DCoeff{k, p}^2 \: 
| \braket{k | \Psi} |^2 
+ \DCoeff{j, p} \: \DCoeff{k, p} \: T_{j, k}.
\end{align}
Since $ R_{j, k, p}^{(2)} $ is the observable for the subspace tomography and $\{ | \braket{j | \Psi} |^2  \}$ can be computed by photon number counting, the expectation value for the photon transfer operator can be computed as
\begin{equation}
T_{j, k}
= \frac{1}{ \DCoeff{j, p} \: \DCoeff{k, p} } \: \Big( 
R_{j, k, p}^{(2)} 
- \DCoeff{j, p}^2 \: 
| \braket{j | \Psi} |^2 
- \DCoeff{k, p}^2 \: 
| \braket{k | \Psi} |^2 \Big).
\end{equation} 

The generalization of the above approach to $N$ number of qumodes is straightforward with one phase space displacement operators acting on each of the qumodes. 
In this case, we want to compute the expectation value of the photon transfer operator 
\begin{equation}
T_{\mathbf{j}, \mathbf{k}}
= \braket{\Psi | \big( 
\ket{\mathbf{j}} \bra{\mathbf{k}} 
+ \HermConj \big) | \Psi},      
\end{equation}
where \textbf{j} is a vector of natural numbers and 
$ \ket{\mathbf{j}} 
\equiv \ket{j_1, \cdots, j_N}_B $
is a bosonic Fock state. 
The corresponding subspace density matrix is
\begin{equation}
\rho_{\mathbf{j}, \mathbf{k}} 
= \Big( \sum_{\mathbf{n} = \mathbf{j}, \textbf{k}} C_{\mathbf{n}} \ket{\mathbf{n}} \Big) \: 
\Big( \sum_{\mathbf{n} = \mathbf{j}, \mathbf{k}} C_{\mathbf{n}}^* \bra{\mathbf{n}} \Big)
= \sum_{\mathbf{n}, \mathbf{m} \: \in \: \mathcal{S}}
C_{\mathbf{n}} C_{\mathbf{m}}^* 
\ket{\mathbf{n}} \bra{\mathbf{m}},   
\end{equation}
and the experimental observables are 
\begin{subequations}
\begin{align} 
R_{\mathbf{j}, \mathbf{k}}^{(1)} 
&\equiv D_N (\alpha) \cdots D_1 (\alpha) \: 
\rho_{\mathbf{j}, \mathbf{k}} \: 
D_1^\dagger (\alpha) \cdots D_N^\dagger (\alpha), 
\\
R_{\mathbf{j}, \mathbf{k}, \mathbf{p}}^{(2)}  
&\equiv \text{Tr} ( 
\ket{\mathbf{p}} \bra{\mathbf{p}} 
R_{\mathbf{j}, \mathbf{k}}^{(1)} 
\ket{\mathbf{p}} \bra{\mathbf{p}} ),
\end{align}    
\end{subequations}
where $D_p (\alpha)$ is the displacement operator acting on the $p$-th qumode and 
$ \ket{\mathbf{p}} \bra{\mathbf{p}} $ is the multimode projection operator. 
Similar to discussion above, $ T_{\mathbf{j}, \mathbf{k}} $ can then be expressed as 
\begin{equation}
T_{\mathbf{j}, \mathbf{k}}
= \frac{1}{ \prod_{i = 1}^N \DCoeff{j_i, p_i} \: \DCoeff{k_i, p_i} } \: \Big[ 
R_{\textbf{j}, \textbf{k}, \textbf{p}}^{(2)} 
- \Big( \prod_{i = 1}^N \: \DCoeff{j_i, p_i}^2 \Big) \: 
| \braket{\mathbf{j} | \Psi} |^2 
- \Big( \prod_{i = 1}^N \: \DCoeff{k_i, p_i}^2 \Big) \: 
| \braket{\mathbf{k} | \Psi} |^2 \Big].   
\end{equation}
Thus, it is possible to compute the expectation value of any photon transfer operator of the form 
$ \ket{\mathbf{j}} \bra{\mathbf{k}} 
+ \HermConj $ using the subspace tomography approach.


\section*{Hybrid variational approach}

We have all the tools needed for applying a hybrid quantum-classical variational algorithm for finding the ground state energy of the dihydrogen molecule 
\begin{equation} \label{eq: variational_si}
\min_{\psi} E 
= \frac{\braket{\psi| \HamB | \psi}}{\braket{\psi | \psi}},
\end{equation} 
where $\ket{\psi}$ is a bosonic trial state and $\HamB$ is the mapped bosonic Hamiltonian in \Eq{\ref{eq: ham_dihydrogen_bosonic_si}}.
The hybrid algorithm can assign the computation of the expectation value in \Eq{\ref{eq: variational_si}} to a bosonic device while the energy function is optimized in a classical processor.
Similar to the hybrid algorithms designed for quantum computers with qubits, one needs a robust ansatz for $\ket{\psi}$ for the minimization in \Eq{\ref{eq: variational_si}}.

\begin{figure}[b!]

\includegraphics[width=0.9\columnwidth]{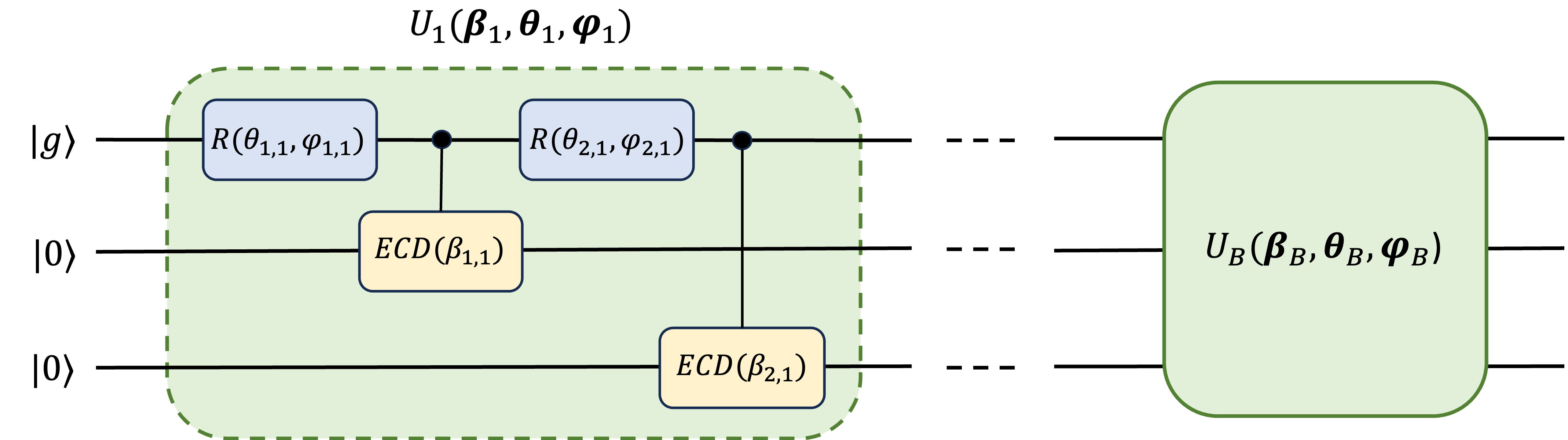}

\caption{
    The universal bosonic ansatz for two qumodes with an ancilla qubit, following \Reference{\citenum{You2024Crosstalk}}. 
    The qubit-qumode circuit is initialized in the state 
    $\ket{g} \otimes \ket{0, 0}_B$ and then a block of ECD gates and qubit rotations are acted sequentially, as defined in \Eq{\ref{eq: bosonic_ansatz_si}}.
}
\label{fig: ansatz_si}
\end{figure}

We explore the universal bosonic ansatz of two qumodes for the mapped bosonic Hamiltonian of \Eq{\ref{eq: ham_dihydrogen_bosonic_si}} here.
Universal control of qumodes requires non-Gaussian resources, \cite{Lloyd1999,Braunstein2005} which in cQED can be provided by the third or higher-order nonlinearity of the ancilla Josephson qubits or couplers. \cite{Krastanov2015,Heeres2015,Fosel2020,Eickbusch2022,Chakram2022multimode,Diringer2024} 
Multiple non-Gaussian elementary gates in cQED can be used to construct a universal gate set, including, most notably, the ancilla-controlled rotation (native for dispersive Hamiltonian), \cite{Vlastakis2013} the selective number-dependent arbitrary phase (SNAP) gate, \cite{Krastanov2015,Heeres2015,Fosel2020} 
and the conditional displacement gate. \cite{Eickbusch2022,You2024Crosstalk} 
For example, one possible way to implement an arbitrary multi-qumode unitary can be achieved by a sequence of echoed conditional displacement (ECD) gates
\begin{equation}
ECD (\beta)
= \ket{e} \bra{g} \otimes D (\beta/2) 
+ \ket{g} \bra{e} \otimes D (- \beta/2),    
\end{equation}
and ancilla rotations 
\begin{equation}
R (\theta, \varphi) 
= e^{  - i \frac{\theta}{2} \: ( 
\sigma_x \cos \varphi 
+ \sigma_y \sin \varphi ) },
\end{equation}
where 
$ D (\alpha) = \exp ( \alpha \BC{} - \alpha^* \BA{} ) $ 
is the one-qumode displacement operator,  $\ket{g}, \ket{e}$ are the ground and excited states of the ancilla qubit, and $\sigma_x, \sigma_y$ are the one-qubit Pauli operators. \cite{You2024Crosstalk}
Thus, a general ansatz for any bosonic Hamiltonian of two qumodes can be written as 
\begin{equation} \label{eq: bosonic_ansatz_si}
\ket{\psi}
= U_B (\bm{\beta}_B, \bm{\theta}_B, \bm{\varphi}_B) \: 
\cdots \:
U_1 (\bm{\beta}_1, \bm{\theta}_1, \bm{\varphi}_1) \:
\big( \ket{g} \otimes \ket{0, 0}_B \big),   
\end{equation}
where the initial state for the one ancilla qubit and the two qumodes is $\ket{g, 0, 0}$ and the $U_j$ unitary is defined as 
\begin{align}
U_j 
&= \Big( \ket{e} \bra{g} \otimes \EYE \otimes D (\beta_{2, j}/2) 
+ \ket{g} \bra{e} \otimes \EYE \otimes D (- \beta_{2, j}/2) \Big) 
\Big( R (\theta_{2, j}, \varphi_{2, j}) \otimes \EYE \otimes \EYE \Big) 
\nonumber
\\
&\times \Big( \ket{e} \bra{g} \otimes D (\beta_{1, j}/2) \otimes \EYE
+ \ket{g} \bra{e} \otimes D (- \beta_{1, j}/2) \otimes \EYE \Big)
\Big( R (\theta_{1, j}, \varphi_{1, j}) \otimes \EYE \otimes \EYE \Big).
\end{align}
The two-qumode ansatz of \Eq{\ref{eq: bosonic_ansatz_si}} has $B$ number of $U_j$ blocks and is illustrated in \Fig{\ref{fig: ansatz_si}}. 
The strategy of using an ancilla qubit rotation and ECD gates can be similarly extended for any number of qumodes. \cite{You2024Crosstalk}
Additional strategies for multi-mode control such as based on the conditional-NOT displacement \cite{Diringer2024} and photon blockade \cite{Chakram2022multimode} have also been demonstrated recently.


\begin{figure}[t!]

\includegraphics[width=0.75\columnwidth]{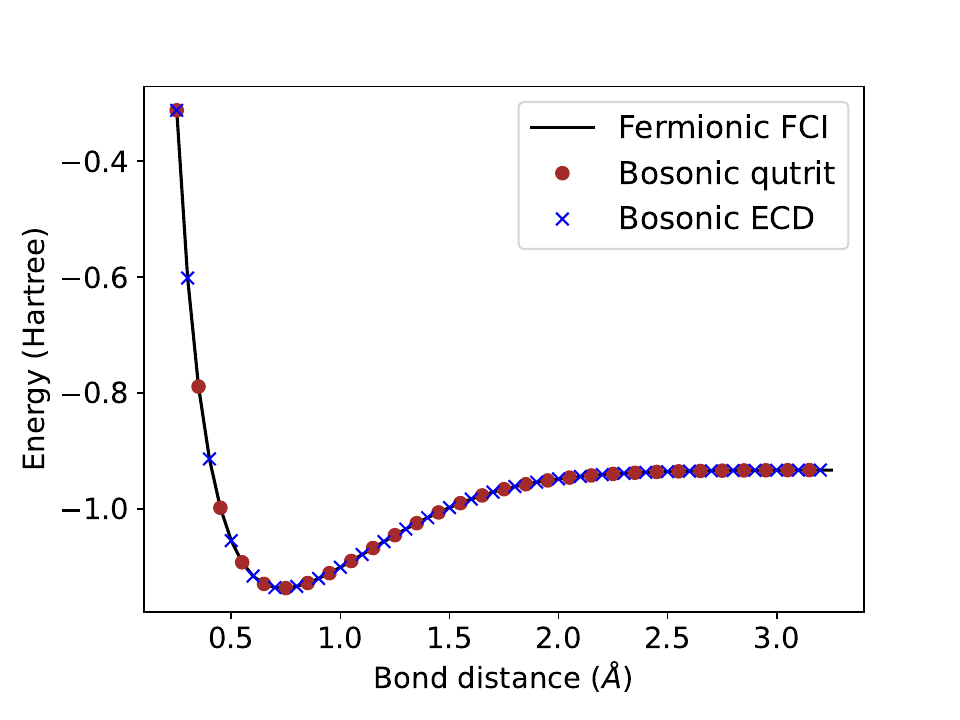}

\caption{
    Ground state energies of the dihydrogen molecule in the STO-3G minimal basis for different H--H bond distances. 
    The black line represents the exact energies in the basis of Slater determinants. 
    The brown dots and blue crosses represent the mapped bosonic variational methods. 
    Both the bosonic qutrit ansatz (brown dots) and the ECD-rotation ansatz (blue crosses) of \Eq{\ref{eq: bosonic_ansatz_dihydrogen_si}} reproduce the exact ground state energies. 
    For the ECD-rotation ansatz, the results are shown for $B = 2$ blocks.
}
\label{fig: variational_dihydrogen_si}
\end{figure}

 
Specifically for the ground state of H$_2$ molecule in a minimal basis, the only relevant Slater determinant basis states are 
$\ket{0, 1}_F$ and $\ket{2, 3}_F$, \cite{Lanyon2010} which becomes the qumode Fock basis states 
$\ket{0, 0}_B$ and $\ket{0, 2}_B$ after the state mapping. 
This means the general ansatz of \Eq{\ref{eq: bosonic_ansatz_si}} can be applied to find the ground state of H$_2$ molecule, even without optimizing the parameters on the first qumode. 
In that case, the following reduced version of \Eq{\ref{eq: bosonic_ansatz_si}} is sufficient for the ground state of H$_2$ molecule \cite{Eickbusch2022}
\begin{subequations} \label{eq: bosonic_ansatz_dihydrogen_si}
\begin{align}
\ket{\psi}
&= U_B (\beta_B, \theta_B, \varphi_B) \: 
\cdots \:
U_1 (\beta_1, \theta_1, \varphi_1) \:
\big( \ket{g} \otimes \ket{0, 0}_B \big),
\\
U_j 
&= \Big( \ket{e} \bra{g} \otimes \EYE \otimes D (\beta_j/2) 
+ \ket{g} \bra{e} \otimes \EYE \otimes D (- \beta_j/2) \Big) \:
\Big( R (\theta_j, \varphi_j) \otimes \EYE \otimes \EYE \Big),  
\end{align}
\end{subequations}
where $B$ is the number of blocks and $\{ \beta_j, \theta_j, \phi_j \}$ needs to be optimized following \Eq{\ref{eq: variational_si}}.
As shown in \Fig{\ref{fig: variational_dihydrogen_si}}, the ECD-rotation ansatz of \Eq{\ref{eq: bosonic_ansatz_dihydrogen_si}} with only $B = 2$ blocks can reproduce the exact ground state energies for different H--H bond distances. 
Since the $\HamB$ defined in \Eq{\ref{eq: ham_dihydrogen_bosonic_si}} can be thought of as a Hamiltonian of two qutrits, starting from the initial state $\ket{0, 0}_B$ and applying
$ e^{- 2 i \theta \: ( - i \ket{0} \bra{2} + \HermConj )} $, which is a qutrit $R_y$ gate, \cite{Arrazola2021,Goss2022} to the second qumode is also an legitimate ansatz for the ground state, where the the scalar $\theta$ is optimized following \Eq{\ref{eq: variational_si}}.
As shown in \Fig{\ref{fig: variational_dihydrogen_si}}, both the general ECD-rotation and the qutrit ansatze can reach the exact ground state energies of the H$_2$ molecule.
It is important to note that the expectation value of $\HamB$ for any Fock basis state that is absent in $\HamB$ is naturally zero, which allows the design of flexible bosonic ansatz without contaminating the trial energy in \Eq{\ref{eq: variational_si}}, while potentially boosting the optimization.

%% file: arxiv_main.bbl
\providecommand{\latin}[1]{#1}
\makeatletter
\providecommand{\doi}
  {\begingroup\let\do\@makeother\dospecials
  \catcode`\{=1 \catcode`\}=2 \doi@aux}
\providecommand{\doi@aux}[1]{\endgroup\texttt{#1}}
\makeatother
\providecommand*\mcitethebibliography{\thebibliography}
\csname @ifundefined\endcsname{endmcitethebibliography}
  {\let\endmcitethebibliography\endthebibliography}{}
\begin{mcitethebibliography}{89}
\providecommand*\natexlab[1]{#1}
\providecommand*\mciteSetBstSublistMode[1]{}
\providecommand*\mciteSetBstMaxWidthForm[2]{}
\providecommand*\mciteBstWouldAddEndPuncttrue
  {\def\EndOfBibitem{\unskip.}}
\providecommand*\mciteBstWouldAddEndPunctfalse
  {\let\EndOfBibitem\relax}
\providecommand*\mciteSetBstMidEndSepPunct[3]{}
\providecommand*\mciteSetBstSublistLabelBeginEnd[3]{}
\providecommand*\EndOfBibitem{}
\mciteSetBstSublistMode{f}
\mciteSetBstMaxWidthForm{subitem}{(\alph{mcitesubitemcount})}
\mciteSetBstSublistLabelBeginEnd
  {\mcitemaxwidthsubitemform\space}
  {\relax}
  {\relax}

\bibitem[Vogiatzis \latin{et~al.}(2017)Vogiatzis, Ma, Olsen, Gagliardi, and
  De~Jong]{Vogiatzis2017}
Vogiatzis,~K.~D.; Ma,~D.; Olsen,~J.; Gagliardi,~L.; De~Jong,~W.~A. Pushing
  configuration-interaction to the limit: {Towards} massively parallel MCSCF
  calculations. \emph{J. Chem. Phys.} \textbf{2017}, \emph{147}, 184111\relax
\mciteBstWouldAddEndPuncttrue
\mciteSetBstMidEndSepPunct{\mcitedefaultmidpunct}
{\mcitedefaultendpunct}{\mcitedefaultseppunct}\relax
\EndOfBibitem
\bibitem[Chan(2024)]{Chan2024spiers}
Chan,~G. K.-L. Spiers Memorial Lecture: {Quantum} chemistry, classical
  heuristics, and quantum advantage. \emph{Faraday Discuss.} \textbf{2024},
  \emph{254}, 11--52\relax
\mciteBstWouldAddEndPuncttrue
\mciteSetBstMidEndSepPunct{\mcitedefaultmidpunct}
{\mcitedefaultendpunct}{\mcitedefaultseppunct}\relax
\EndOfBibitem
\bibitem[Peruzzo \latin{et~al.}(2014)Peruzzo, McClean, Shadbolt, Yung, Zhou,
  Love, Aspuru-Guzik, and O'Brien]{Peruzzo2014}
Peruzzo,~A.; McClean,~J.; Shadbolt,~P.; Yung,~M.-H.; Zhou,~X.-Q.; Love,~P.~J.;
  Aspuru-Guzik,~A.; O'Brien,~J.~L. A Variational Eigenvalue Solver on a
  Photonic Quantum Processor. \emph{Nat. Commun.} \textbf{2014}, \emph{5},
  4213\relax
\mciteBstWouldAddEndPuncttrue
\mciteSetBstMidEndSepPunct{\mcitedefaultmidpunct}
{\mcitedefaultendpunct}{\mcitedefaultseppunct}\relax
\EndOfBibitem
\bibitem[Grimsley \latin{et~al.}(2019)Grimsley, Economou, Barnes, and
  Mayhall]{Grimsley2019}
Grimsley,~H.~R.; Economou,~S.~E.; Barnes,~E.; Mayhall,~N.~J. An adaptive
  variational algorithm for exact molecular simulations on a quantum computer.
  \emph{Nat. Commun.} \textbf{2019}, \emph{10}\relax
\mciteBstWouldAddEndPuncttrue
\mciteSetBstMidEndSepPunct{\mcitedefaultmidpunct}
{\mcitedefaultendpunct}{\mcitedefaultseppunct}\relax
\EndOfBibitem
\bibitem[McArdle \latin{et~al.}(2019)McArdle, Jones, Endo, Li, Benjamin, and
  Yuan]{Mcardle2019}
McArdle,~S.; Jones,~T.; Endo,~S.; Li,~Y.; Benjamin,~S.~C.; Yuan,~X. Variational
  ansatz-based quantum simulation of imaginary time evolution. \emph{npj
  Quantum Inf.} \textbf{2019}, \emph{5}, 75\relax
\mciteBstWouldAddEndPuncttrue
\mciteSetBstMidEndSepPunct{\mcitedefaultmidpunct}
{\mcitedefaultendpunct}{\mcitedefaultseppunct}\relax
\EndOfBibitem
\bibitem[Motta \latin{et~al.}(2020)Motta, Sun, Tan, O'Rourke, Ye, Minnich,
  Brandao, and Chan]{Motta2020}
Motta,~M.; Sun,~C.; Tan,~A.~T.; O'Rourke,~M.~J.; Ye,~E.; Minnich,~A.~J.;
  Brandao,~F.~G.; Chan,~G. K.-L. Determining eigenstates and thermal states on
  a quantum computer using quantum imaginary time evolution. \emph{Nat. Phys.}
  \textbf{2020}, \emph{16}, 205\relax
\mciteBstWouldAddEndPuncttrue
\mciteSetBstMidEndSepPunct{\mcitedefaultmidpunct}
{\mcitedefaultendpunct}{\mcitedefaultseppunct}\relax
\EndOfBibitem
\bibitem[Smart and Mazziotti(2021)Smart, and Mazziotti]{Smart2021}
Smart,~S.~E.; Mazziotti,~D.~A. Quantum solver of contracted eigenvalue
  equations for scalable molecular simulations on quantum computing devices.
  \emph{Phys. Rev. Lett.} \textbf{2021}, \emph{126}, 070504\relax
\mciteBstWouldAddEndPuncttrue
\mciteSetBstMidEndSepPunct{\mcitedefaultmidpunct}
{\mcitedefaultendpunct}{\mcitedefaultseppunct}\relax
\EndOfBibitem
\bibitem[Kyaw \latin{et~al.}(2023)Kyaw, Soley, Allen, Bergold, Sun, Batista,
  and Aspuru-Guzik]{Kyaw2023}
Kyaw,~T.~H.; Soley,~M.~B.; Allen,~B.; Bergold,~P.; Sun,~C.; Batista,~V.~S.;
  Aspuru-Guzik,~A. Boosting quantum amplitude exponentially in variational
  quantum algorithms. \emph{Quantum Sci. Technol.} \textbf{2023}, \emph{9},
  01LT01\relax
\mciteBstWouldAddEndPuncttrue
\mciteSetBstMidEndSepPunct{\mcitedefaultmidpunct}
{\mcitedefaultendpunct}{\mcitedefaultseppunct}\relax
\EndOfBibitem
\bibitem[Whitfield \latin{et~al.}(2011)Whitfield, Biamonte, and
  Aspuru-Guzik]{Whitfield2011}
Whitfield,~J.~D.; Biamonte,~J.; Aspuru-Guzik,~A. Simulation of electronic
  structure {Hamiltonians} using quantum computers. \emph{Mol. Phys.}
  \textbf{2011}, \emph{109}, 735\relax
\mciteBstWouldAddEndPuncttrue
\mciteSetBstMidEndSepPunct{\mcitedefaultmidpunct}
{\mcitedefaultendpunct}{\mcitedefaultseppunct}\relax
\EndOfBibitem
\bibitem[Seeley \latin{et~al.}(2012)Seeley, Richard, and Love]{Seeley2012}
Seeley,~J.~T.; Richard,~M.~J.; Love,~P.~J. The {Bravyi-Kitaev} transformation
  for Quantum Computation of Electronic Structure. \emph{J. Chem. Phys.}
  \textbf{2012}, \emph{137}, 224109\relax
\mciteBstWouldAddEndPuncttrue
\mciteSetBstMidEndSepPunct{\mcitedefaultmidpunct}
{\mcitedefaultendpunct}{\mcitedefaultseppunct}\relax
\EndOfBibitem
\bibitem[Dutta \latin{et~al.}(2024)Dutta, Cabral, Lyu, Vu, Wang, Allen, Dan,
  Corti{\~n}as, Khazaei, Smart, \latin{et~al.} others]{Dutta2024perspective}
Dutta,~R.; Cabral,~D.~G.; Lyu,~N.; Vu,~N.~P.; Wang,~Y.; Allen,~B.; Dan,~X.;
  Corti{\~n}as,~R.~G.; Khazaei,~P.; Smart,~S.~E., \latin{et~al.}  Simulating
  Chemistry on Bosonic Quantum Devices. \emph{J. Chem. Theory Comput.}
  \textbf{2024}, \emph{20}, 6426\relax
\mciteBstWouldAddEndPuncttrue
\mciteSetBstMidEndSepPunct{\mcitedefaultmidpunct}
{\mcitedefaultendpunct}{\mcitedefaultseppunct}\relax
\EndOfBibitem
\bibitem[Huh \latin{et~al.}(2015)Huh, Guerreschi, Peropadre, McClean, and
  Aspuru-Guzik]{Huh2015}
Huh,~J.; Guerreschi,~G.; Peropadre,~B.; McClean,~J.~R.; Aspuru-Guzik,~A. Boson
  sampling for molecular vibronic spectra. \emph{Nature Photon.} \textbf{2015},
  \emph{9}, 615\relax
\mciteBstWouldAddEndPuncttrue
\mciteSetBstMidEndSepPunct{\mcitedefaultmidpunct}
{\mcitedefaultendpunct}{\mcitedefaultseppunct}\relax
\EndOfBibitem
\bibitem[Wang \latin{et~al.}(2020)Wang, Curtis, Lester, Zhang, Gao, Freeze,
  Batista, Vaccaro, Chuang, Frunzio, Jiang, Girvin, and Schoelkopf]{Wang2020}
Wang,~C.~S.; Curtis,~J.~C.; Lester,~B.~J.; Zhang,~Y.; Gao,~Y.~Y.; Freeze,~J.;
  Batista,~V.~S.; Vaccaro,~P.~H.; Chuang,~I.~L.; Frunzio,~L.; Jiang,~L.;
  Girvin,~S.~M.; Schoelkopf,~R.~J. Efficient Multiphoton Sampling of Molecular
  Vibronic Spectra on a Superconducting Bosonic Processor. \emph{Phys. Rev. X}
  \textbf{2020}, \emph{10}, 021060\relax
\mciteBstWouldAddEndPuncttrue
\mciteSetBstMidEndSepPunct{\mcitedefaultmidpunct}
{\mcitedefaultendpunct}{\mcitedefaultseppunct}\relax
\EndOfBibitem
\bibitem[Malpathak \latin{et~al.}(2024)Malpathak, Kallullathil, and
  Izmaylov]{Malpathak2024simulating}
Malpathak,~S.; Kallullathil,~S.~D.; Izmaylov,~A.~F. Simulating Vibrational
  Dynamics on Bosonic Quantum Devices. \emph{arXiv preprint arXiv:2411.17950}
  \textbf{2024}, \relax
\mciteBstWouldAddEndPunctfalse
\mciteSetBstMidEndSepPunct{\mcitedefaultmidpunct}
{}{\mcitedefaultseppunct}\relax
\EndOfBibitem
\bibitem[Wang \latin{et~al.}(2023)Wang, Frattini, Chapman, Puri, Girvin,
  Devoret, and Schoelkopf]{Wang2023}
Wang,~C.~S.; Frattini,~N.~E.; Chapman,~B.~J.; Puri,~S.; Girvin,~S.~M.;
  Devoret,~M.~H.; Schoelkopf,~R.~J. Observation of wave-packet branching
  through an engineered conical intersection. \emph{Phys. Rev. X}
  \textbf{2023}, \emph{13}, 011008\relax
\mciteBstWouldAddEndPuncttrue
\mciteSetBstMidEndSepPunct{\mcitedefaultmidpunct}
{\mcitedefaultendpunct}{\mcitedefaultseppunct}\relax
\EndOfBibitem
\bibitem[Lyu \latin{et~al.}(2023)Lyu, Miano, Tsioutsios, Corti{\~n}as, Jung,
  Wang, Hu, Geva, Kais, and Batista]{Lyu2023}
Lyu,~N.; Miano,~A.; Tsioutsios,~I.; Corti{\~n}as,~R.~G.; Jung,~K.; Wang,~Y.;
  Hu,~Z.; Geva,~E.; Kais,~S.; Batista,~V.~S. Mapping Molecular {Hamiltonians}
  into {Hamiltonians} of Modular {cQED} Processors. \emph{J. Chem. Theory
  Comput.} \textbf{2023}, \emph{19}, 6564\relax
\mciteBstWouldAddEndPuncttrue
\mciteSetBstMidEndSepPunct{\mcitedefaultmidpunct}
{\mcitedefaultendpunct}{\mcitedefaultseppunct}\relax
\EndOfBibitem
\bibitem[Cabral \latin{et~al.}(2024)Cabral, Khazaei, Allen, Videla,
  Sch{\"a}fer, Corti{\~n}as, Carrillo~de Albornoz, Ch{\'a}vez-Carlos, Santos,
  Geva, \latin{et~al.} others]{Cabral2024roadmap}
Cabral,~D.~G.; Khazaei,~P.; Allen,~B.~C.; Videla,~P.~E.; Sch{\"a}fer,~M.;
  Corti{\~n}as,~R.~G.; Carrillo~de Albornoz,~A.~C.; Ch{\'a}vez-Carlos,~J.;
  Santos,~L.~F.; Geva,~E., \latin{et~al.}  A Roadmap for Simulating Chemical
  Dynamics on a Parametrically Driven Bosonic Quantum Device. \emph{J. Phys.
  Chem. Lett.} \textbf{2024}, \emph{15}, 12042--12050\relax
\mciteBstWouldAddEndPuncttrue
\mciteSetBstMidEndSepPunct{\mcitedefaultmidpunct}
{\mcitedefaultendpunct}{\mcitedefaultseppunct}\relax
\EndOfBibitem
\bibitem[Copetudo \latin{et~al.}(2024)Copetudo, Fontaine, Valadares, and
  Gao]{Copetudo2024}
Copetudo,~A.; Fontaine,~C.~Y.; Valadares,~F.; Gao,~Y.~Y. Shaping photons:
  {Quantum} information processing with bosonic {cQED}. \emph{Appl. Phys.
  Lett.} \textbf{2024}, \emph{124}, 080502\relax
\mciteBstWouldAddEndPuncttrue
\mciteSetBstMidEndSepPunct{\mcitedefaultmidpunct}
{\mcitedefaultendpunct}{\mcitedefaultseppunct}\relax
\EndOfBibitem
\bibitem[Deleglise \latin{et~al.}(2008)Deleglise, Dotsenko, Sayrin, Bernu,
  Brune, Raimond, and Haroche]{Deleglise2008}
Deleglise,~S.; Dotsenko,~I.; Sayrin,~C.; Bernu,~J.; Brune,~M.; Raimond,~J.-M.;
  Haroche,~S. Reconstruction of non-classical cavity field states with
  snapshots of their decoherence. \emph{Nature} \textbf{2008}, \emph{455},
  510\relax
\mciteBstWouldAddEndPuncttrue
\mciteSetBstMidEndSepPunct{\mcitedefaultmidpunct}
{\mcitedefaultendpunct}{\mcitedefaultseppunct}\relax
\EndOfBibitem
\bibitem[Hacker \latin{et~al.}(2019)Hacker, Welte, Daiss, Shaukat, Ritter, Li,
  and Rempe]{Hacker2019}
Hacker,~B.; Welte,~S.; Daiss,~S.; Shaukat,~A.; Ritter,~S.; Li,~L.; Rempe,~G.
  Deterministic creation of entangled atom--light {Schr{\"o}dinger-cat} states.
  \emph{Nature Photon.} \textbf{2019}, \emph{13}, 110\relax
\mciteBstWouldAddEndPuncttrue
\mciteSetBstMidEndSepPunct{\mcitedefaultmidpunct}
{\mcitedefaultendpunct}{\mcitedefaultseppunct}\relax
\EndOfBibitem
\bibitem[Bruzewicz \latin{et~al.}(2019)Bruzewicz, Chiaverini, McConnell, and
  Sage]{Bruzewicz2019}
Bruzewicz,~C.~D.; Chiaverini,~J.; McConnell,~R.; Sage,~J.~M. Trapped-ion
  quantum computing: {Progress} and challenges. \emph{Appl. Phys. Rev.}
  \textbf{2019}, \emph{6}, 021314\relax
\mciteBstWouldAddEndPuncttrue
\mciteSetBstMidEndSepPunct{\mcitedefaultmidpunct}
{\mcitedefaultendpunct}{\mcitedefaultseppunct}\relax
\EndOfBibitem
\bibitem[Araz \latin{et~al.}(2024)Araz, Grau, Montgomery, and
  Ringer]{Araz2024toward}
Araz,~J.~Y.; Grau,~M.; Montgomery,~J.; Ringer,~F. Toward hybrid quantum
  simulations with qubits and qumodes on trapped-ion platforms. \emph{arXiv
  preprint arXiv:2410.07346} \textbf{2024}, \relax
\mciteBstWouldAddEndPunctfalse
\mciteSetBstMidEndSepPunct{\mcitedefaultmidpunct}
{}{\mcitedefaultseppunct}\relax
\EndOfBibitem
\bibitem[Joshi \latin{et~al.}(2021)Joshi, Noh, and Gao]{Joshi2021}
Joshi,~A.; Noh,~K.; Gao,~Y.~Y. Quantum information processing with bosonic
  qubits in circuit {QED}. \emph{Quantum Sci. Technol.} \textbf{2021},
  \emph{6}, 033001\relax
\mciteBstWouldAddEndPuncttrue
\mciteSetBstMidEndSepPunct{\mcitedefaultmidpunct}
{\mcitedefaultendpunct}{\mcitedefaultseppunct}\relax
\EndOfBibitem
\bibitem[Blais \latin{et~al.}(2021)Blais, Grimsmo, Girvin, and
  Wallraff]{Blais2021}
Blais,~A.; Grimsmo,~A.~L.; Girvin,~S.~M.; Wallraff,~A. Circuit quantum
  electrodynamics. \emph{Rev. Mod. Phys.} \textbf{2021}, \emph{93},
  025005\relax
\mciteBstWouldAddEndPuncttrue
\mciteSetBstMidEndSepPunct{\mcitedefaultmidpunct}
{\mcitedefaultendpunct}{\mcitedefaultseppunct}\relax
\EndOfBibitem
\bibitem[Liu \latin{et~al.}(2024)Liu, Singh, Smith, Crane, Martyn, Eickbusch,
  Schuckert, Li, Sinanan-Singh, Soley, \latin{et~al.}
  others]{Liu2024qumodequbitreview}
Liu,~Y.; Singh,~S.; Smith,~K.~C.; Crane,~E.; Martyn,~J.~M.; Eickbusch,~A.;
  Schuckert,~A.; Li,~R.~D.; Sinanan-Singh,~J.; Soley,~M.~B., \latin{et~al.}
  Hybrid oscillator-qubit quantum processors: Instruction set architectures,
  abstract machine models, and applications. \emph{arXiv preprint
  arXiv:2407.10381} \textbf{2024}, \relax
\mciteBstWouldAddEndPunctfalse
\mciteSetBstMidEndSepPunct{\mcitedefaultmidpunct}
{}{\mcitedefaultseppunct}\relax
\EndOfBibitem
\bibitem[Crane \latin{et~al.}(2024)Crane, Smith, Tomesh, Eickbusch, Martyn,
  K{\"u}hn, Funcke, DeMarco, Chuang, Wiebe, \latin{et~al.}
  others]{Crane2024hybrid}
Crane,~E.; Smith,~K.~C.; Tomesh,~T.; Eickbusch,~A.; Martyn,~J.~M.;
  K{\"u}hn,~S.; Funcke,~L.; DeMarco,~M.~A.; Chuang,~I.~L.; Wiebe,~N.,
  \latin{et~al.}  Hybrid Oscillator-Qubit Quantum Processors: Simulating
  Fermions, Bosons, and Gauge Fields. \emph{arXiv preprint arXiv:2409.03747}
  \textbf{2024}, \relax
\mciteBstWouldAddEndPunctfalse
\mciteSetBstMidEndSepPunct{\mcitedefaultmidpunct}
{}{\mcitedefaultseppunct}\relax
\EndOfBibitem
\bibitem[Romanenko \latin{et~al.}(2020)Romanenko, Pilipenko, Zorzetti, Frolov,
  Awida, Belomestnykh, Posen, and Grassellino]{Romanenko2020}
Romanenko,~A.; Pilipenko,~R.; Zorzetti,~S.; Frolov,~D.; Awida,~M.;
  Belomestnykh,~S.; Posen,~S.; Grassellino,~A. Three-Dimensional
  Superconducting Resonators at {$T < 20$ mK} with Photon Lifetimes up to
  $\ensuremath{\tau}=2$ s. \emph{Phys. Rev. Appl.} \textbf{2020}, \emph{13},
  034032\relax
\mciteBstWouldAddEndPuncttrue
\mciteSetBstMidEndSepPunct{\mcitedefaultmidpunct}
{\mcitedefaultendpunct}{\mcitedefaultseppunct}\relax
\EndOfBibitem
\bibitem[Ring and Schuck(1980)Ring, and Schuck]{RingBook}
Ring,~P.; Schuck,~P. \emph{The Nuclear Many-Body Problem}; Springer-Verlag,
  1980\relax
\mciteBstWouldAddEndPuncttrue
\mciteSetBstMidEndSepPunct{\mcitedefaultmidpunct}
{\mcitedefaultendpunct}{\mcitedefaultseppunct}\relax
\EndOfBibitem
\bibitem[Garbaczewski(1975)]{Garbaczewski1975}
Garbaczewski,~P. Representations of the {CAR} Generated by Representations of
  the {CCR}. {Fock} Case. \emph{Commun. Math. Phys.} \textbf{1975}, \emph{43},
  131\relax
\mciteBstWouldAddEndPuncttrue
\mciteSetBstMidEndSepPunct{\mcitedefaultmidpunct}
{\mcitedefaultendpunct}{\mcitedefaultseppunct}\relax
\EndOfBibitem
\bibitem[Garbaczewski(1978)]{Garbaczewski1978}
Garbaczewski,~P. The method of Boson expansions in quantum theory. \emph{Phys.
  Rep.} \textbf{1978}, \emph{36}, 65\relax
\mciteBstWouldAddEndPuncttrue
\mciteSetBstMidEndSepPunct{\mcitedefaultmidpunct}
{\mcitedefaultendpunct}{\mcitedefaultseppunct}\relax
\EndOfBibitem
\bibitem[Klein and Marshalek(1991)Klein, and Marshalek]{Klein1991}
Klein,~A.; Marshalek,~E. Boson realizations of {Lie} algebras with applications
  to nuclear physics. \emph{Rev. Mod. Phys.} \textbf{1991}, \emph{63},
  375\relax
\mciteBstWouldAddEndPuncttrue
\mciteSetBstMidEndSepPunct{\mcitedefaultmidpunct}
{\mcitedefaultendpunct}{\mcitedefaultseppunct}\relax
\EndOfBibitem
\bibitem[Ginocchio and Johnson(1996)Ginocchio, and Johnson]{Ginocchio1996}
Ginocchio,~J.~N.; Johnson,~C.~W. Fermion to boson mappings revisited.
  \emph{Phys. Rep.} \textbf{1996}, \emph{264}, 153\relax
\mciteBstWouldAddEndPuncttrue
\mciteSetBstMidEndSepPunct{\mcitedefaultmidpunct}
{\mcitedefaultendpunct}{\mcitedefaultseppunct}\relax
\EndOfBibitem
\bibitem[Von~Delft and Schoeller(1998)Von~Delft, and Schoeller]{VonDelft1998}
Von~Delft,~J.; Schoeller,~H. Bosonization for beginners—refermionization for
  experts. \emph{Ann. Phys.} \textbf{1998}, \emph{510}, 225\relax
\mciteBstWouldAddEndPuncttrue
\mciteSetBstMidEndSepPunct{\mcitedefaultmidpunct}
{\mcitedefaultendpunct}{\mcitedefaultseppunct}\relax
\EndOfBibitem
\bibitem[Scuseria \latin{et~al.}(2013)Scuseria, Henderson, and
  Bulik]{Scuseria2013}
Scuseria,~G.~E.; Henderson,~T.~M.; Bulik,~I.~W. Particle-particle and
  quasiparticle random phase approximations: {Connections} to coupled cluster
  theory. \emph{J. Chem. Phys.} \textbf{2013}, \emph{139}, 104113\relax
\mciteBstWouldAddEndPuncttrue
\mciteSetBstMidEndSepPunct{\mcitedefaultmidpunct}
{\mcitedefaultendpunct}{\mcitedefaultseppunct}\relax
\EndOfBibitem
\bibitem[Liu(2016)]{Liu2016}
Liu,~J. A unified theoretical framework for mapping models for the multi-state
  {Hamiltonian}. \emph{J. Chem. Phys.} \textbf{2016}, \emph{145}, 204105\relax
\mciteBstWouldAddEndPuncttrue
\mciteSetBstMidEndSepPunct{\mcitedefaultmidpunct}
{\mcitedefaultendpunct}{\mcitedefaultseppunct}\relax
\EndOfBibitem
\bibitem[Montoya-Castillo and Markland(2018)Montoya-Castillo, and
  Markland]{Montoya2018}
Montoya-Castillo,~A.; Markland,~T.~E. On the exact continuous mapping of
  fermions. \emph{Sci. Rep.} \textbf{2018}, \emph{8}, 1\relax
\mciteBstWouldAddEndPuncttrue
\mciteSetBstMidEndSepPunct{\mcitedefaultmidpunct}
{\mcitedefaultendpunct}{\mcitedefaultseppunct}\relax
\EndOfBibitem
\bibitem[Ohta(1998)]{Ohta1998}
Ohta,~K. New bosonic excitation operators in many-electron wave functions.
  \emph{Int. J. Quantum Chem.} \textbf{1998}, \emph{67}, 71\relax
\mciteBstWouldAddEndPuncttrue
\mciteSetBstMidEndSepPunct{\mcitedefaultmidpunct}
{\mcitedefaultendpunct}{\mcitedefaultseppunct}\relax
\EndOfBibitem
\bibitem[Dhar \latin{et~al.}(2006)Dhar, Mandal, and Suryanarayana]{Dhar2006}
Dhar,~A.; Mandal,~G.; Suryanarayana,~N.~V. Exact operator bosonization of
  finite number of fermions in one space dimension. \emph{J. High Energy Phys.}
  \textbf{2006}, \emph{2006}, 118\relax
\mciteBstWouldAddEndPuncttrue
\mciteSetBstMidEndSepPunct{\mcitedefaultmidpunct}
{\mcitedefaultendpunct}{\mcitedefaultseppunct}\relax
\EndOfBibitem
\bibitem[Qin \latin{et~al.}(2022)Qin, Sch{\"a}fer, Andergassen, Corboz, and
  Gull]{Qin2022}
Qin,~M.; Sch{\"a}fer,~T.; Andergassen,~S.; Corboz,~P.; Gull,~E. The {Hubbard}
  model: {A} computational perspective. \emph{Annu. Rev. Condens. Matter Phys.}
  \textbf{2022}, \emph{13}, 275\relax
\mciteBstWouldAddEndPuncttrue
\mciteSetBstMidEndSepPunct{\mcitedefaultmidpunct}
{\mcitedefaultendpunct}{\mcitedefaultseppunct}\relax
\EndOfBibitem
\bibitem[Szabo and Ostlund(1996)Szabo, and Ostlund]{SzaboBook}
Szabo,~A.; Ostlund,~N.~S. \emph{Modern Quantum Chemistry}; Dover Publications,
  1996\relax
\mciteBstWouldAddEndPuncttrue
\mciteSetBstMidEndSepPunct{\mcitedefaultmidpunct}
{\mcitedefaultendpunct}{\mcitedefaultseppunct}\relax
\EndOfBibitem
\bibitem[Helgaker \latin{et~al.}(2000)Helgaker, J{\o}rgensen, and
  Olsen]{HelgakerBook}
Helgaker,~T.; J{\o}rgensen,~P.; Olsen,~J. \emph{Molecular Electronic Structure
  Theory}; John Wiley and Sons, 2000\relax
\mciteBstWouldAddEndPuncttrue
\mciteSetBstMidEndSepPunct{\mcitedefaultmidpunct}
{\mcitedefaultendpunct}{\mcitedefaultseppunct}\relax
\EndOfBibitem
\bibitem[Jordan(1935)]{Jordan1935}
Jordan,~P. Der Zusammenhang der symmetrischen und linearen Gruppen und das
  Mehrk{\"o}rperproblem. \emph{Zeitschrift f{\"u}r Physik} \textbf{1935},
  \emph{94}, 531\relax
\mciteBstWouldAddEndPuncttrue
\mciteSetBstMidEndSepPunct{\mcitedefaultmidpunct}
{\mcitedefaultendpunct}{\mcitedefaultseppunct}\relax
\EndOfBibitem
\bibitem[Liu \latin{et~al.}(2024)Liu, Che, Zhou, Shi, and
  Li]{Liu2024fermihedral}
Liu,~Y.; Che,~S.; Zhou,~J.; Shi,~Y.; Li,~G. Fermihedral: {On} the Optimal
  Compilation for Fermion-to-Qubit Encoding. Proceedings of the 29th ACM
  International Conference on Architectural Support for Programming Languages
  and Operating Systems, Volume 3. 2024; pp 382--397\relax
\mciteBstWouldAddEndPuncttrue
\mciteSetBstMidEndSepPunct{\mcitedefaultmidpunct}
{\mcitedefaultendpunct}{\mcitedefaultseppunct}\relax
\EndOfBibitem
\bibitem[Babbush \latin{et~al.}(2018)Babbush, Wiebe, McClean, McClain, Neven,
  and Chan]{Babbush2018}
Babbush,~R.; Wiebe,~N.; McClean,~J.; McClain,~J.; Neven,~H.; Chan,~G. K.-L.
  Low-Depth Quantum Simulation of Materials. \emph{Phys. Rev. X} \textbf{2018},
  \emph{8}, 011044\relax
\mciteBstWouldAddEndPuncttrue
\mciteSetBstMidEndSepPunct{\mcitedefaultmidpunct}
{\mcitedefaultendpunct}{\mcitedefaultseppunct}\relax
\EndOfBibitem
\bibitem[Eickbusch \latin{et~al.}(2022)Eickbusch, Sivak, Ding, Elder, Jha,
  Venkatraman, Royer, Girvin, Schoelkopf, and Devoret]{Eickbusch2022}
Eickbusch,~A.; Sivak,~V.; Ding,~A.~Z.; Elder,~S.~S.; Jha,~S.~R.;
  Venkatraman,~J.; Royer,~B.; Girvin,~S.~M.; Schoelkopf,~R.~J.; Devoret,~M.~H.
  Fast universal control of an oscillator with weak dispersive coupling to a
  qubit. \emph{Nat. Phys.} \textbf{2022}, \emph{18}, 1464\relax
\mciteBstWouldAddEndPuncttrue
\mciteSetBstMidEndSepPunct{\mcitedefaultmidpunct}
{\mcitedefaultendpunct}{\mcitedefaultseppunct}\relax
\EndOfBibitem
\bibitem[You \latin{et~al.}(2024)You, Lu, Kim, Kurkcuoglu, Zhu, van Zanten,
  Roy, Lu, Chakram, Grassellino, Romanenko, Koch, and
  Zorzetti]{You2024Crosstalk}
You,~X.; Lu,~Y.; Kim,~T.; Kurkcuoglu,~D.~M.; Zhu,~S.; van Zanten,~D.; Roy,~T.;
  Lu,~Y.; Chakram,~S.; Grassellino,~A.; Romanenko,~A.; Koch,~J.; Zorzetti,~S.
  Crosstalk-Robust Quantum Control in Multimode Bosonic Systems. \emph{Phys.
  Rev. Appl.} \textbf{2024}, \emph{22}, 044072\relax
\mciteBstWouldAddEndPuncttrue
\mciteSetBstMidEndSepPunct{\mcitedefaultmidpunct}
{\mcitedefaultendpunct}{\mcitedefaultseppunct}\relax
\EndOfBibitem
\bibitem[Krastanov \latin{et~al.}(2015)Krastanov, Albert, Shen, Zou, Heeres,
  Vlastakis, Schoelkopf, and Jiang]{Krastanov2015}
Krastanov,~S.; Albert,~V.~V.; Shen,~C.; Zou,~C.-L.; Heeres,~R.~W.;
  Vlastakis,~B.; Schoelkopf,~R.~J.; Jiang,~L. Universal control of an
  oscillator with dispersive coupling to a qubit. \emph{Phys. Rev. A}
  \textbf{2015}, \emph{92}, 040303\relax
\mciteBstWouldAddEndPuncttrue
\mciteSetBstMidEndSepPunct{\mcitedefaultmidpunct}
{\mcitedefaultendpunct}{\mcitedefaultseppunct}\relax
\EndOfBibitem
\bibitem[Diringer \latin{et~al.}(2024)Diringer, Blumenthal, Grinberg, Jiang,
  and Hacohen-Gourgy]{Diringer2024}
Diringer,~A.~A.; Blumenthal,~E.; Grinberg,~A.; Jiang,~L.; Hacohen-Gourgy,~S.
  Conditional-not Displacement: {Fast} Multioscillator Control with a Single
  Qubit. \emph{Phys. Rev. X} \textbf{2024}, \emph{14}, 011055\relax
\mciteBstWouldAddEndPuncttrue
\mciteSetBstMidEndSepPunct{\mcitedefaultmidpunct}
{\mcitedefaultendpunct}{\mcitedefaultseppunct}\relax
\EndOfBibitem
\bibitem[Zhang and Zhuang(2024)Zhang, and Zhuang]{Zhang2023energy}
Zhang,~B.; Zhuang,~Q. Energy-dependent barren plateau in bosonic variational
  quantum circuits. \emph{Quantum Sci. Technol.} \textbf{2024}, \emph{10},
  015009\relax
\mciteBstWouldAddEndPuncttrue
\mciteSetBstMidEndSepPunct{\mcitedefaultmidpunct}
{\mcitedefaultendpunct}{\mcitedefaultseppunct}\relax
\EndOfBibitem
\bibitem[Job(2023)]{Job2023efficient}
Job,~J. Efficient, direct compilation of SU (N) operations into SNAP \&
  Displacement gates. \emph{arXiv preprint arXiv:2307.11900} \textbf{2023},
  \relax
\mciteBstWouldAddEndPunctfalse
\mciteSetBstMidEndSepPunct{\mcitedefaultmidpunct}
{}{\mcitedefaultseppunct}\relax
\EndOfBibitem
\bibitem[Shang \latin{et~al.}(2024)Shang, Zhong, Zhang, Yu, Yuan, Lu, Pan, and
  Chen]{Shang2024boson}
Shang,~Z.-X.; Zhong,~H.-S.; Zhang,~Y.-K.; Yu,~C.-C.; Yuan,~X.; Lu,~C.-Y.;
  Pan,~J.-W.; Chen,~M.-C. Boson sampling enhanced quantum chemistry.
  \emph{arXiv preprint arXiv:2403.16698} \textbf{2024}, \relax
\mciteBstWouldAddEndPunctfalse
\mciteSetBstMidEndSepPunct{\mcitedefaultmidpunct}
{}{\mcitedefaultseppunct}\relax
\EndOfBibitem
\bibitem[Ryabinkin \latin{et~al.}(2018)Ryabinkin, Genin, and
  Izmaylov]{Ryabinkin2018constrained}
Ryabinkin,~I.~G.; Genin,~S.~N.; Izmaylov,~A.~F. Constrained variational quantum
  eigensolver: {Quantum} computer search engine in the {Fock} space. \emph{J.
  Chem. Theory Comput.} \textbf{2018}, \emph{15}, 249\relax
\mciteBstWouldAddEndPuncttrue
\mciteSetBstMidEndSepPunct{\mcitedefaultmidpunct}
{\mcitedefaultendpunct}{\mcitedefaultseppunct}\relax
\EndOfBibitem
\bibitem[Hehre \latin{et~al.}(1969)Hehre, Stewart, and Pople]{Hehre1969}
Hehre,~W.~J.; Stewart,~R.~F.; Pople,~J.~A. Self-consistent molecular-orbital
  methods. {I} {Use} of {Gaussian} expansions of {Slater}-type atomic orbitals.
  \emph{J. Chem. Phys.} \textbf{1969}, \emph{51}, 2657\relax
\mciteBstWouldAddEndPuncttrue
\mciteSetBstMidEndSepPunct{\mcitedefaultmidpunct}
{\mcitedefaultendpunct}{\mcitedefaultseppunct}\relax
\EndOfBibitem
\bibitem[Virtanen \latin{et~al.}(2020)Virtanen, Gommers, Oliphant, Haberland,
  Reddy, Cournapeau, Burovski, Peterson, Weckesser, Bright, {van der Walt},
  Brett, Wilson, Millman, Mayorov, Nelson, Jones, Kern, Larson, Carey, Polat,
  Feng, Moore, {VanderPlas}, Laxalde, Perktold, Cimrman, Henriksen, Quintero,
  Harris, Archibald, Ribeiro, Pedregosa, {van Mulbregt}, and {SciPy 1.0
  Contributors}]{2020SciPy-NMeth}
Virtanen,~P.; Gommers,~R.; Oliphant,~T.~E.; Haberland,~M.; Reddy,~T.;
  Cournapeau,~D.; Burovski,~E.; Peterson,~P.; Weckesser,~W.; Bright,~J.; {van
  der Walt},~S.~J.; Brett,~M.; Wilson,~J.; Millman,~K.~J.; Mayorov,~N.;
  Nelson,~A. R.~J.; Jones,~E.; Kern,~R.; Larson,~E.; Carey,~C.~J.;
  Polat,~{\.I}.; Feng,~Y.; Moore,~E.~W.; {VanderPlas},~J.; Laxalde,~D.;
  Perktold,~J.; Cimrman,~R.; Henriksen,~I.; Quintero,~E.~A.; Harris,~C.~R.;
  Archibald,~A.~M.; Ribeiro,~A.~H.; Pedregosa,~F.; {van Mulbregt},~P.; {SciPy
  1.0 Contributors}, {{SciPy} 1.0: Fundamental Algorithms for Scientific
  Computing in Python}. \emph{Nat. Methods} \textbf{2020}, \emph{17},
  261--272\relax
\mciteBstWouldAddEndPuncttrue
\mciteSetBstMidEndSepPunct{\mcitedefaultmidpunct}
{\mcitedefaultendpunct}{\mcitedefaultseppunct}\relax
\EndOfBibitem
\bibitem[Lambert \latin{et~al.}(2024)Lambert, Gigu{\`e}re, Menczel, Li, Hopf,
  Su{\'a}rez, Gali, Lishman, Gadhvi, Agarwal, \latin{et~al.}
  others]{Lambert2024qutip}
Lambert,~N.; Gigu{\`e}re,~E.; Menczel,~P.; Li,~B.; Hopf,~P.; Su{\'a}rez,~G.;
  Gali,~M.; Lishman,~J.; Gadhvi,~R.; Agarwal,~R., \latin{et~al.}  QuTiP 5: The
  Quantum Toolbox in Python. \emph{arXiv preprint arXiv:2412.04705}
  \textbf{2024}, \relax
\mciteBstWouldAddEndPunctfalse
\mciteSetBstMidEndSepPunct{\mcitedefaultmidpunct}
{}{\mcitedefaultseppunct}\relax
\EndOfBibitem
\bibitem[McClean \latin{et~al.}(2022)McClean, Sung, Kivlichan, Cao, Dai, Fried,
  Gidney, Gimby, Gokhale, H{\"a}ner, \latin{et~al.}
  others]{Mcclean2022openfermion}
McClean,~J.~R.; Sung,~K.; Kivlichan,~I.; Cao,~Y.; Dai,~C.; Fried,~E.;
  Gidney,~C.; Gimby,~B.; Gokhale,~P.; H{\"a}ner,~T., \latin{et~al.}
  Openfermion: The electronic structure package for quantum computers (2017).
  \emph{arXiv preprint arXiv:1710.07629} \textbf{2022}, \emph{1710}\relax
\mciteBstWouldAddEndPuncttrue
\mciteSetBstMidEndSepPunct{\mcitedefaultmidpunct}
{\mcitedefaultendpunct}{\mcitedefaultseppunct}\relax
\EndOfBibitem
\bibitem[F{\"o}sel \latin{et~al.}(2020)F{\"o}sel, Krastanov, Marquardt, and
  Jiang]{Fosel2020}
F{\"o}sel,~T.; Krastanov,~S.; Marquardt,~F.; Jiang,~L. Efficient cavity control
  with {SNAP} gates. \emph{arXiv preprint arXiv:2004.14256} \textbf{2020},
  \relax
\mciteBstWouldAddEndPunctfalse
\mciteSetBstMidEndSepPunct{\mcitedefaultmidpunct}
{}{\mcitedefaultseppunct}\relax
\EndOfBibitem
\bibitem[Valadares \latin{et~al.}(2024)Valadares, Huang, Chu, Dorogov, Chua,
  Kong, Song, and Gao]{Valadares2024demand}
Valadares,~F.; Huang,~N.-N.; Chu,~K. T.~N.; Dorogov,~A.; Chua,~W.; Kong,~L.;
  Song,~P.; Gao,~Y.~Y. On-demand transposition across light-matter interaction
  regimes in bosonic {cQED}. \emph{Nat. Commun.} \textbf{2024}, \emph{15},
  5816\relax
\mciteBstWouldAddEndPuncttrue
\mciteSetBstMidEndSepPunct{\mcitedefaultmidpunct}
{\mcitedefaultendpunct}{\mcitedefaultseppunct}\relax
\EndOfBibitem
\bibitem[Gao \latin{et~al.}(2018)Gao, Lester, Zhang, Wang, Rosenblum, Frunzio,
  Jiang, Girvin, and Schoelkopf]{Gao2018}
Gao,~Y.~Y.; Lester,~B.~J.; Zhang,~Y.; Wang,~C.; Rosenblum,~S.; Frunzio,~L.;
  Jiang,~L.; Girvin,~S.~M.; Schoelkopf,~R.~J. Programmable Interference between
  Two Microwave Quantum Memories. \emph{Phys. Rev. X} \textbf{2018}, \emph{8},
  021073\relax
\mciteBstWouldAddEndPuncttrue
\mciteSetBstMidEndSepPunct{\mcitedefaultmidpunct}
{\mcitedefaultendpunct}{\mcitedefaultseppunct}\relax
\EndOfBibitem
\bibitem[Zhang \latin{et~al.}(2019)Zhang, Lester, Gao, Jiang, Schoelkopf, and
  Girvin]{Zhang2019}
Zhang,~Y.; Lester,~B.~J.; Gao,~Y.~Y.; Jiang,~L.; Schoelkopf,~R.~J.;
  Girvin,~S.~M. Engineering bilinear mode coupling in circuit QED: Theory and
  experiment. \emph{Phys. Rev. A} \textbf{2019}, \emph{99}, 012314\relax
\mciteBstWouldAddEndPuncttrue
\mciteSetBstMidEndSepPunct{\mcitedefaultmidpunct}
{\mcitedefaultendpunct}{\mcitedefaultseppunct}\relax
\EndOfBibitem
\bibitem[Lu \latin{et~al.}(2023)Lu, Maiti, Garmon, Ganjam, Zhang, Claes,
  Frunzio, Girvin, and Schoelkopf]{Lu2023high}
Lu,~Y.; Maiti,~A.; Garmon,~J.~W.; Ganjam,~S.; Zhang,~Y.; Claes,~J.;
  Frunzio,~L.; Girvin,~S.~M.; Schoelkopf,~R.~J. High-fidelity parametric
  beamsplitting with a parity-protected converter. \emph{Nat. Commun.}
  \textbf{2023}, \emph{14}, 5767\relax
\mciteBstWouldAddEndPuncttrue
\mciteSetBstMidEndSepPunct{\mcitedefaultmidpunct}
{\mcitedefaultendpunct}{\mcitedefaultseppunct}\relax
\EndOfBibitem
\bibitem[Chapman \latin{et~al.}(2023)Chapman, de~Graaf, Xue, Zhang, Teoh,
  Curtis, Tsunoda, Eickbusch, Read, Koottandavida, Mundhada, Frunzio, Devoret,
  Girvin, and Schoelkopf]{Chapman2023StrongWeakDispersive}
Chapman,~B.~J.; de~Graaf,~S.~J.; Xue,~S.~H.; Zhang,~Y.; Teoh,~J.;
  Curtis,~J.~C.; Tsunoda,~T.; Eickbusch,~A.; Read,~A.~P.; Koottandavida,~A.;
  Mundhada,~S.~O.; Frunzio,~L.; Devoret,~M.; Girvin,~S.; Schoelkopf,~R.
  High-On-Off-Ratio Beam-Splitter Interaction for Gates on Bosonically Encoded
  Qubits. \emph{PRX Quantum} \textbf{2023}, \emph{4}, 020355\relax
\mciteBstWouldAddEndPuncttrue
\mciteSetBstMidEndSepPunct{\mcitedefaultmidpunct}
{\mcitedefaultendpunct}{\mcitedefaultseppunct}\relax
\EndOfBibitem
\bibitem[Gargiulo \latin{et~al.}(2021)Gargiulo, Oleschko, Prat-Camps, Zanner,
  and Kirchmair]{Gargiulo2021fast}
Gargiulo,~O.; Oleschko,~S.; Prat-Camps,~J.; Zanner,~M.; Kirchmair,~G. Fast flux
  control of 3D transmon qubits using a magnetic hose. \emph{Appl. Phys. Lett.}
  \textbf{2021}, \emph{118}\relax
\mciteBstWouldAddEndPuncttrue
\mciteSetBstMidEndSepPunct{\mcitedefaultmidpunct}
{\mcitedefaultendpunct}{\mcitedefaultseppunct}\relax
\EndOfBibitem
\bibitem[Fujii(2003)]{Fujii2003exchange}
Fujii,~K. Exchange gate on the qudit space and Fock space. \emph{Journal of
  Optics B: Quantum and Semiclassical Optics} \textbf{2003}, \emph{5},
  S613\relax
\mciteBstWouldAddEndPuncttrue
\mciteSetBstMidEndSepPunct{\mcitedefaultmidpunct}
{\mcitedefaultendpunct}{\mcitedefaultseppunct}\relax
\EndOfBibitem
\bibitem[Frattini \latin{et~al.}(2017)Frattini, Vool, Shankar, Narla, Sliwa,
  and Devoret]{Frattini2017}
Frattini,~N.; Vool,~U.; Shankar,~S.; Narla,~A.; Sliwa,~K.; Devoret,~M. 3-wave
  mixing {Josephson} dipole element. \emph{Appl. Phys. Lett.} \textbf{2017},
  \emph{110}\relax
\mciteBstWouldAddEndPuncttrue
\mciteSetBstMidEndSepPunct{\mcitedefaultmidpunct}
{\mcitedefaultendpunct}{\mcitedefaultseppunct}\relax
\EndOfBibitem
\bibitem[Vaquero-Sabater \latin{et~al.}(2024)Vaquero-Sabater, Carreras,
  Or{\'u}s, Mayhall, and Casanova]{Vaquero2024physically}
Vaquero-Sabater,~N.; Carreras,~A.; Or{\'u}s,~R.; Mayhall,~N.~J.; Casanova,~D.
  Physically Motivated Improvements of Variational Quantum Eigensolvers.
  \emph{J. Chem. Theory Comput.} \textbf{2024}, \relax
\mciteBstWouldAddEndPunctfalse
\mciteSetBstMidEndSepPunct{\mcitedefaultmidpunct}
{}{\mcitedefaultseppunct}\relax
\EndOfBibitem
\bibitem[Motta \latin{et~al.}(2017)Motta, Ceperley, Chan, Gomez, Gull, Guo,
  Jim\'enez-Hoyos, Lan, Li, Ma, Millis, Prokof'ev, Ray, Scuseria, Sorella,
  Stoudenmire, Sun, Tupitsyn, White, Zgid, and Zhang]{Motta2017Simons}
Motta,~M.; Ceperley,~D.~M.; Chan,~G. K.-L.; Gomez,~J.~A.; Gull,~E.; Guo,~S.;
  Jim\'enez-Hoyos,~C.~A.; Lan,~T.~N.; Li,~J.; Ma,~F.; Millis,~A.~J.;
  Prokof'ev,~N.~V.; Ray,~U.; Scuseria,~G.~E.; Sorella,~S.; Stoudenmire,~E.~M.;
  Sun,~Q.; Tupitsyn,~I.~S.; White,~S.~R.; Zgid,~D.; Zhang,~S. Towards the
  Solution of the Many-Electron Problem in Real Materials: Equation of State of
  the Hydrogen Chain with State-of-the-Art Many-Body Methods. \emph{Phys. Rev.
  X} \textbf{2017}, \emph{7}, 031059\relax
\mciteBstWouldAddEndPuncttrue
\mciteSetBstMidEndSepPunct{\mcitedefaultmidpunct}
{\mcitedefaultendpunct}{\mcitedefaultseppunct}\relax
\EndOfBibitem
\bibitem[Abadi \latin{et~al.}(2015)Abadi, Agarwal, Barham, Brevdo, Chen, Citro,
  Corrado, Davis, Dean, Devin, Ghemawat, Goodfellow, Harp, Irving, Isard, Jia,
  Jozefowicz, Kaiser, Kudlur, Levenberg, Man\'{e}, Monga, Moore, Murray, Olah,
  Schuster, Shlens, Steiner, Sutskever, Talwar, Tucker, Vanhoucke, Vasudevan,
  Vi\'{e}gas, Vinyals, Warden, Wattenberg, Wicke, Yu, and
  Zheng]{tensorflow2015-whitepaper}
Abadi,~M.; Agarwal,~A.; Barham,~P.; Brevdo,~E.; Chen,~Z.; Citro,~C.;
  Corrado,~G.~S.; Davis,~A.; Dean,~J.; Devin,~M.; Ghemawat,~S.; Goodfellow,~I.;
  Harp,~A.; Irving,~G.; Isard,~M.; Jia,~Y.; Jozefowicz,~R.; Kaiser,~L.;
  Kudlur,~M.; Levenberg,~J.; Man\'{e},~D.; Monga,~R.; Moore,~S.; Murray,~D.;
  Olah,~C.; Schuster,~M.; Shlens,~J.; Steiner,~B.; Sutskever,~I.; Talwar,~K.;
  Tucker,~P.; Vanhoucke,~V.; Vasudevan,~V.; Vi\'{e}gas,~F.; Vinyals,~O.;
  Warden,~P.; Wattenberg,~M.; Wicke,~M.; Yu,~Y.; Zheng,~X. {TensorFlow}:
  Large-Scale Machine Learning on Heterogeneous Systems. 2015;
  \url{https://www.tensorflow.org/}, Software available from
  tensorflow.org\relax
\mciteBstWouldAddEndPuncttrue
\mciteSetBstMidEndSepPunct{\mcitedefaultmidpunct}
{\mcitedefaultendpunct}{\mcitedefaultseppunct}\relax
\EndOfBibitem
\bibitem[Javadi-Abhari \latin{et~al.}(2024)Javadi-Abhari, Treinish, Krsulich,
  Wood, Lishman, Gacon, Martiel, Nation, Bishop, Cross, \latin{et~al.}
  others]{Javadi2024quantum}
Javadi-Abhari,~A.; Treinish,~M.; Krsulich,~K.; Wood,~C.~J.; Lishman,~J.;
  Gacon,~J.; Martiel,~S.; Nation,~P.~D.; Bishop,~L.~S.; Cross,~A.~W.,
  \latin{et~al.}  Quantum computing with Qiskit. \emph{arXiv preprint
  arXiv:2405.08810} \textbf{2024}, \relax
\mciteBstWouldAddEndPunctfalse
\mciteSetBstMidEndSepPunct{\mcitedefaultmidpunct}
{}{\mcitedefaultseppunct}\relax
\EndOfBibitem
\bibitem[Ryabinkin \latin{et~al.}(2018)Ryabinkin, Yen, Genin, and
  Izmaylov]{Ryabinkin2018}
Ryabinkin,~I.~G.; Yen,~T.-C.; Genin,~S.~N.; Izmaylov,~A.~F. Qubit Coupled
  Cluster Method: A Systematic Approach to Quantum Chemistry on a Quantum
  Computer. \emph{J. Chem. Theory Comput.} \textbf{2018}, \emph{14}, 6317\relax
\mciteBstWouldAddEndPuncttrue
\mciteSetBstMidEndSepPunct{\mcitedefaultmidpunct}
{\mcitedefaultendpunct}{\mcitedefaultseppunct}\relax
\EndOfBibitem
\bibitem[Nakatsuji(2000)]{Nakatsuji2000}
Nakatsuji,~H. Structure of the exact wave function. \emph{J. Chem. Phys.}
  \textbf{2000}, \emph{113}, 2949\relax
\mciteBstWouldAddEndPuncttrue
\mciteSetBstMidEndSepPunct{\mcitedefaultmidpunct}
{\mcitedefaultendpunct}{\mcitedefaultseppunct}\relax
\EndOfBibitem
\bibitem[Fukutome(1981)]{Fukutome1981}
Fukutome,~H. The Group Theoretical Structure of Fermion Many-Body Systems
  Arising from the Canonical Anticommutation Relation. {I}. \emph{Prog. Theor.
  Phys.} \textbf{1981}, \emph{56}, 809\relax
\mciteBstWouldAddEndPuncttrue
\mciteSetBstMidEndSepPunct{\mcitedefaultmidpunct}
{\mcitedefaultendpunct}{\mcitedefaultseppunct}\relax
\EndOfBibitem
\bibitem[Suryanarayana(2006)]{Suryanarayana2006}
Suryanarayana,~N.~V. Half-{BPS} giants, free fermions and microstates of
  superstars. \emph{J. High Energy Phys.} \textbf{2006}, \emph{2006}, 82\relax
\mciteBstWouldAddEndPuncttrue
\mciteSetBstMidEndSepPunct{\mcitedefaultmidpunct}
{\mcitedefaultendpunct}{\mcitedefaultseppunct}\relax
\EndOfBibitem
\bibitem[Wang \latin{et~al.}(2020)Wang, Hu, Sanders, and Kais]{Wang2020Qudits}
Wang,~Y.; Hu,~Z.; Sanders,~B.~C.; Kais,~S. Qudits and high-dimensional quantum
  computing. \emph{Front. Phys.} \textbf{2020}, \emph{8}, 589504\relax
\mciteBstWouldAddEndPuncttrue
\mciteSetBstMidEndSepPunct{\mcitedefaultmidpunct}
{\mcitedefaultendpunct}{\mcitedefaultseppunct}\relax
\EndOfBibitem
\bibitem[Lloyd and Braunstein(1999)Lloyd, and Braunstein]{Lloyd1999}
Lloyd,~S.; Braunstein,~S.~L. Quantum Computation over Continuous Variables.
  \emph{Phys. Rev. Lett.} \textbf{1999}, \emph{82}, 1784\relax
\mciteBstWouldAddEndPuncttrue
\mciteSetBstMidEndSepPunct{\mcitedefaultmidpunct}
{\mcitedefaultendpunct}{\mcitedefaultseppunct}\relax
\EndOfBibitem
\bibitem[Braunstein and van Loock(2005)Braunstein, and van
  Loock]{Braunstein2005}
Braunstein,~S.~L.; van Loock,~P. Quantum information with continuous variables.
  \emph{Rev. Mod. Phys.} \textbf{2005}, \emph{77}, 513\relax
\mciteBstWouldAddEndPuncttrue
\mciteSetBstMidEndSepPunct{\mcitedefaultmidpunct}
{\mcitedefaultendpunct}{\mcitedefaultseppunct}\relax
\EndOfBibitem
\bibitem[Heeres \latin{et~al.}(2015)Heeres, Vlastakis, Holland, Krastanov,
  Albert, Frunzio, Jiang, and Schoelkopf]{Heeres2015}
Heeres,~R.~W.; Vlastakis,~B.; Holland,~E.; Krastanov,~S.; Albert,~V.~V.;
  Frunzio,~L.; Jiang,~L.; Schoelkopf,~R.~J. Cavity State Manipulation Using
  Photon-Number Selective Phase Gates. \emph{Phys. Rev. Lett.} \textbf{2015},
  \emph{115}, 137002\relax
\mciteBstWouldAddEndPuncttrue
\mciteSetBstMidEndSepPunct{\mcitedefaultmidpunct}
{\mcitedefaultendpunct}{\mcitedefaultseppunct}\relax
\EndOfBibitem
\bibitem[Leghtas \latin{et~al.}(2013)Leghtas, Kirchmair, Vlastakis, Devoret,
  Schoelkopf, and Mirrahimi]{Leghtas2013CD}
Leghtas,~Z.; Kirchmair,~G.; Vlastakis,~B.; Devoret,~M.~H.; Schoelkopf,~R.~J.;
  Mirrahimi,~M. Deterministic protocol for mapping a qubit to coherent state
  superpositions in a cavity. \emph{Phys. Rev. A} \textbf{2013}, \emph{87},
  042315\relax
\mciteBstWouldAddEndPuncttrue
\mciteSetBstMidEndSepPunct{\mcitedefaultmidpunct}
{\mcitedefaultendpunct}{\mcitedefaultseppunct}\relax
\EndOfBibitem
\bibitem[Divochiy \latin{et~al.}(2008)Divochiy, Marsili, Bitauld, Gaggero,
  Leoni, Mattioli, Korneev, Seleznev, Kaurova, Minaeva, \latin{et~al.}
  others]{Divochiy2008}
Divochiy,~A.; Marsili,~F.; Bitauld,~D.; Gaggero,~A.; Leoni,~R.; Mattioli,~F.;
  Korneev,~A.; Seleznev,~V.; Kaurova,~N.; Minaeva,~O., \latin{et~al.}
  Superconducting nanowire photon-number-resolving detector at
  telecommunication wavelengths. \emph{Nature Photon.} \textbf{2008}, \emph{2},
  302\relax
\mciteBstWouldAddEndPuncttrue
\mciteSetBstMidEndSepPunct{\mcitedefaultmidpunct}
{\mcitedefaultendpunct}{\mcitedefaultseppunct}\relax
\EndOfBibitem
\bibitem[Kardyna{\l} \latin{et~al.}(2008)Kardyna{\l}, Yuan, and
  Shields]{Kardynal2008}
Kardyna{\l},~B.; Yuan,~Z.; Shields,~A. An avalanche-photodiode-based
  photon-number-resolving detector. \emph{Nature Photon.} \textbf{2008},
  \emph{2}, 425\relax
\mciteBstWouldAddEndPuncttrue
\mciteSetBstMidEndSepPunct{\mcitedefaultmidpunct}
{\mcitedefaultendpunct}{\mcitedefaultseppunct}\relax
\EndOfBibitem
\bibitem[Xu \latin{et~al.}(2024)Xu, Zhou, Tao, Zhong, Wang, Cao, Xia, Wang,
  Zhan, Zhang, Yu, Xu, Dong, Ren, and Zhang]{Xu2024DensityMatrix}
Xu,~L.; Zhou,~M.; Tao,~R.; Zhong,~Z.; Wang,~B.; Cao,~Z.; Xia,~H.; Wang,~Q.;
  Zhan,~H.; Zhang,~A.; Yu,~S.; Xu,~N.; Dong,~Y.; Ren,~C.; Zhang,~L.
  Resource-Efficient Direct Characterization of General Density Matrix.
  \emph{Phys. Rev. Lett.} \textbf{2024}, \emph{132}, 030201\relax
\mciteBstWouldAddEndPuncttrue
\mciteSetBstMidEndSepPunct{\mcitedefaultmidpunct}
{\mcitedefaultendpunct}{\mcitedefaultseppunct}\relax
\EndOfBibitem
\bibitem[Hofheinz \latin{et~al.}(2008)Hofheinz, Weig, Ansmann, Bialczak,
  Lucero, Neeley, O'connell, Wang, Martinis, and Cleland]{Hofheinz2008}
Hofheinz,~M.; Weig,~E.; Ansmann,~M.; Bialczak,~R.~C.; Lucero,~E.; Neeley,~M.;
  O'connell,~A.; Wang,~H.; Martinis,~J.~M.; Cleland,~A. Generation of {Fock}
  states in a superconducting quantum circuit. \emph{Nature} \textbf{2008},
  \emph{454}, 310\relax
\mciteBstWouldAddEndPuncttrue
\mciteSetBstMidEndSepPunct{\mcitedefaultmidpunct}
{\mcitedefaultendpunct}{\mcitedefaultseppunct}\relax
\EndOfBibitem
\bibitem[Gertler \latin{et~al.}(2023)Gertler, van Geldern, Shirol, Jiang, and
  Wang]{Gertler2023}
Gertler,~J.~M.; van Geldern,~S.; Shirol,~S.; Jiang,~L.; Wang,~C. Experimental
  Realization and Characterization of Stabilized Pair-Coherent States.
  \emph{PRX Quantum} \textbf{2023}, \emph{4}, 020319\relax
\mciteBstWouldAddEndPuncttrue
\mciteSetBstMidEndSepPunct{\mcitedefaultmidpunct}
{\mcitedefaultendpunct}{\mcitedefaultseppunct}\relax
\EndOfBibitem
\bibitem[Chakram \latin{et~al.}(2022)Chakram, He, Dixit, Oriani, Naik, Leung,
  Kwon, Ma, Jiang, and Schuster]{Chakram2022multimode}
Chakram,~S.; He,~K.; Dixit,~A.~V.; Oriani,~A.~E.; Naik,~R.~K.; Leung,~N.;
  Kwon,~H.; Ma,~W.-L.; Jiang,~L.; Schuster,~D.~I. Multimode photon blockade.
  \emph{Nat. Phys.} \textbf{2022}, \emph{18}, 879\relax
\mciteBstWouldAddEndPuncttrue
\mciteSetBstMidEndSepPunct{\mcitedefaultmidpunct}
{\mcitedefaultendpunct}{\mcitedefaultseppunct}\relax
\EndOfBibitem
\bibitem[Vlastakis \latin{et~al.}(2013)Vlastakis, Kirchmair, Leghtas, Nigg,
  Frunzio, Girvin, Mirrahimi, Devoret, and Schoelkopf]{Vlastakis2013}
Vlastakis,~B.; Kirchmair,~G.; Leghtas,~Z.; Nigg,~S.~E.; Frunzio,~L.;
  Girvin,~S.~M.; Mirrahimi,~M.; Devoret,~M.~H.; Schoelkopf,~R.~J.
  Deterministically encoding quantum information using 100-photon
  {Schr{\"o}dinger} cat states. \emph{Science} \textbf{2013}, \emph{342},
  607\relax
\mciteBstWouldAddEndPuncttrue
\mciteSetBstMidEndSepPunct{\mcitedefaultmidpunct}
{\mcitedefaultendpunct}{\mcitedefaultseppunct}\relax
\EndOfBibitem
\bibitem[Lanyon \latin{et~al.}(2010)Lanyon, Whitfield, Gillett, Goggin,
  Almeida, Kassal, Biamonte, Mohseni, Powell, Barbieri, \latin{et~al.}
  others]{Lanyon2010}
Lanyon,~B.~P.; Whitfield,~J.~D.; Gillett,~G.~G.; Goggin,~M.~E.; Almeida,~M.~P.;
  Kassal,~I.; Biamonte,~J.~D.; Mohseni,~M.; Powell,~B.~J.; Barbieri,~M.,
  \latin{et~al.}  Towards quantum chemistry on a quantum computer. \emph{Nature
  Chem.} \textbf{2010}, \emph{2}, 106\relax
\mciteBstWouldAddEndPuncttrue
\mciteSetBstMidEndSepPunct{\mcitedefaultmidpunct}
{\mcitedefaultendpunct}{\mcitedefaultseppunct}\relax
\EndOfBibitem
\bibitem[Arrazola \latin{et~al.}(2021)Arrazola, Jahangiri, Delgado, Ceroni,
  Izaac, Sz{\'a}va, Azad, Lang, Niu, Di~Matteo, \latin{et~al.}
  others]{Arrazola2021}
Arrazola,~J.~M.; Jahangiri,~S.; Delgado,~A.; Ceroni,~J.; Izaac,~J.;
  Sz{\'a}va,~A.; Azad,~U.; Lang,~R.~A.; Niu,~Z.; Di~Matteo,~O., \latin{et~al.}
  Differentiable quantum computational chemistry with {PennyLane}. \emph{arXiv
  preprint arXiv:2111.09967} \textbf{2021}, \relax
\mciteBstWouldAddEndPunctfalse
\mciteSetBstMidEndSepPunct{\mcitedefaultmidpunct}
{}{\mcitedefaultseppunct}\relax
\EndOfBibitem
\bibitem[Goss \latin{et~al.}(2022)Goss, Morvan, Marinelli, Mitchell, Nguyen,
  Naik, Chen, J{\"u}nger, Kreikebaum, Santiago, Wallman, and Siddiqi]{Goss2022}
Goss,~N.; Morvan,~A.; Marinelli,~B.; Mitchell,~B.~K.; Nguyen,~L.~B.;
  Naik,~R.~K.; Chen,~L.; J{\"u}nger,~C.; Kreikebaum,~J.~M.; Santiago,~D.~I.;
  Wallman,~J.~J.; Siddiqi,~I. High-fidelity qutrit entangling gates for
  superconducting circuits. \emph{Nat. Commun.} \textbf{2022}, \emph{13},
  7481\relax
\mciteBstWouldAddEndPuncttrue
\mciteSetBstMidEndSepPunct{\mcitedefaultmidpunct}
{\mcitedefaultendpunct}{\mcitedefaultseppunct}\relax
\EndOfBibitem
\end{mcitethebibliography}
